\documentclass[useAMS,usenatbib]{mn2e}

\newif\ifAMStwofonts

\usepackage{graphicx}
\usepackage{longtable}
\usepackage[table,dvipsnames]{xcolor}
\usepackage[flushleft]{threeparttable}

\title[On the nuclear orientation of AGN]
      {Are there reliable methods to estimate the nuclear orientation of Seyfert galaxies?}
\author[F. Marin]
      {F.~Marin\thanks{frederic.marin@astro.unistra.fr} \\
       Observatoire Astronomique de Strasbourg, Universit\'e de Strasbourg, CNRS, UMR 7550,\\
       11 rue de l'Universit\'e, 67000 Strasbourg, France}
\date{Accepted 2016 May 10. 
      Received 2016 May 10; 
      in original form 2016 February 29}

\pagerange{\pageref{firstpage}--\pageref{lastpage}}
\pubyear{2016}

\begin{document}

\maketitle

\label{firstpage}

\begin{abstract}
Orientation, together with accretion and evolution, is one of the three main drivers in the Grand Unification of Active 
Galactic Nuclei (AGN). Being unresolved, determining the true inclination of those powerful sources is always difficult 
and indirect, yet it remains a vital clue to apprehend the numerous, panchromatic and complex spectroscopic features we 
detect. There are only a hundred inclinations derived so far; in this context, can we be sure that we measure the true 
orientation of AGN? To answer this question, four methods to estimate the nuclear inclination of AGN are investigated 
and compared to inclination-dependent observables (hydrogen column density, Balmer linewidth, optical polarization, and 
flux ratios within the IR and relative to X-rays). Among these orientation indicators, the method developed by Fisher, 
Crenshaw, Kraemer et al., mapping and modeling the radial velocities of the [O~{\sc iii}] emission region in AGN, is the 
most successful. The [O~{\sc iii}]-mapping technique shows highly statistically significant correlations at $>$~95\% 
confidence level for rejecting null hypothesis for all the test cases. Such results confirm that the Unified Model is 
correct at a scale ranging from kiloparsec to a fraction of a parsec. However, at a radial distance less than 0.01~pc 
from the central black hole, warps and misalignments may change this picture.
\end{abstract}


\section{Introduction}
\label{Intro}
The question of the geometric properties (positions, orientations and shapes) of cosmic objects concerns every field of astrophysics, 
from galaxies to gas/dust filaments, binary stars to planet rings, accreting stellar mass black holes to ionized protoplanetary disks 
(proplyds). To figure out the composition, morphology and kinematics of a source, determining its three-dimensional geometry with respect 
to the observer is mandatory. It leads to understanding complex line profiles such as double-peaked Balmer lines in AGN \citep{Storchi1997},
the absolute orientation of the binary orbit of extremely massive stars \citep{Madura2011}, or the structure of Keplerian disks around 
classical Be stars \citep{Carciofi2006,Carciofi2008}. Without measuring and understanding the importance of inclination, dilemma such 
as the apparent superluminal (faster-than-light) motion in quasi-stellar objects (see e.g., \citealt{Porcas1983}) would still hold. 
\citet{Rees1966} was the first to predict the possibility of superluminal motion in quasars, a result of high bulk Lorentz factor 
jets viewed at angles very close to the line-of-sight, i.e., blazars. This effect has been detected and studied thoroughly afterwards, 
and the importance of Doppler orientation bias was highlighted by \citet{Orr1982}.

Measuring the inclination of the observed target is important but not always easy nor direct. The powerful radiation processes
occurring in quasars \citep{Mortlock2011}, galaxies \citep{Zitrin2015}) and Gamma-ray bursts \citep{Salvaterra2009} allow the detection
of distant astronomical sources but their internal structure is almost completely unresolved. Focusing on the most stable emitters (quasars), 
the host galaxy around their central luminous core may be detected, but the host galaxy's inclination does not necessarily correspond to 
the inclination of the nucleus \citep{Schmitt1997}. In the nearby Universe, it was found that type-1 AGN preferentially reside in pole-on 
on galaxies, but type-2 systems presumably reside in galaxies with a random orientation \citep{Keel1980,Simcoe1997,Kinney2000}, even though 
optically selected type-2 AGN samples tend to avoid edge-on systems \citep{Maiolino1995,Lagos2011}. Historically, it is possible to estimate 
whether the AGN we detect are type-1 or type-2 objects through optical classification, the difference relying on the presence/absence of 
broad Balmer lines in the total flux \citep{Osterbrock1977}. A type-2 lacks those broad emission lines due to dust obscuration along the 
observer's line-of-sight; the system is most probably seen at an equatorial viewing angle. Optical polarization measurements successfully 
confirmed this hypothesis with the detection of broad Balmer lines in polarized flux, indicating that type-2 Seyfert galaxies are in fact 
type-1s seen at a specific angle \citep{Miller1983,Antonucci1985}, but the exact inclination is almost impossible to determine. 

In order to shed light on the growth mechanism of black holes, the physical condition of the early Universe and the formation of galaxies, 
it is vital to understand the true nature of AGN, which cannot be achieved without the prior knowledge of \textit{how} we see them
\citep{Shen2014}. Misclassification of an AGN type can lead to false interpretations of the physics that govern its internal region \citep{Woo2014}. 
As an example, the presence of equatorial structures around a supermassive black hole (SMBH) can be tested thanks to their spectroscopic 
signatures in the X-ray, ultraviolet (UV), optical and near-infrared (IR) bands. If their geometry is similar to a disk, their
observed line emission should be proportional to the cosine of the disk inclination angle with respect to the observer's line-of-sight 
(see, e.g., \citealt{Wills1986}). This is a potential method to extract the true inclination of an AGN through their emission line 
properties, but one has to be careful as not all emission lines correlate with orientation. The Boroson \& Green eigenvector~1 
stipulates that the dominant source of variation in the observed properties of low-redshift quasi-stellar objects (QSO) emission lines 
is a physical parameter that is not always driven by the viewing angle \citep{Boroson1992}. In particular, the Boroson \& Green 
eigenvector~1 is anticorrelated with the Fe~{\sc ii}~$\lambda$4570 strength (equivalent width and Fe~{\sc ii}/H$\beta$ ratio), 
anticorrelated with the blue asymmetry of the H$\beta$ line, but correlated with [O~{\sc iii}]~$\lambda$5007 strength (luminosity 
and peak) and H$\beta$ linewidth.

Estimating the true inclination (with reasonable uncertainty) of Seyfert galaxies and quasars is thus challenging, but necessary, 
to progress beyond the basic assertions of the Unified Model of AGN, such as it was proposed by \citet{Lawrence1991}, \citet{Antonucci1993}, 
and \citet{Urry1995}. To achieve this, the identification of a good orientation indicator in quasars is crucial. There are potential 
indicators to estimate the viewing angle of radio-loud objects \citep{VanGorkom2015}, such as the radio-core dominance parameter 
\citep{Orr1982}, the continuum optical flux density \citep{Wills1995}, or the luminosity of the narrow-line region (NLR, \citealt{Rawlings1991}). 
However, none of these techniques can be applied to radio-quiet AGN as they intrinsically miss a relativistic, beamed, parsec scale jet. 

The aim of this paper is to explore the diverse techniques used in literature to estimate the nuclear inclination of radio-quiet Seyfert 
galaxies and to identify the best orientation indicator. To achieve this, the catalog of inclinations used in the sample is described in 
Sect.~\ref{Comp}, together with the four main Seyfert inclination indicators. The key question is how well a candidate inclination indicator 
separates the Seyfert 1s from the Seyfert 2s. This is investigated in Sect.~\ref{Exploiting}, where the orientation indicators are 
compared with inclination-dependent observables. The existence of statistically significant correlations is investigated using efficient 
rank correlation statistics; the evidence for the Unified Model is very strong, and while it does not require for an inclination indicator 
to separate the two types perfectly, a near-perfect separation is unlikely to be a coincidence, suggesting a very good indicator. Results 
and limits are then discussed in Sect.~\ref{Discussion}, and Sect.~\ref{Conclusions} concludes the paper by listing the most important 
outcomes of the comparisons.

\section{Compiling the catalog}
\label{Comp}
Since the goal of this paper is to achieve a comparison between different orientation indicators, the existence of quantitative evaluations
of AGN inclinations is the main driver of the selection process. Only Seyfert galaxies with estimated inclinations were selected, regardless 
of their redshift, black hole mass, bolometric luminosity, accretion rate or any spectroscopic features.

\subsection{Inclinations from literature}
\label{Comp:Inclination}
Roughly 161 AGN inclination values were found during data mining, among which 37 are duplicates. In total, 124 unique radio-quiet objects 
have an inclination estimation reported in Tab.~\ref{Table:Data}\footnote{Due to their multiple-pages length, the tables compiling the various 
parameters of the sample are shifted to the end of this paper, after the references section.}. This table only accounts for one inclination
per AGN; in the case of multiple values, the most probable orientation angle was kept and duplicates were rejected according to two criteria: 
1) if the uncertainty of the inclination is larger than 25$^\circ$, the inclination is discarded since it would cover the whole permitted range 
of inclination for a given AGN type, and 2) in the case of two different inclinations referring to the same object, the value with the uncertainty 
encompassing the less constraint inclination was chosen in order to be conservative. The details of the selected/rejected inclinations are 
given in \citet{Marin2014}. 

Almost all these inclination values belong to one of the four main categories of orientation indicators that have emerged while collecting these 
data. They are classified based on the different mechanisms they use to extract an orientation parameter from their sets of observations and are 
listed in Tab.~\ref{Table:VEL}, \ref{Table:X}, \ref{Table:IR}, and \ref{Table:NLR}. The duplicates will be discussed in Sect.~\ref{Discussion:Duplicates},
while the four classes of inclination indicators are reviewed in the following subsections. A fifth class, gathering all the inclinations emerging 
from singular techniques that were employed in isolated papers, is also mentioned for completeness.

\subsubsection{Method~I: ``M-$\sigma$''}
\label{Comp:Inclination:BH}
The M-$\sigma$ relationship (or M$_{\rm BH}$-$\sigma$) is an empirical, significantly tight, correlation between the velocity dispersion $\sigma$ 
measured in the bulb of a galaxy and the mass of the supermassive black hole situated at the center of this galaxy found \citep{Gebhardt2000,Ferrrarese2000}. 
In a limited number of AGN, the SMBH mass can be retrieved thanks to reverberation mapping techniques \citep[e.g.][]{Blandford1982,Wandel1999,Bentz2006,Bentz2010}, 
where the mass of the compact source can be estimated from the broad line region (BLR) size and the characteristic velocity of low ionization, 
broad, emission lines (LIL, such as H$\alpha$, H$\beta$, H$\gamma$, He~{\sc i} or He~{\sc ii}). This velocity, determined by the full width at 
half maximum (FWHM) of the emission line, is strongly dependent on the inclination of the BLR. Thus, by assuming a Keplerian motion of the LIL 
BLR and a similar M-$\sigma$ relationship between Seyfert-1s and regular galaxies, \citet{Wu2001} and \citet{Zhang2002} estimated the orientation 
angles $i$ for a variety of type-1 AGN with known black hole masses and measured FWHM. This resulted in 19 unique inclination estimations, reported 
in Tab.~\ref{Table:VEL}. Note that the technique, requiring the measurement of the FWHM of low ionization broad emission lines, is intrinsically 
limited to type-1 objects.

\subsubsection{Method~II: ``X-ray''}
\label{Comp:Inclination:X}
X-ray spectroscopy is a valuable tool that can probe the few inner gravitational radii around a singularity. In AGN, an accretion disk 
around the SMBH \citep{Pringle1972,Shakura1973,Novikov1973} acts like a mirror reflecting/absorbing part of the X-ray radiation that 
is isotropically produced by a hot corona situated above the disk. The corona up-scatters thermally emitted, ultraviolet (UV), 
disk photons to higher energies \citep{Haardt1991,Haardt1993}, producing the observed power-law spectrum. The intense gravitational field 
around the potential well will affect the re-emitted disk fluorescent emission by broadening the lines due to Doppler effects and gravitational 
plus transverse redshifts. It will result in a strong asymmetrically blurred emission feature at 6.4~keV, associated with iron fluorescence 
in near-neutral material \citep{Reeves2006}. Since this fluorescent line is emitted in a disk, its line width will be characteristic of the 
inclination of the system. By applying X-ray spectral fits accounting for a non-Euclidean space time, it becomes possible to constrain the 
orientation of the accretion disk, which is usually tied to the AGN inclination \citep{Nandra1997,Nandra2007}. However, this technique 
is intrinsically limited to bright AGN. The compilation of X-ray fitted AGN inclination results in 54 unique objects reported in 
Tab.~\ref{Table:X}.

\subsubsection{Method~III: ``IR''}
\label{Comp:Inclination:IR}
AGN act like calorimeters \citep{Antonucci2012}. By absorbing the optical and UV light thermally produced by the accretion disk, the 
dust embedding the nuclear region will be heated and will re-emit the stored energy at larger wavelengths, principally in the mid-infrared 
(MIR). The fact that AGN are surrounded by an asymmetrically distributed amount of dust grains, with a predominance of dust along the 
equatorial region (the seminal dusty torus\footnote{The real morphology of the circumnuclear region, either compact, clumpy or windy is 
not of interest here. The only important characteristic of this obscuring region is that it is close to the equatorial plane.}), 
allows a determination of the inclination of the MIR emitting region by looking at the amount of re-emitted radiation and the spectral 
features in the NIR and MIR spectra \citep[e.g.][]{Mor2009,Alonso2011,Sales2011,Ruschel2014}. To achieve this, clumpy torus models 
are applied to observed data in order to retrieve several characteristics (such as the spectral energy distribution, SED, or emission and 
absorption features). Detailed fitting procedures, such as masking the emission lines and the telluric band region \citep{Ruschel2014}, or 
implementing more complex reprocessing geometries \citep{Mor2009}, also help to better estimate the inclination of the torus. In total, 
37 individual objects have been observed and modeled, and the final compilation of inclination values is listed in Tab.~\ref{Table:IR}.

\subsubsection{Method~IV: ``NLR''}
\label{Comp:Inclination:NLR}
NASA/ESA Hubble Space Telescope observations of the radial velocities of the [O~{\sc iii}]-emitting gas in a sample of nearby Seyfert galaxies
(e.g. NGC~4151 by \citealt{Crenshaw2000c} or NGC~1068 by \citealt{Crenshaw2000b}) have shown that the kinematics of the extended narrow 
line region (NLR) of AGN tend to be dominated by radial outflows in the approximate morphology of an hourglass. By matching several 
observed radial velocities to their kinematic model, \citet{Crenshaw2000} postulated that the orientation of the AGN nuclei could be
determined from kinematic mapping. This work was undertaken by \citet{Fischer2013}, who used [O~{\sc iii}] imaging and long-slit spectra
of 53 Seyfert galaxies to extract the inclination of the bicone axis, and hence of the obscuring torus. Using uniform, hollow, bi-conical 
models with sharp edges, \citet{Fischer2013} found that out of the 53 AGN they observed, 17 objects had clear enough signatures to retrieve 
their potential inclination. Those 17 objects are listed in Tab.~\ref{Table:NLR}. Note that, to be able to retrieve an inclination,
this technique requires bright, nearby AGN with resolved NLR structures.

\subsubsection{Method~V: ``Other''}
\label{Comp:Inclination:Other}
Under the label ``other'' are gathered all the techniques used by a variety of authors to estimate the inclination from one, seldomly 
more, object(s). It includes spectropolarimetric observations and modeling of highly polarized type-1 objects such as ESO~323-G077 
\citep{Schmid2003} or Fairall~51 \citep{Schmid2001}, fits of the observed broad, double-peaked Balmer emission lines in NGC~1097
\citep{Storchi1997} using an eccentric accretion ring model, and several other techniques that are detailed in \citet{Marin2014}. 
The inclinations derived from this mix of approaches are included in Tab.~\ref{Table:Data} and contain 25 Seyfert galaxies. Since 
those orientation indicators do not share a common method, the inclinations listed as ``other'' will only be used in the global 
sample. 

\begin{figure*}
    \centering
    \includegraphics[trim = 0mm 0mm 0mm 0mm, clip, width=15cm]{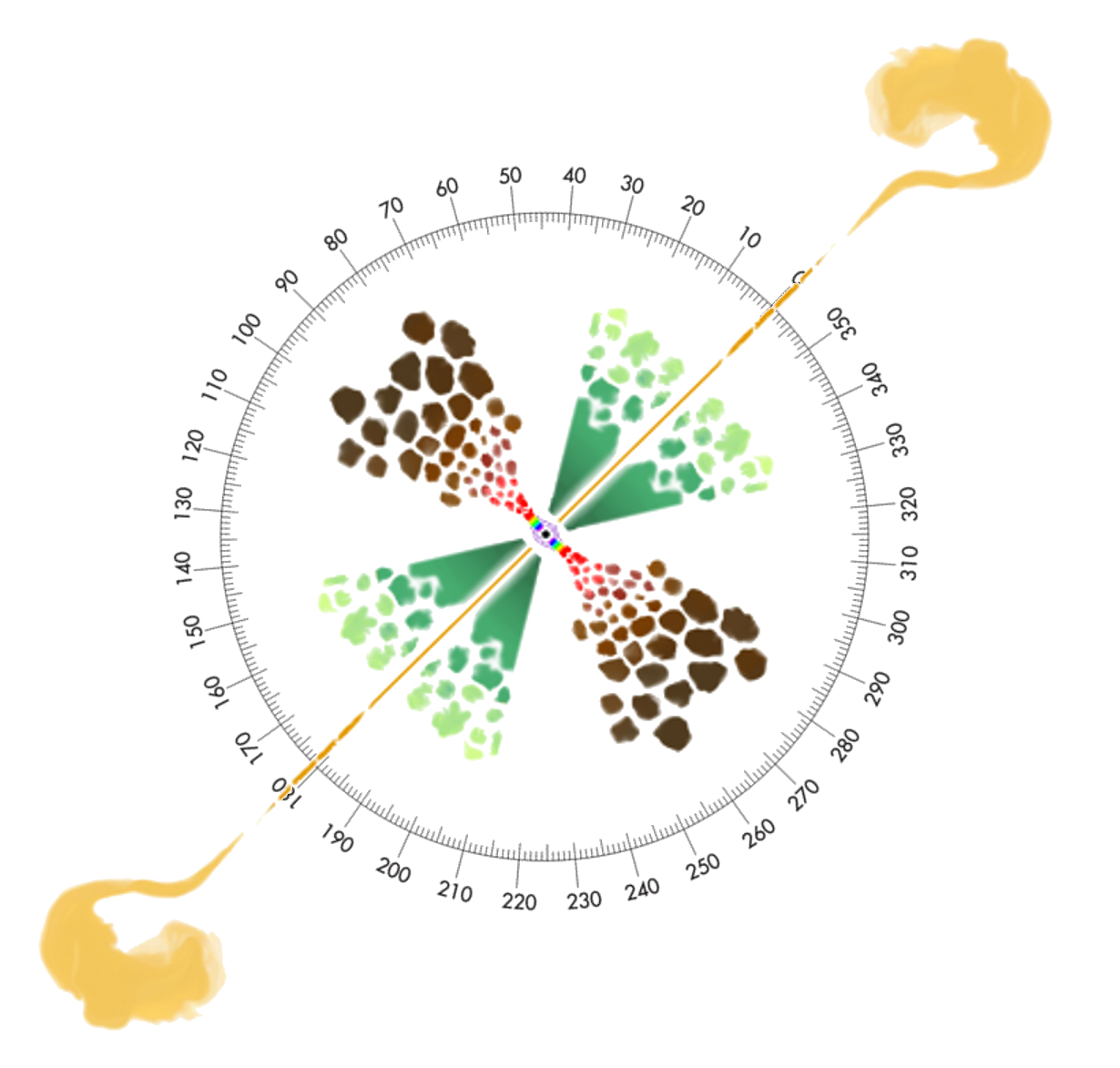}
    \caption{Unscaled sketch of the AGN unification theory. 
	     A type-1 AGN is seen at inclinations 0 -- 60$^\circ$
	     while a type-2 AGN is seen at 60 -- 90$^\circ$, 
	     approximately. Color code: the central supermassive 
	     black hole is in black, the surrounding X-ray corona is
	     in violet, the multi-temperature accretion disk is shown 
	     with the color pattern of a rainbow, the BLR is in red and 
	     light brown, the circumnuclear dust in dark brown, the polar
	     ionized winds in dark green and the final extension of 
	     the NLR in yellow-green. A double-sided, kilo-parsec jet 
	     is added to account for radio-loud AGN. Details about 
	     the composition and spatial scales are given in the 
	     text.}
    \label{Fig:Scheme}            
\end{figure*}

~\

In total, there are four main indicators: M-$\sigma$ relationship, X-ray reflection spectroscopy, IR modeling, and [O~{\sc iii}]-mapping. 
Interestingly these four methods focus on different and distinct physical scales. In increasing radial distance from the central 
supermassive black hole: 1) the X-ray method probes the inclination of the inner part of the accretion disk at a couple of gravitational radii 
\citep{Dovciak2015}, 2) the M-$\sigma$ indicator focuses on the BLR emission, spanning from 10$^{-4}$ to 10$^{-1}$~pc, \citep{Hansen2014}, 3) the IR 
method models the dusty torus whose radius is estimated between 10$^{-1}$ and 10$^{1}$~pc \citep{Burtscher2013}, and 4) the [O~{\sc iii}] kinematic 
modeling of the NLR probes physical scales ranging from a parsec up to hundreds of parsecs \citep{Crenshaw2000}. A color-coded sketch of the Unified 
Model is presented in Fig.~\ref{Fig:Scheme} in order to show the different AGN components targeted by those inclination indicators. It becomes 
clear that the concept of a global AGN orientation angle is a complicated matter, as the four indicators are meant to measure the inclination of 
separate components. In the following, the reader is cautioned to remember that the investigations are intended to see if the inclination derived for 
a given region (X-ray: innermost AGN components, M-$\sigma$: BLR, IR fitting: torus, and [O~{\sc iii}]-mapping: NLR) can be valid over a wider range
of physical scales.

\begin{figure*}
    \begin{center}
      \begin{tabular}{cc}
	\includegraphics[trim = 0mm 0mm 0mm 0mm, clip, width=8cm]{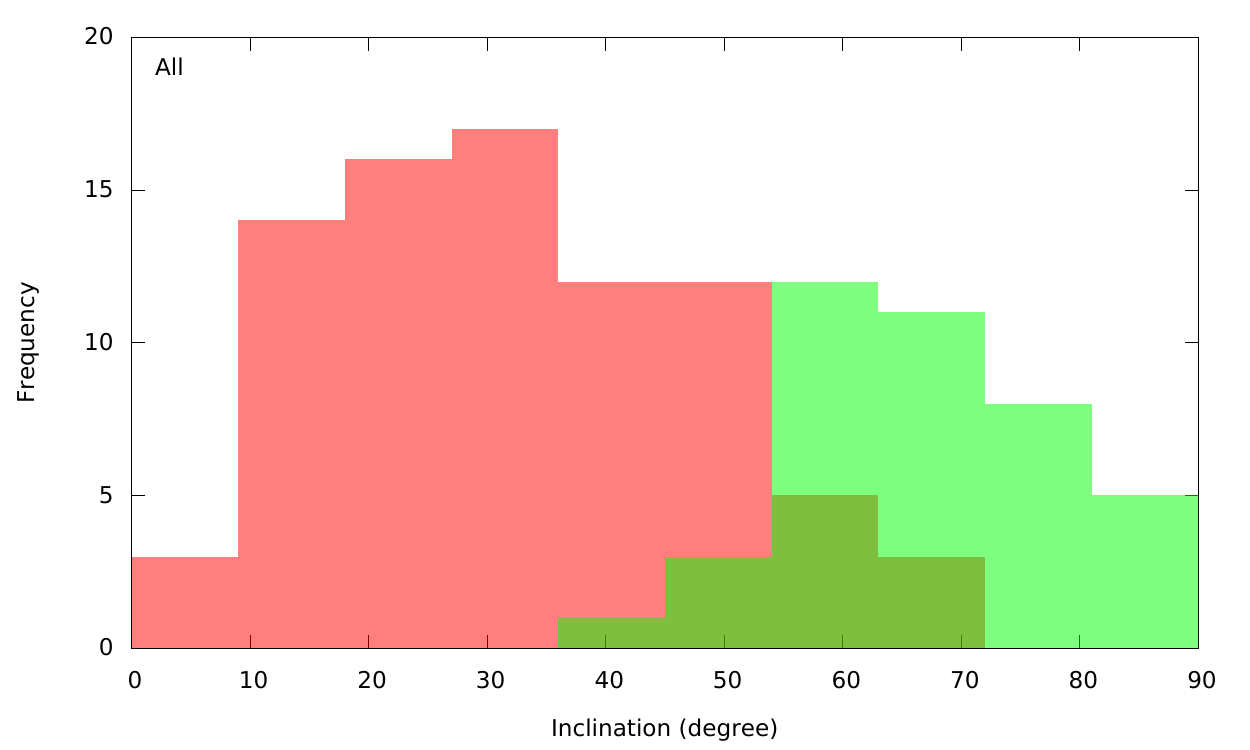} \\
      \end{tabular}
      \begin{tabular}{cc}
	\includegraphics[trim = 0mm 0mm 0mm 0mm, clip, width=8cm]{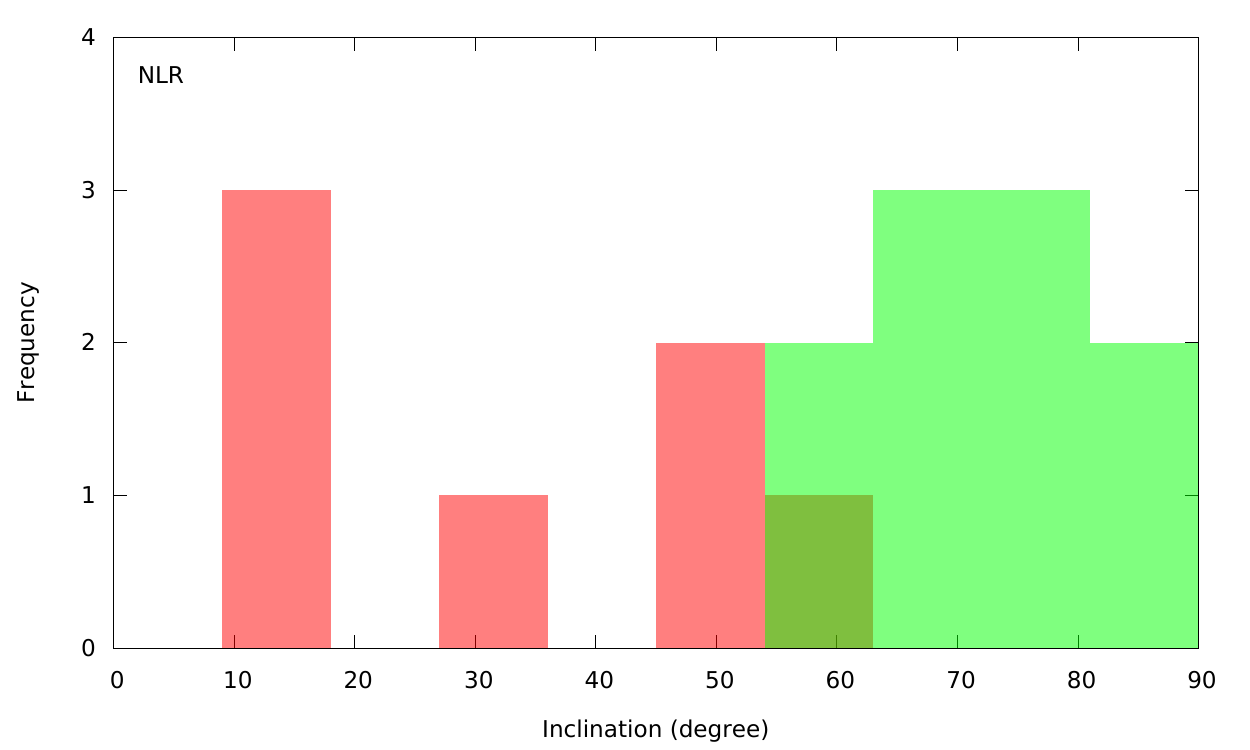} & \includegraphics[trim = 0mm 0mm 0mm 0mm, clip, width=8cm]{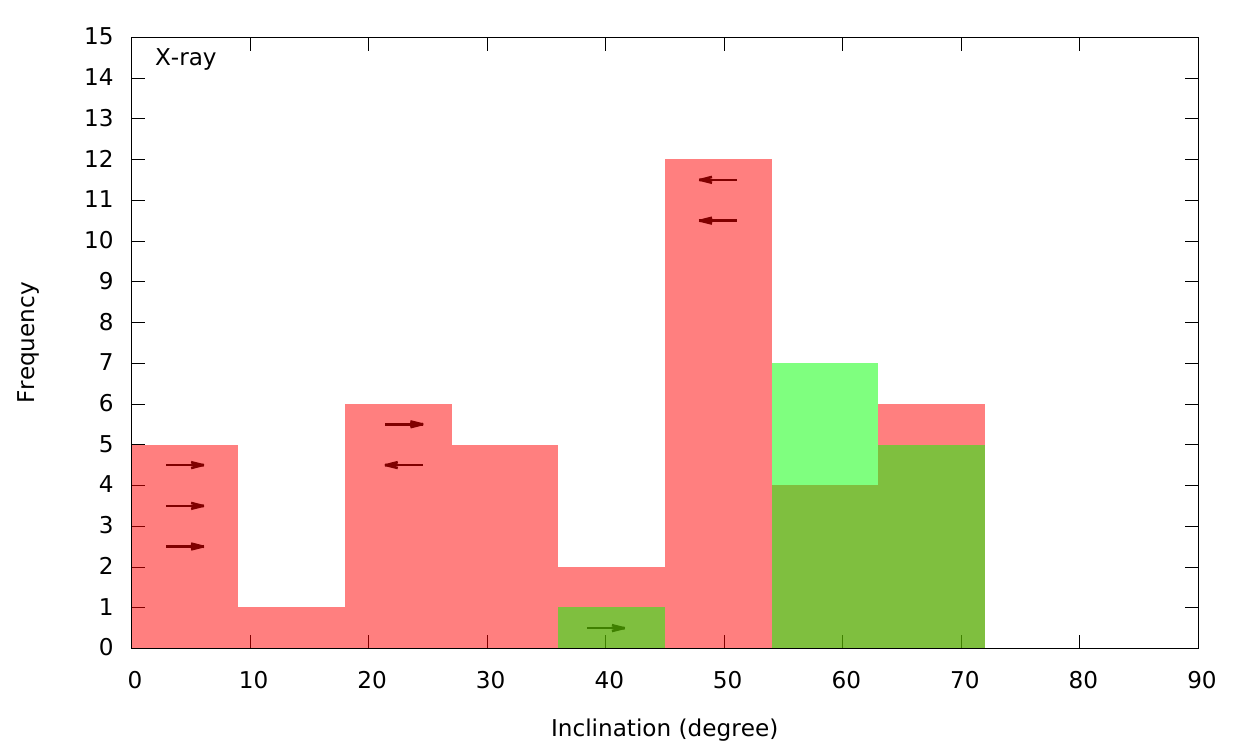} \\
      \end{tabular}
      \begin{tabular}{cc}
	\includegraphics[trim = 0mm 0mm 0mm 0mm, clip, width=8cm]{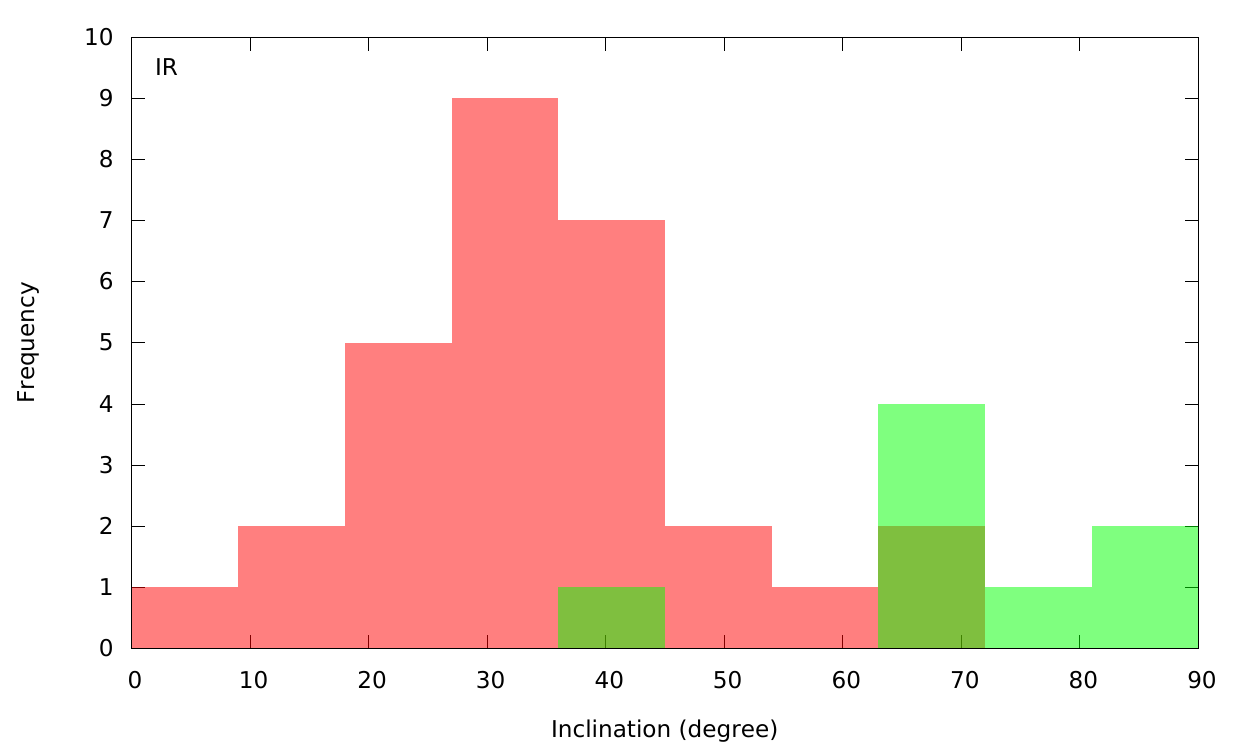} & \includegraphics[trim = 0mm 0mm 0mm 0mm, clip, width=8cm]{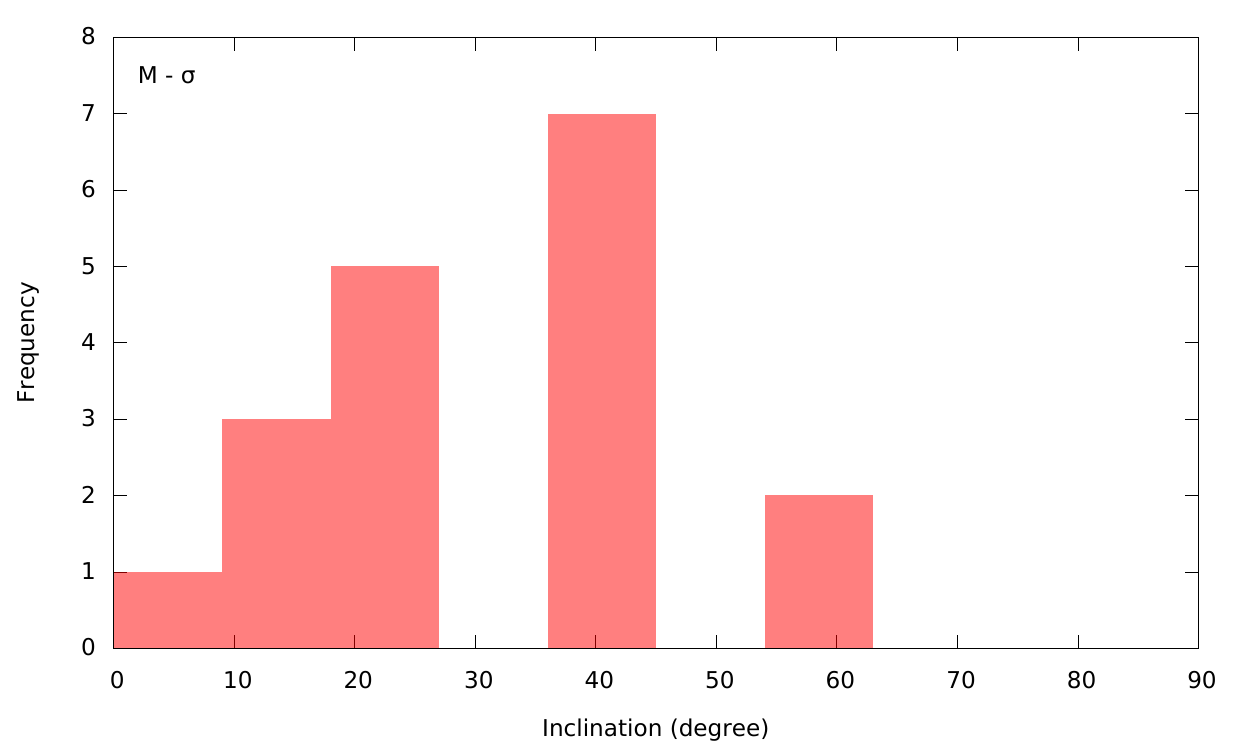} \\
      \end{tabular}
    \end{center}
    \caption{Frequency distributions of AGN inclinations according to 
	     the method used to retrieve the nuclear orientation. The top 
	     histogram is the cumulated histogram of all methods,
	     minus inclinations that did not pass the selection
	     criteria. Middle-left: NLR method; middle-right: X-ray method;
	     bottom-left: IR method; bottom-right: M-$\sigma$ method. 
	     Type-1 AGN are in red, type-2s in green. The dark-green
	     color results from the superposition of both type-1 and 
	     type-2 Seyfert galaxies.}
    \label{Fig:Histograms}%
\end{figure*}

\subsection{Distribution of inclinations}
\label{Comp:Inclination:Histograms}

The final distributions of inclinations are shown in Fig.~\ref{Fig:Histograms}. The top figure presents the histogram of the full sample 
of 124 Seyfert galaxies, including orientation measurements from all the different methods. The four other histograms show the distribution 
of inclinations per orientation indicator (middle-left: NLR, middle-right: X-ray, bottom-left: IR, bottom-right: M-$\sigma$). Type-1 AGN 
are shown in red, type-2s in green\footnote{For the remainder of this paper, it is assumed that Seyfert-1s show some evidence of a BLR, 
therefore all sub types (type 1, 1.2, 1.5, 1.8 and 1.9) belong to the type-1 category. Type-2s are AGN without any sign of BLR in total 
flux spectra.}. This graphical ordering and color-coding will be the same for all the following figures comparing the different inclination 
indicators. 

The distribution of inclinations in the whole sample shows a lack of extreme type-1 objects, as expected from the Unified Model: if
the solid angle at which we can detect pole-on AGN is small, the observational number count should also be small. The total number of 
detected sources per solid angle increases with inclination, up to a maximum value at 25$^\circ$ -- 35$^\circ$, and the frequency 
distribution shows a constant diminution until edge-on line-of-sights. It appears that the inclinations derived from type-2 Seyferts 
do not succeed to fill their solid angle uniformly, otherwise the number count of edge-on AGN should be higher. This suggests that the 
inclination indicators might not be suited to retrieve extreme nuclear orientations. Finally, there is only a narrow band of inclinations 
where type-1 and type-2 AGN overlap. This range, extending from $i$ = 36$^\circ$ to $i$ = 72$^\circ$, corresponds to the transition 
region between the two classes of AGN, where ``changing-look'' AGN\footnote{Changing look AGN are characterized by rapid variation in 
the line-of-sight of cold absorber. These eclipses, mostly observed in X-rays, suggest that these absorbers are located on compact scales 
consistent within the inner wall of the torus, the BLR region and the outer part of the accretion disk, and seen at a line of sight that 
is grazing the circumnuclear obscuring dust.} are detected \citep[e.g.][]{Elvis2004,Risaliti2005,Matt2009}. This range of inclinations 
is consistent with the type-1/type-2 transition limits ($\ge$~45$^\circ$) found by torus-obscuration modeling of the INTEGRAL all-sky 
hard X-ray survey by \citet{Sazonov2015}, and is also consistent with optical polarimetric compilation and modeling \citep{Marin2014}, 
where a transition region between 45$^\circ$ and 60$^\circ$ was found. At first glance, the inclination properties of the global sample 
are in agreement with past deductions. 

Looking at the histograms of the four main orientation indicators, both the ``IR'' and ``M-$\sigma$'' methods are able to reproduce 
the expected number count of AGN per solid angle at type-1 inclinations, but the ``NLR'' method by \citet{Fischer2013} lacks the 
statistics to draw any conclusions. The transition region between the obscured and unobscured nuclei is at $\sim$~60$^\circ$ in the 
case of the ``NLR'' method, between 44$^\circ$ and 72$^\circ$ for the ``X-ray'' method, and at $\sim$~68$^\circ$ for the ``IR'' fitting
method. The last orientation indicator, only targeting type-1 AGN, give a lower limit of 62$^\circ$. Overall, the four methods agree 
relatively well.

\subsection{Summary of the inclination-independent characteristics of the sample}
\label{Comp:characteristics}

The AGN selection process, purely based on the existence of an orientation indicator, results in a final catalog that might be biased with 
respect to some intrinsic properties. While not directly related to the topic of inclination of Seyfert galaxies, it is necessary to investigate 
whether those characteristics are likely to bias the analysis.

\subsubsection{Redshift}
\label{Comp:characteristics:Redshift}

\begin{figure*}
    \begin{center}
      \begin{tabular}{cc}
	\includegraphics[trim = 0mm 0mm 0mm 0mm, clip, width=8cm]{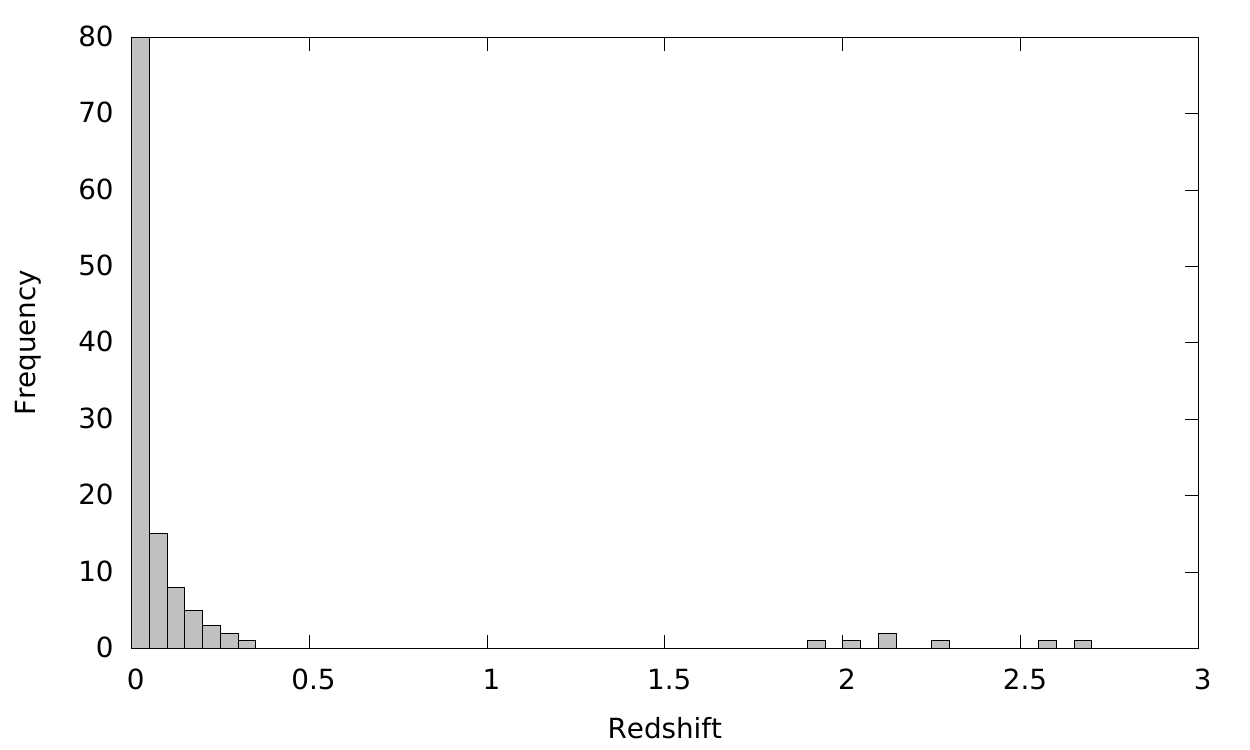} \\
      \end{tabular}
      \begin{tabular}{cc}
	\includegraphics[trim = 0mm 0mm 0mm 0mm, clip, width=8cm]{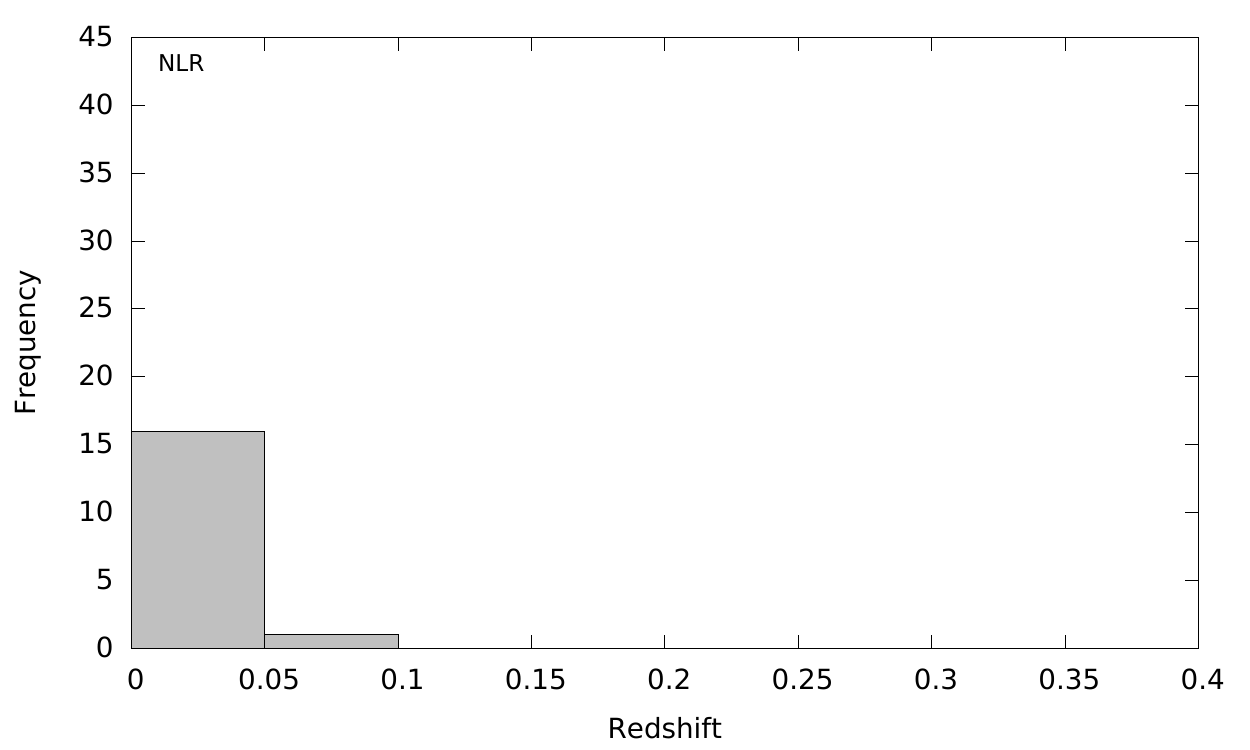} & \includegraphics[trim = 0mm 0mm 0mm 0mm, clip, width=8cm]{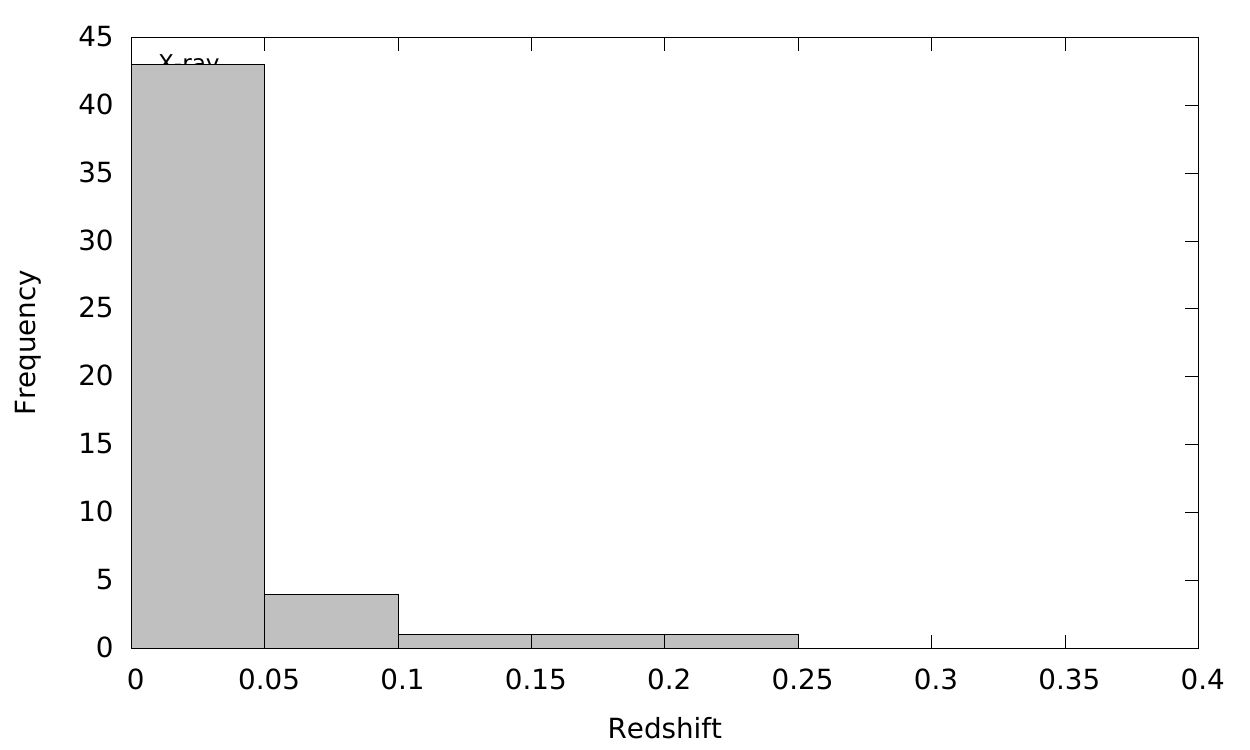} \\
      \end{tabular}
      \begin{tabular}{cc}
	\includegraphics[trim = 0mm 0mm 0mm 0mm, clip, width=8cm]{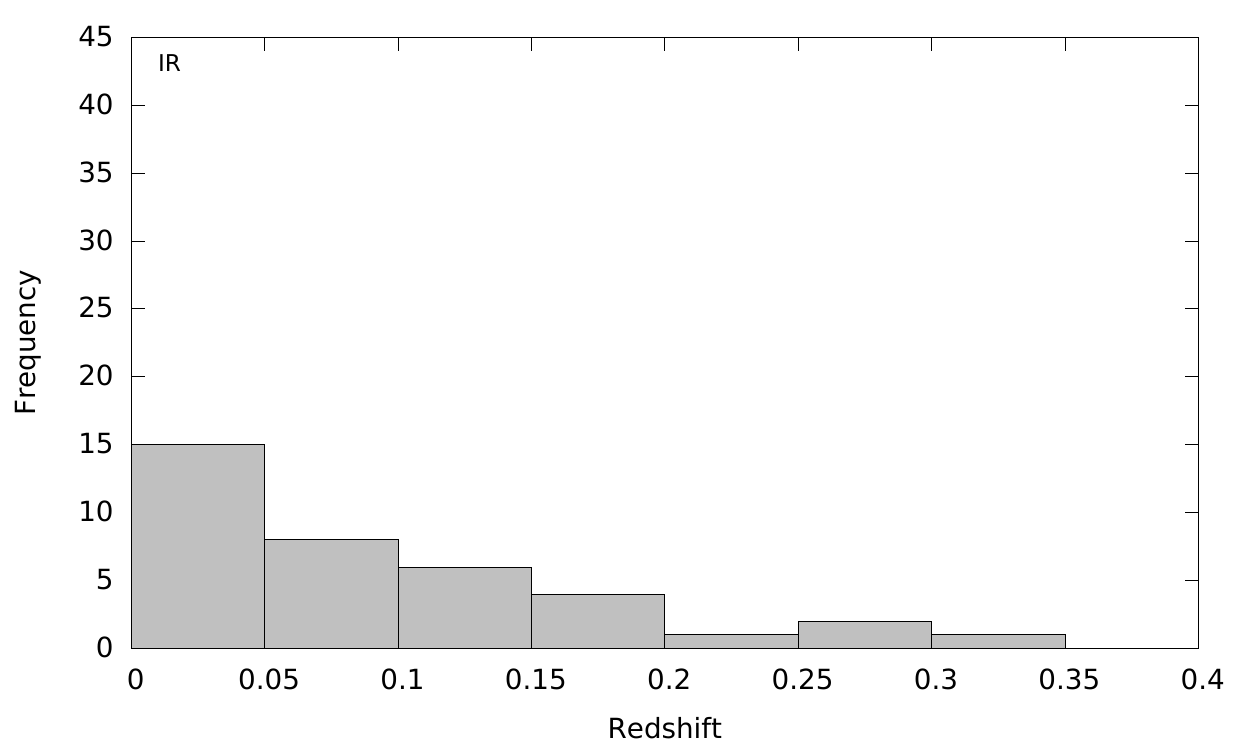} & \includegraphics[trim = 0mm 0mm 0mm 0mm, clip, width=8cm]{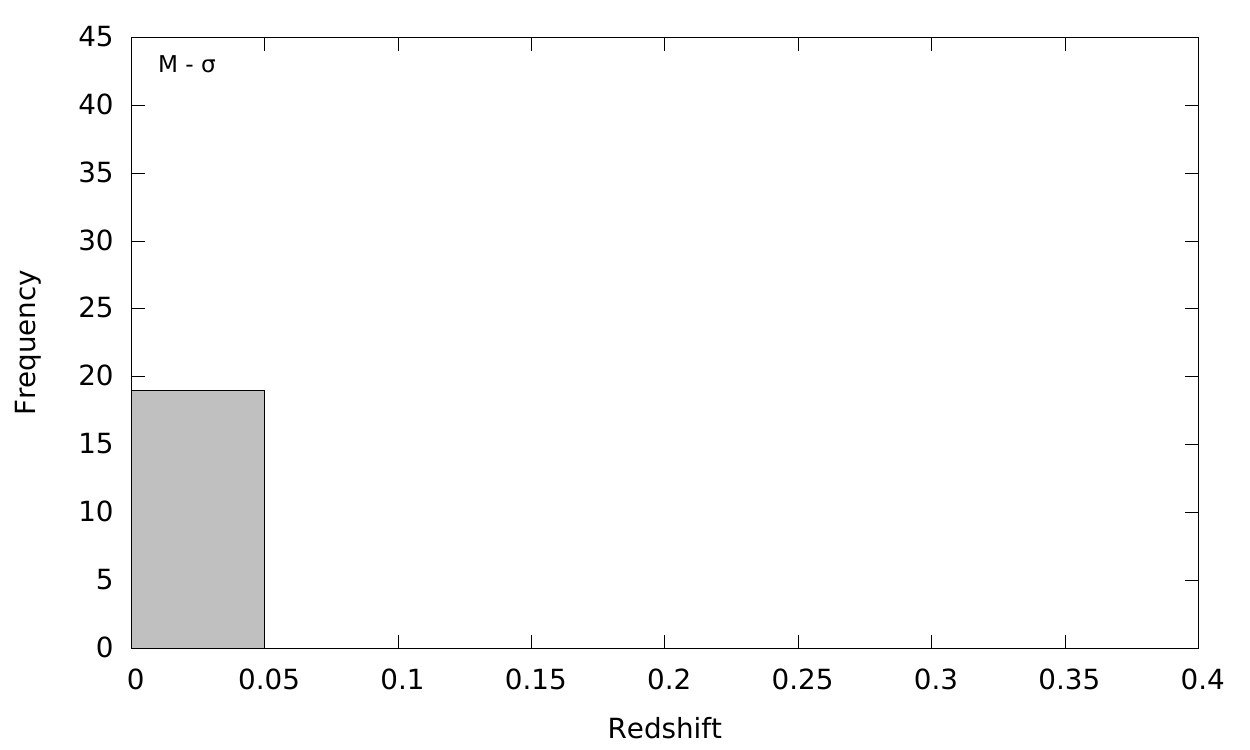} \\
      \end{tabular}
    \end{center}
    \caption{Frequency distribution of the redshifts of the sample. Legend is the same
	     as in Fig.~\ref{Fig:Histograms}.}
    \label{Fig:Hist_z}%
\end{figure*}

Fig~\ref{Fig:Hist_z} (top) shows the redshifts of the 124 Seyfert galaxies in the global sample, and the redshifts of the four sub-catalogs.
Redshifts are taken from the NASA/IPAC Extragalactic Database\footnote{http://ned.ipac.caltech.edu/}, where the parameters for distances and cosmology 
are H$_0$= 73.0~km.s$^{-1}$.Mpc$^{-1}$, $\Omega_{\rm matter}$ = 0.27, and $\Omega_{\rm vacuum}$ = 0.73. It appears that the global sample consists 
of nearby ($z <$~0.33) AGN at 94.4\%, as well as seven broad absorption line QSO (BAL QSO, $z >$~1). All the Seyferts investigated within the framework 
of this paper, looking at four orientation indicators, are closer than $z$ = 0.35. Given the relatively close redshift range, cosmological effects 
(evolution) can be considered as negligible. None of the four AGN sub-samples show significant dependence between inclination and redshift, 
such as expected from studies of radio-loud quasars \citep{Drouart2012}.

\subsubsection{Black hole masses}
\label{Comp:characteristics:BHM}

\begin{figure*}
    \begin{center}
      \begin{tabular}{cc}
	\includegraphics[trim = 0mm 0mm 0mm 0mm, clip, width=8cm]{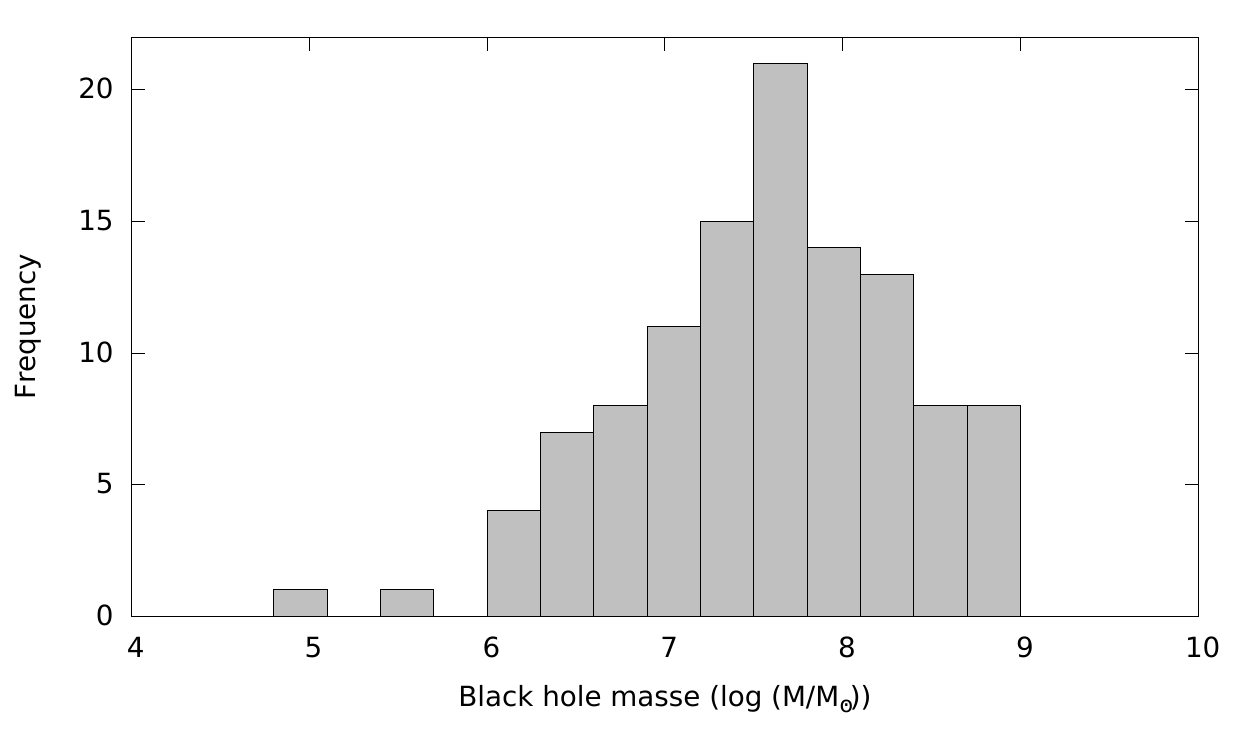} \\
      \end{tabular}
      \begin{tabular}{cc}
	\includegraphics[trim = 0mm 0mm 0mm 0mm, clip, width=8cm]{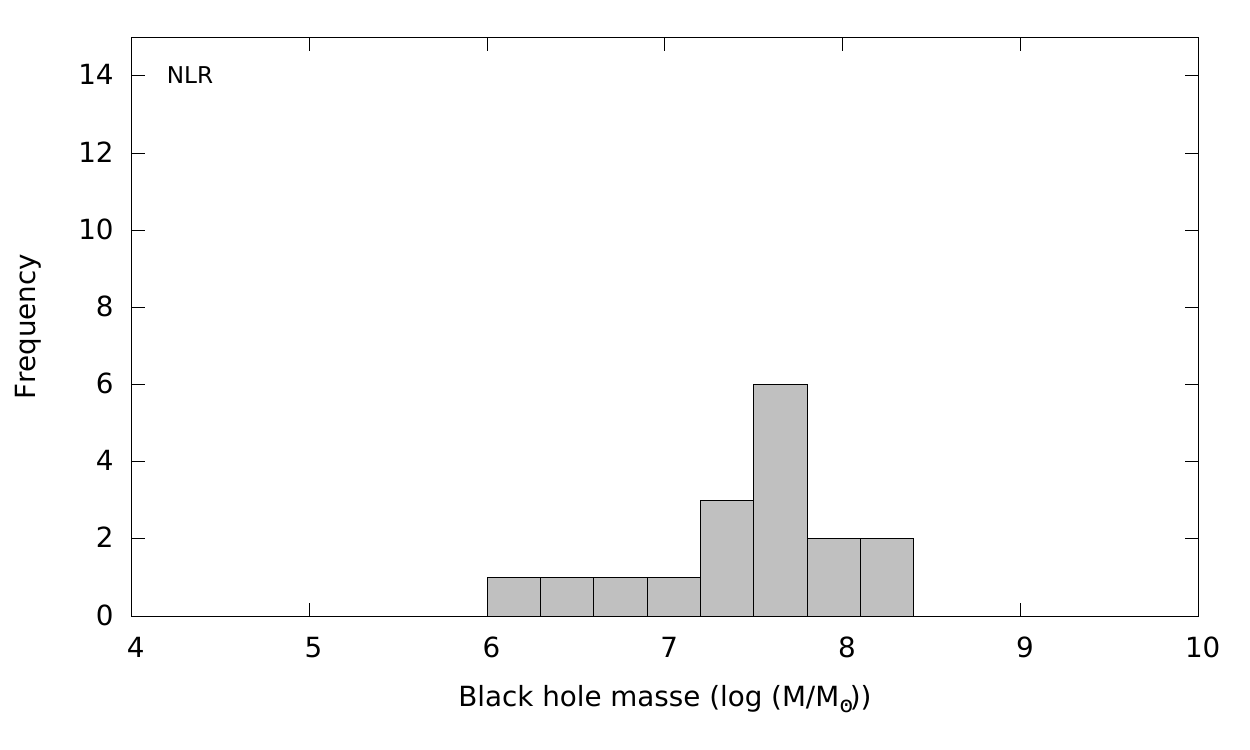} & \includegraphics[trim = 0mm 0mm 0mm 0mm, clip, width=8cm]{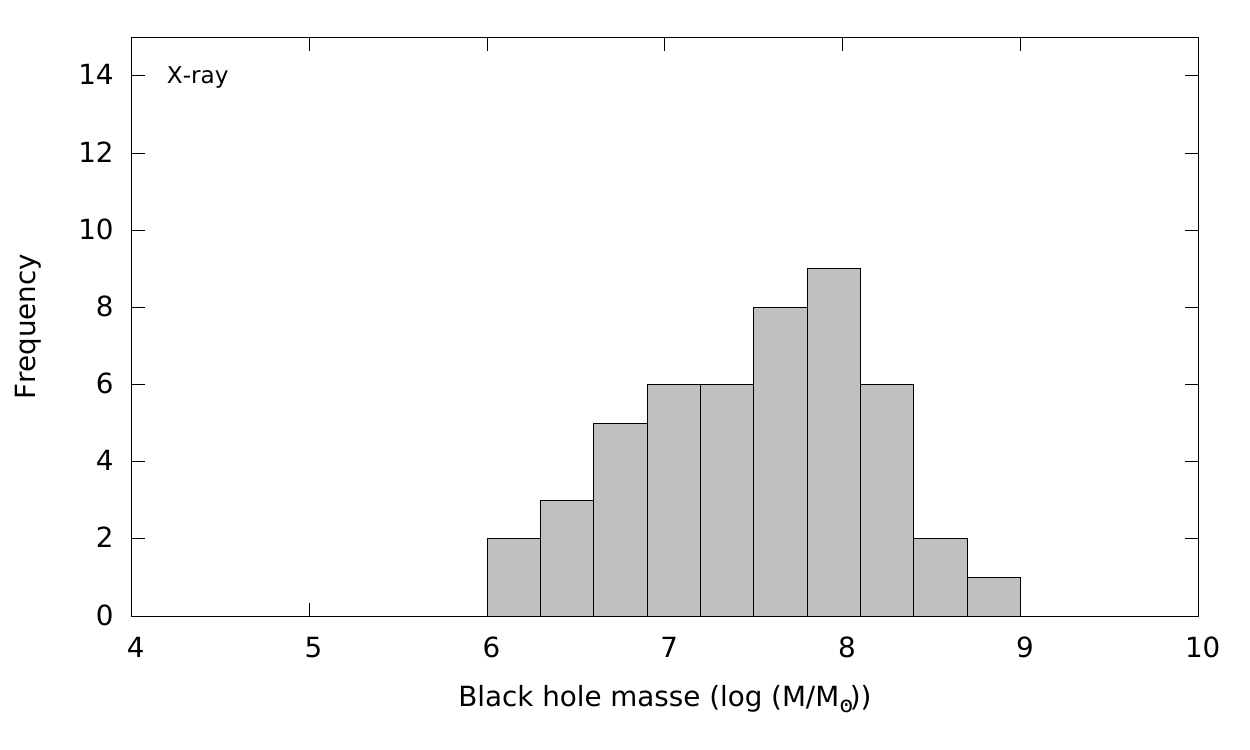} \\
      \end{tabular}
      \begin{tabular}{cc}
	\includegraphics[trim = 0mm 0mm 0mm 0mm, clip, width=8cm]{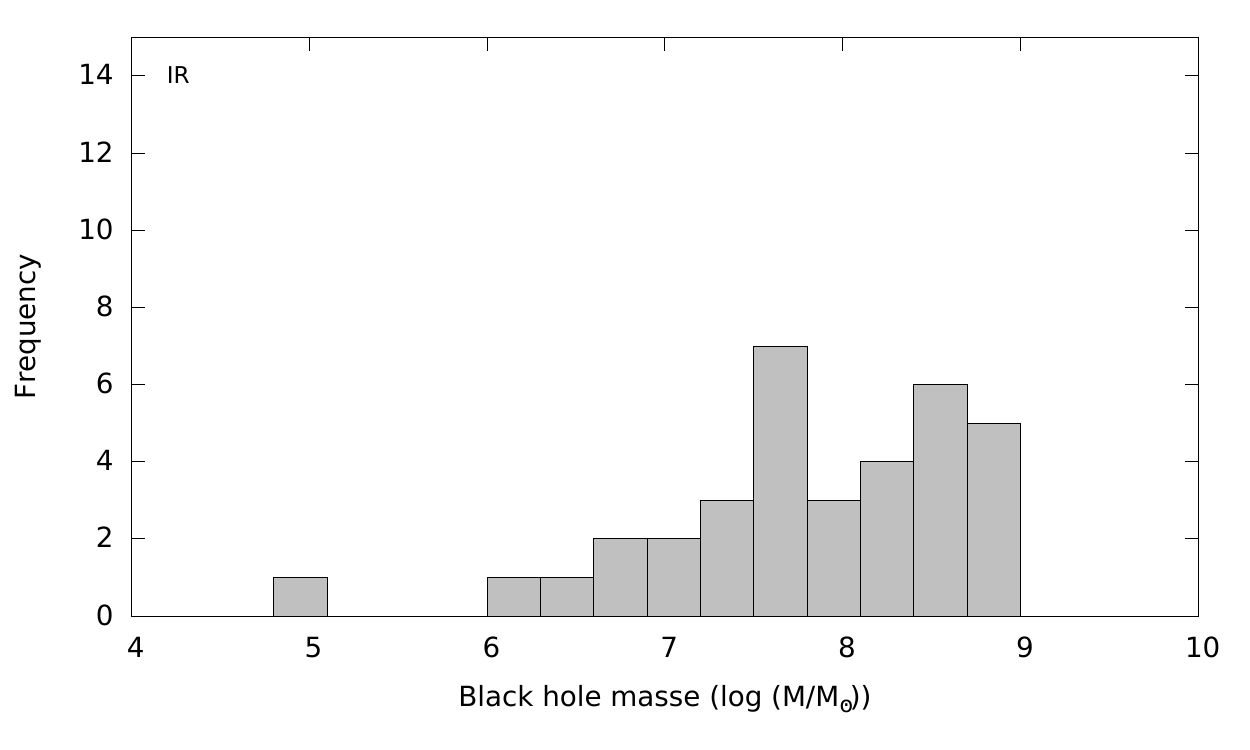} & \includegraphics[trim = 0mm 0mm 0mm 0mm, clip, width=8cm]{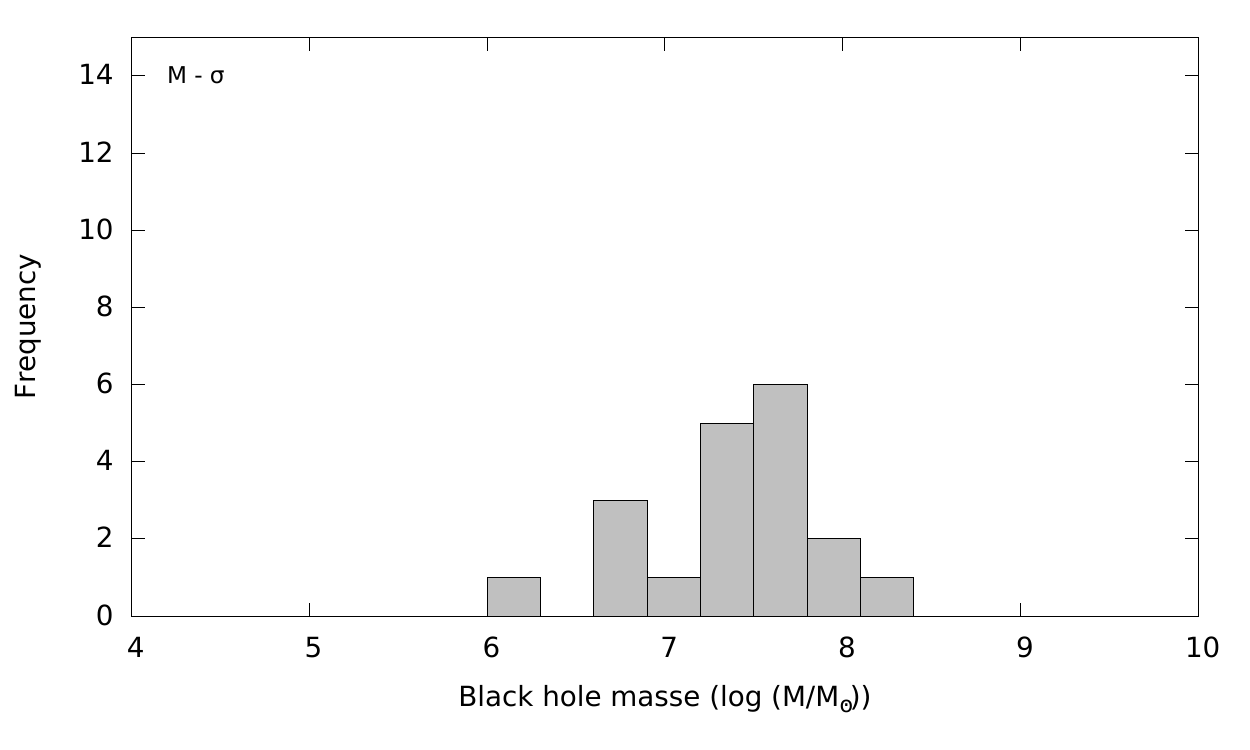} \\
      \end{tabular}
    \end{center}
    \caption{Frequency distribution of the black hole masses of the sample. Legend is the same
	     as in Fig.~\ref{Fig:Histograms}.}
    \label{Fig:Hist_BH}%
\end{figure*}

Since almost all the AGN to be investigated are situated in the nearby Universe, the black hole masses of the catalogs (being a fundamental property 
of AGN that governs the accretion rate) should not have significantly varied between the different lowest and largest redshifts of this sample. 
It was found by \citet{Woo2002}, in a sample of 377 radio-quiet and radio-loud AGN, that the mass distribution is narrow in the case of Seyfert 
galaxies (with average masses $\sim$~10$^{8}$~M$_\odot$), with no black holes masses greater than 10$^{9}$~M$_\odot$. Comparing the sample of 124 objects
in this paper with their results (see Fig~\ref{Fig:Hist_BH}, top), we also find a sharp cut-off at 10$^{9}$~M$_\odot$ and a distribution that peaks at 
$\sim$~10$^{7.7}$~M$_\odot$. Those conclusions also apply to the black hole masses of the four inclination indicators, which all peak at 
10$^{7}$~M$_\odot <$ M$_{\rm BH} <$ 10$^{8}$~M$_\odot$. These agreements confirm that the catalogs of AGN used in this paper are neither biased 
towards light nor very massive SMBH.

\subsubsection{Bolometric luminosities}
\label{Comp:characteristics:Luminosity}

\begin{figure*}
    \begin{center}
      \begin{tabular}{cc}
	\includegraphics[trim = 0mm 0mm 0mm 0mm, clip, width=8cm]{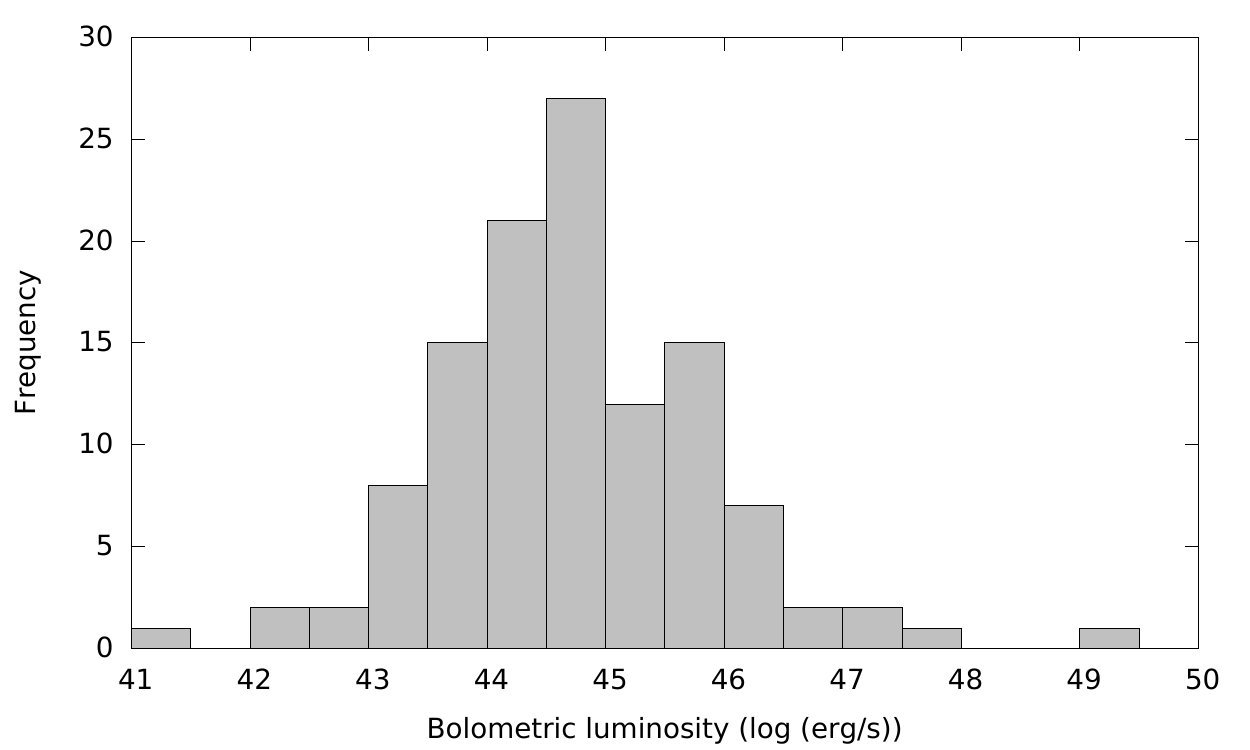} \\
      \end{tabular}
      \begin{tabular}{cc}
	\includegraphics[trim = 0mm 0mm 0mm 0mm, clip, width=8cm]{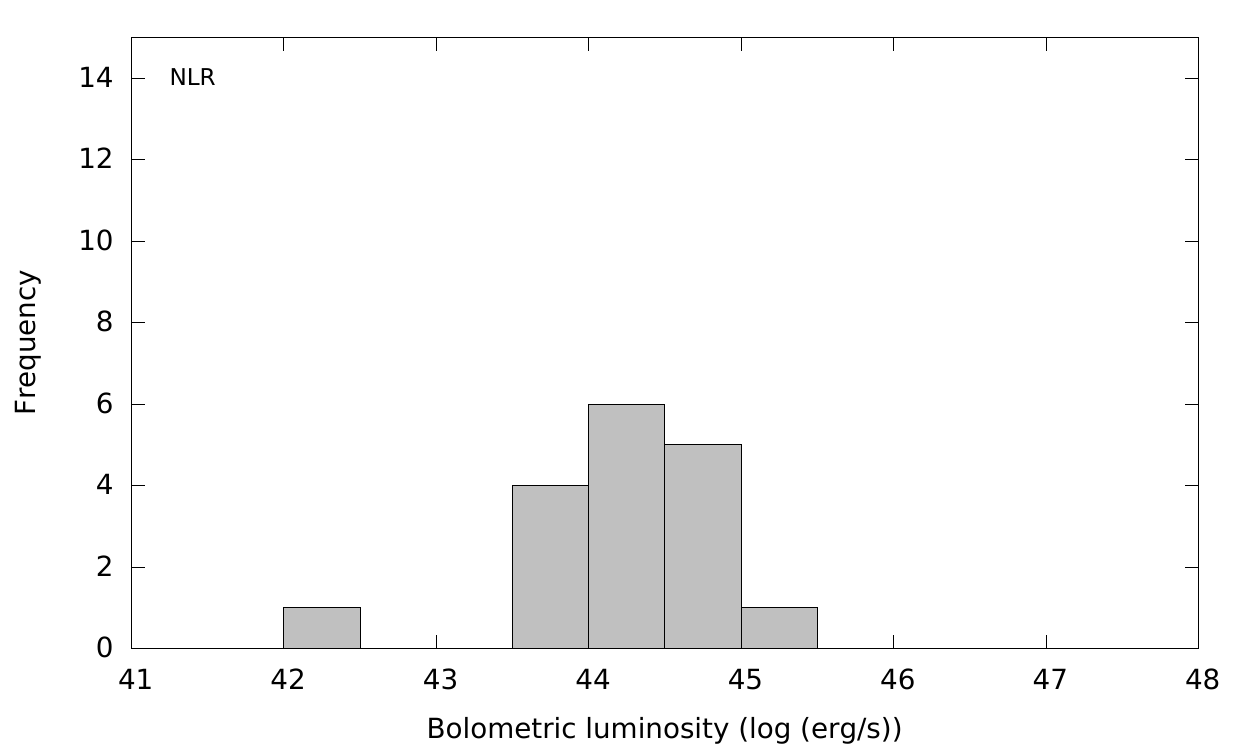} & \includegraphics[trim = 0mm 0mm 0mm 0mm, clip, width=8cm]{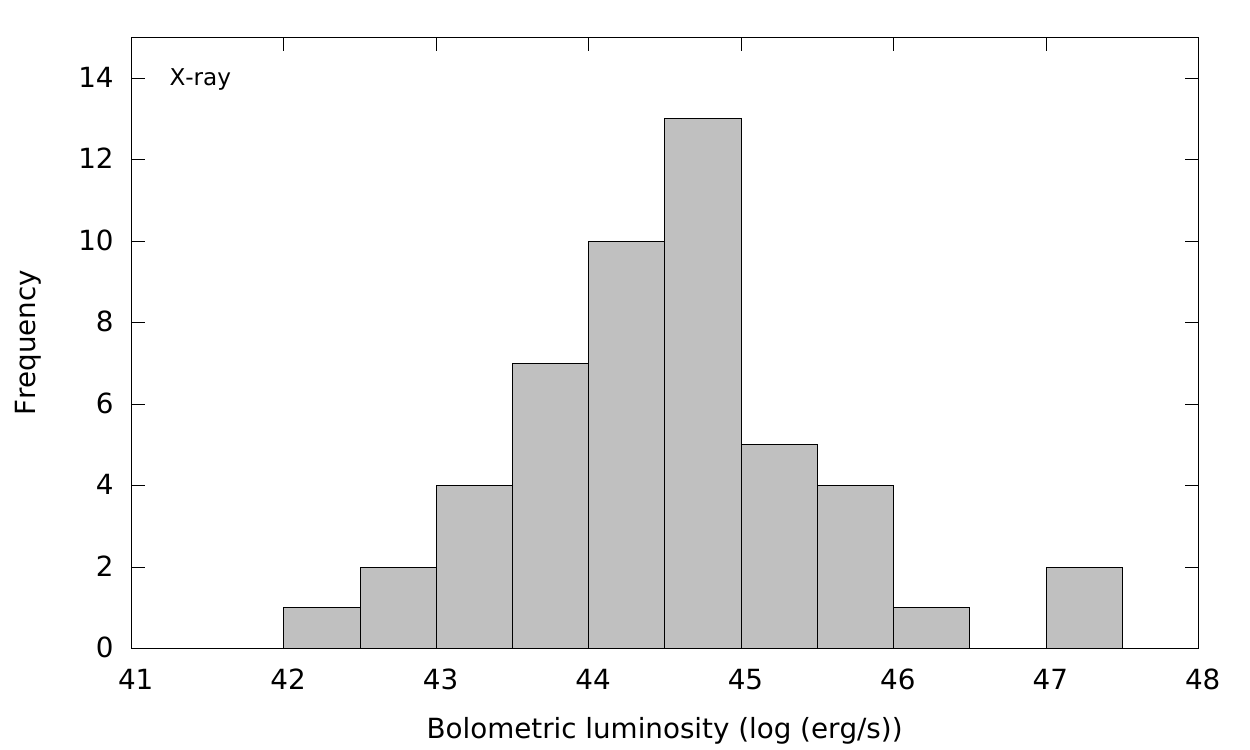} \\
      \end{tabular}
      \begin{tabular}{cc}
	\includegraphics[trim = 0mm 0mm 0mm 0mm, clip, width=8cm]{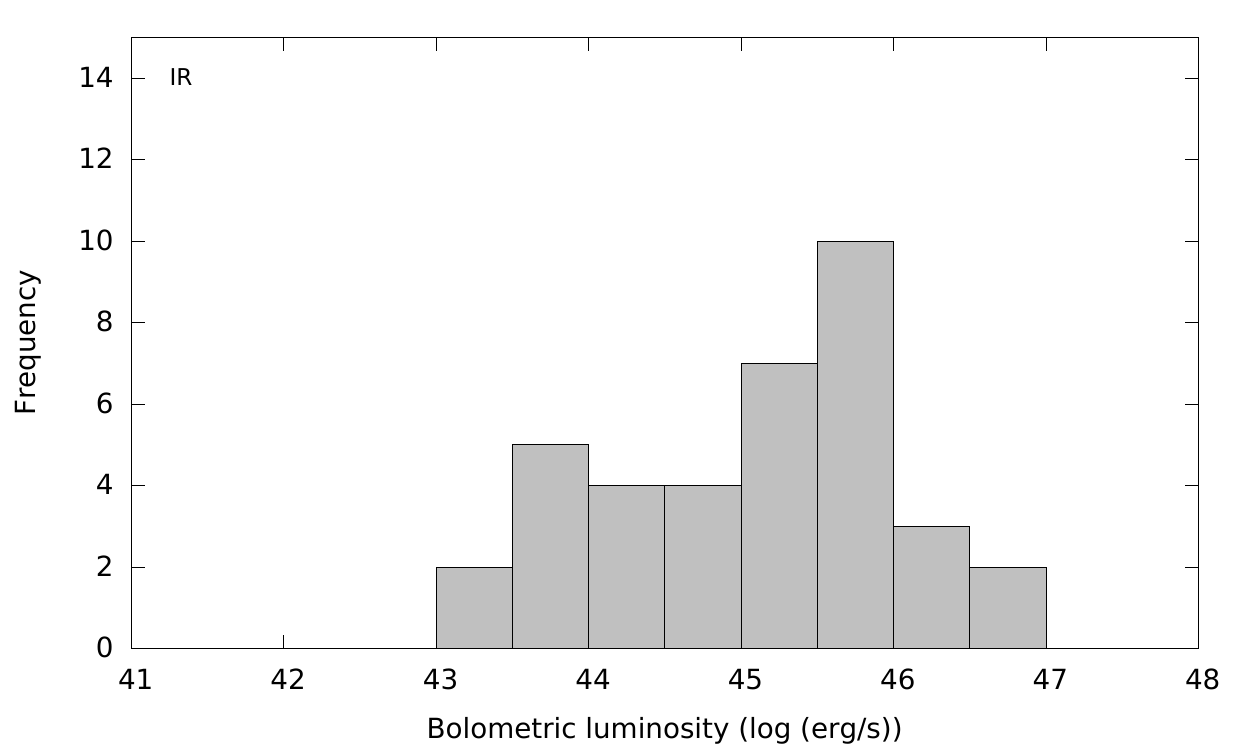} & \includegraphics[trim = 0mm 0mm 0mm 0mm, clip, width=8cm]{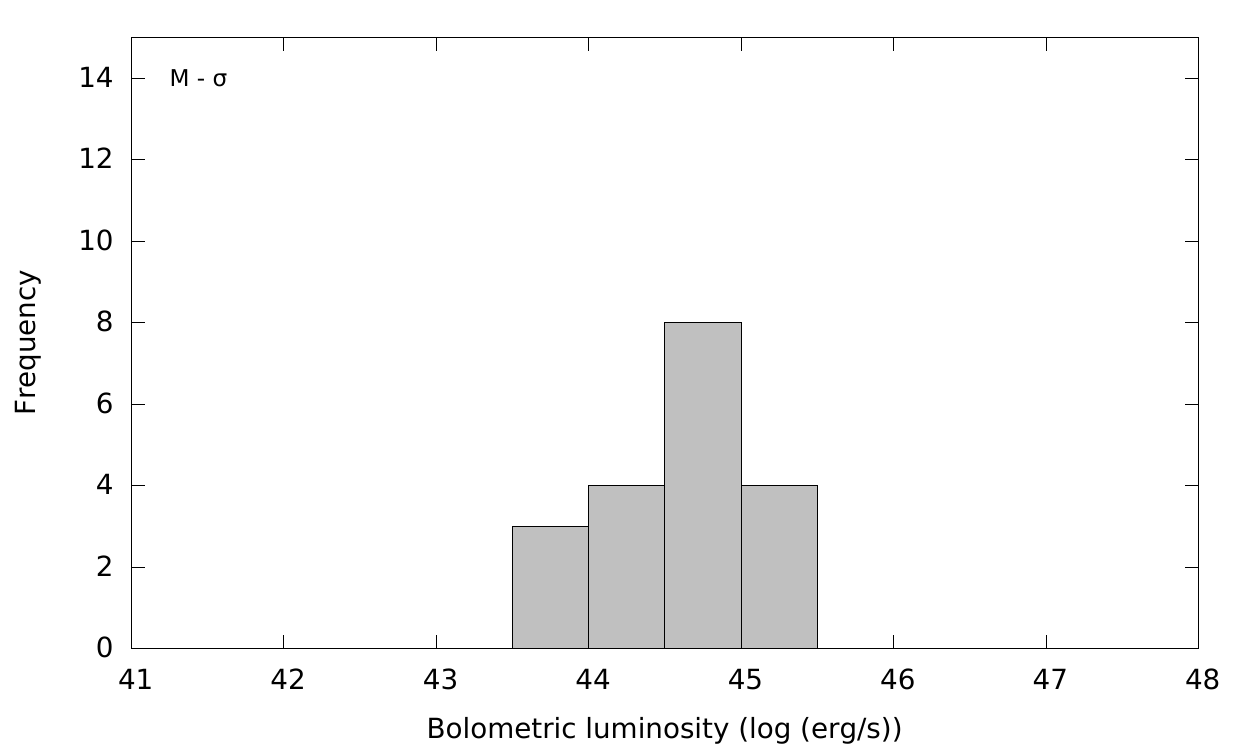} \\
      \end{tabular}
    \end{center}
    \caption{Frequency distribution of the bolometric luminosities of the sample.
	     Legend is the same as in Fig.~\ref{Fig:Histograms}.}
    \label{Fig:Hist_Bol}%
\end{figure*}

The distribution of bolometric luminosities in the global sample (Fig~\ref{Fig:Hist_Bol}, top) presents a general bell curve centered around 
log(L$_{\rm bol}$) = 44.73 (standard deviation: 1.15), the typical signature of a Gaussian distribution. The power output of the 124 AGN 
is therefore normally distributed and coherent with the average luminosity of radio-quiet AGN samples (10$^{44 - 45}$~erg/s, see \citealt{Zakamska2014} 
or \citealt{Comerford2014}).

Similarly to the global sample, the sub-catalogs of the X-ray, NLR and M-$\sigma$ indicators peak at 44 $<$ log(L$_{\rm bol}$) $<$ 45, with the 
exception of the IR fitting technique that peaks at 45 $<$ log(L$_{\rm bol}$) $<$ 46 (average luminosity log(L$_{\rm bol}$) = 45.07, standard 
deviation: 0.94). Thus the IR sample is slightly biased towards luminous AGN. This is due to the inclusion of the \citet{Mor2009} sample of 
mid- and far-IR selected QSO and ultraluminous infrared galaxies (ULIRG) using observations with the $Spitzer$ Space Telescope \citep{Schweitzer2006}. 
The higher power output of the IR sample might increase the resulting torus mass \citep{Mor2009}, be anticorrelated with the torus covering factor 
\citep{Mor2009} or alter the radial size of the torus inner's wall \citep{Simpson2005}, but should not change the nuclear inclination of the 
system. Only a fraction of detected type-1 AGN versus type-2 objects will vary with higher power outputs, as is observed in Fig.~\ref{Fig:Histograms} 
(bottom-left): the averaged half-opening angle of the torus is of the order of 70$^\circ$.

It is then safe to conclude that the different samples investigated in this paper are not strongly biased towards a characteristic parameter
that could theoretically have an effect on this work. It was also confirmed that the sample is not biased by selection effects (in the sense that 
the different methods would apply to intrinsically different classes of AGN): each sub-sample contains narrow and broad line Seyfert-1 (NLS1 and BLS1), 
and type-1.5, 1.8, 1.9 and type-2 AGNs. The presence of 3 LINER (Low Ionization Nuclear Emission Region) AGN is not quantitatively enough to 
tilt the balance towards a specific Seyfert class, and the 7 BAL QSO are only included in the global sample, which is not the main focus 
of this paper.

\section{Correlation with measurable quantities}
\label{Exploiting}
In Sect.~\ref{Comp}, four main methods to retrieve the nuclear inclination of Seyfert galaxies have been identified. It is now
of a prime importance to identify the methods that can be considered as reliable and which are the dubious ones, in order to 
improve our fitting and modeling tools. Since the Unified Model is characterized by a net anisotropy between pole-on and edge-on 
views, it is logical to expect different observed properties for those two extremes. However, two points of comparison ($\sim$~0$^\circ$
and $\sim$~90$^\circ$) are not enough to quantitatively assess the quality of a method. Instead, it is necessary to use multi-wavelength 
observables that are known to vary with inclination. Therefore the following observable quantities are compared to the different 
orientation indicators\footnote{Note that the best way to properly test the reliability of each method would be to compare the same 
list of objects with inclinations derived by each of the four indicators. However, this is hampered by the fact that only NGC~3227 
and NGC~4151 have an orientation estimation evaluated from the four methods.}: X-ray column density (Sect.~\ref{Exploiting:nH}), 
Balmer H$\beta$ line widths (Sect.~\ref{Exploiting:Balmer}), optical continuum polarization (Sect.~\ref{Exploiting:Pol}), and flux 
ratios (Sect.~\ref{Exploiting:Fluxes}).

\subsection{X-ray column density}
\label{Exploiting:nH}

\begin{figure*}
    \begin{center}
      \begin{tabular}{cc}
	\includegraphics[trim = 0mm 0mm 0mm 0mm, clip, width=8cm]{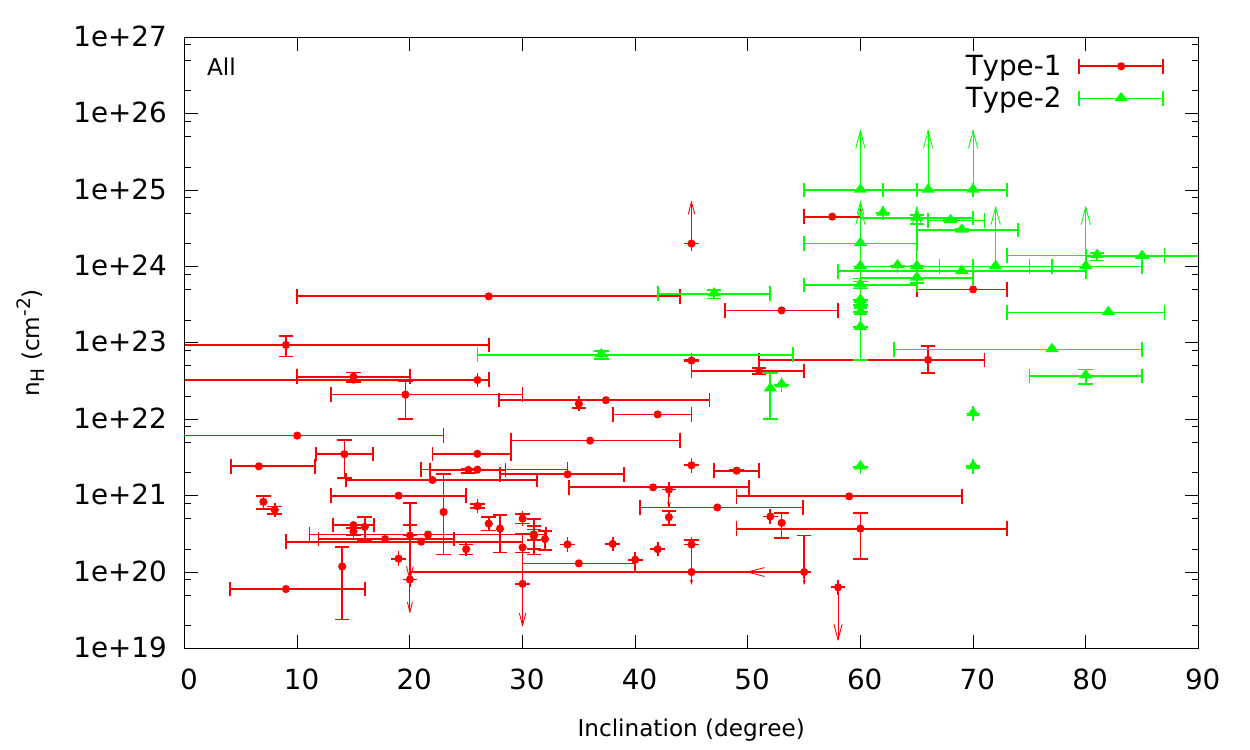} \\
      \end{tabular}
      \begin{tabular}{cc}
	\includegraphics[trim = 0mm 0mm 0mm 0mm, clip, width=8cm]{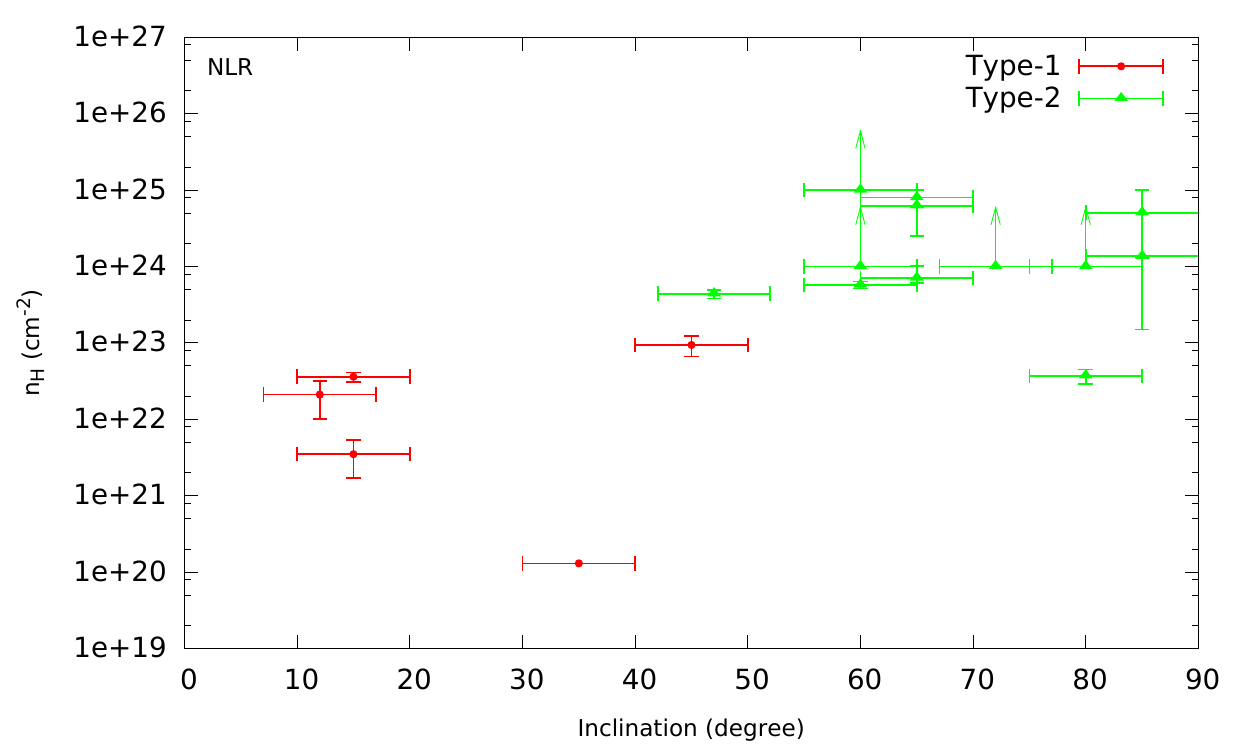} & \includegraphics[trim = 0mm 0mm 0mm 0mm, clip, width=8cm]{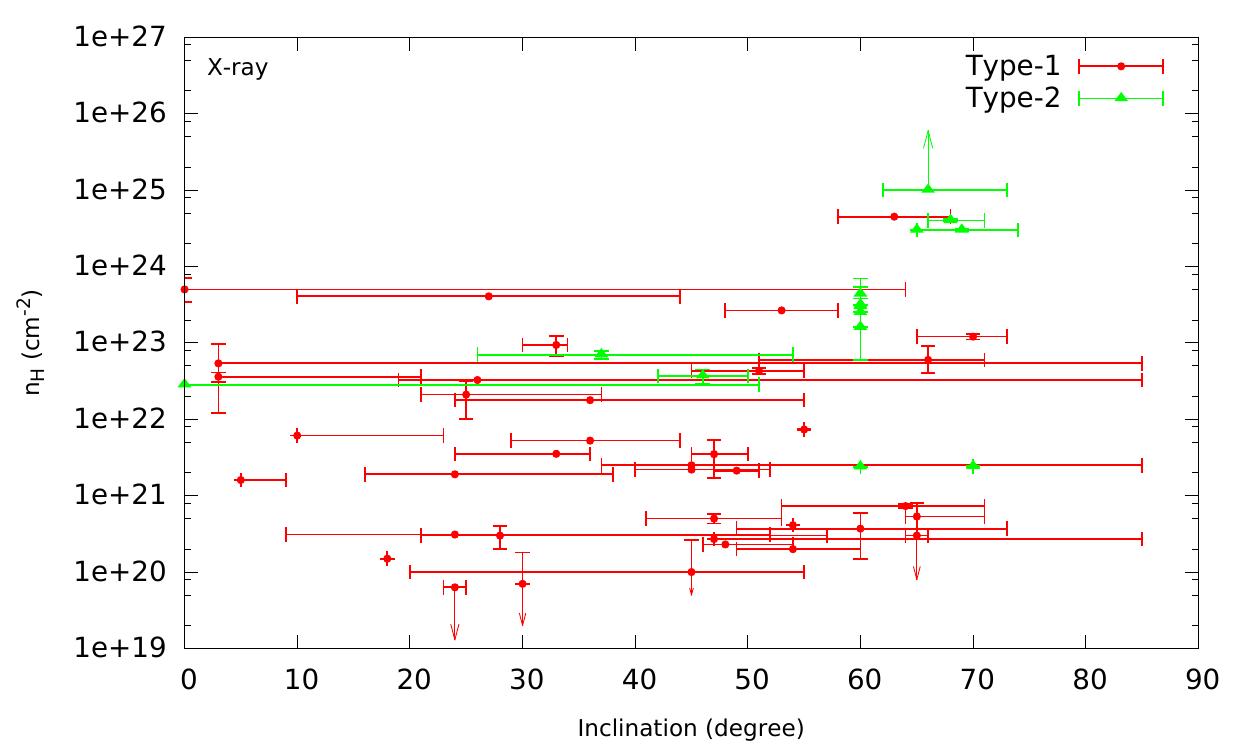} \\
      \end{tabular}
      \begin{tabular}{cc}
	\includegraphics[trim = 0mm 0mm 0mm 0mm, clip, width=8cm]{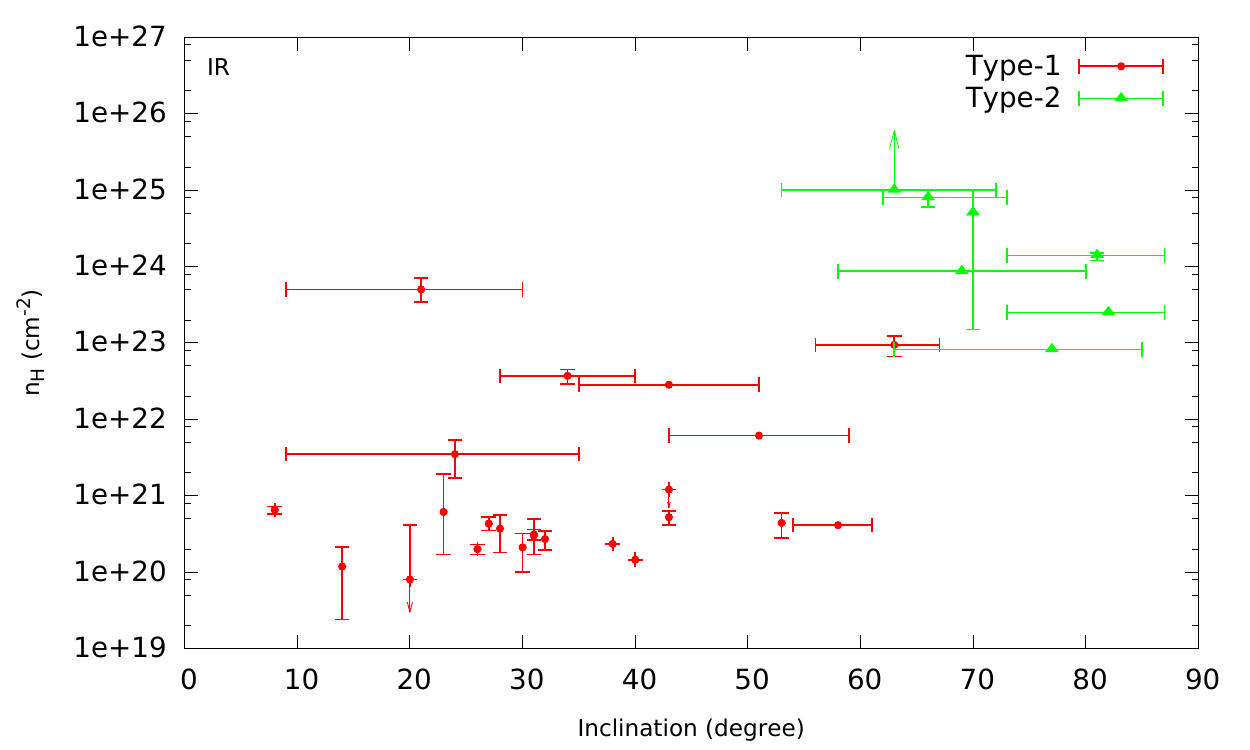} & \includegraphics[trim = 0mm 0mm 0mm 0mm, clip, width=8cm]{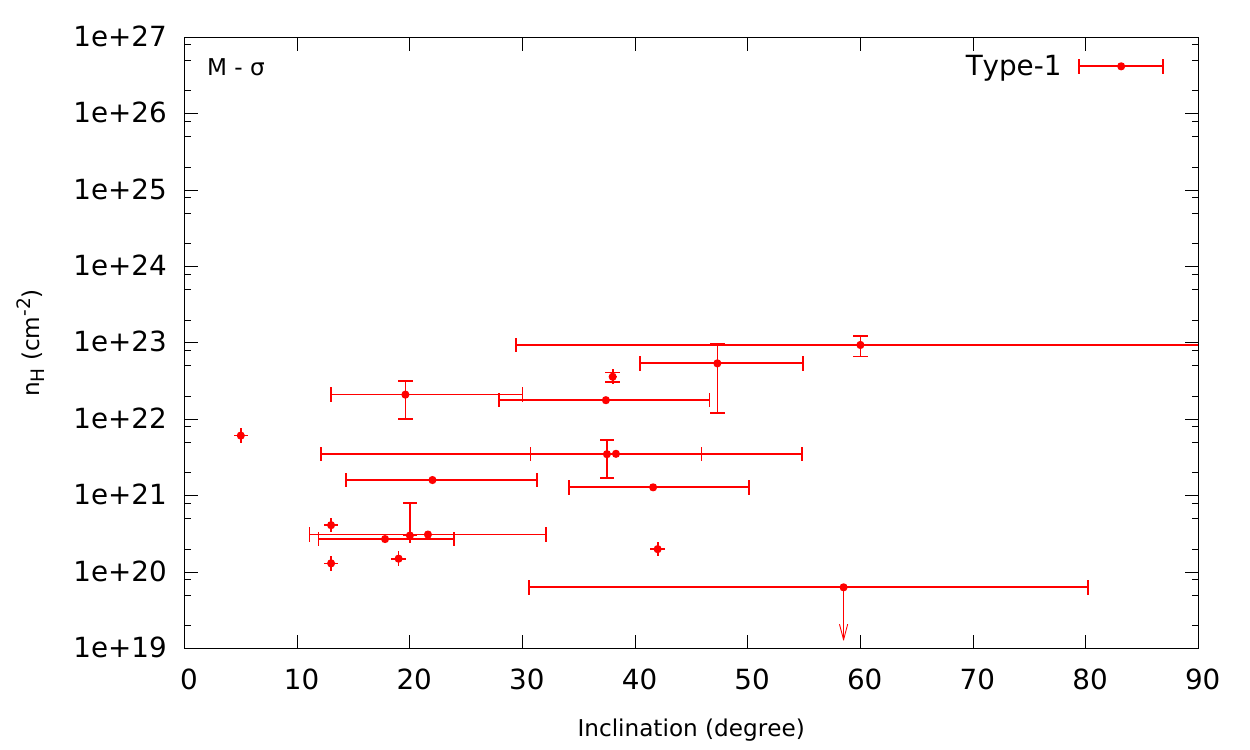} \\
      \end{tabular}
    \end{center}
    \caption{Intrinsic hydrogen column density resulting from X-ray spectral 
	     fitting as a function of AGN inclination derived from the indicated 
	     method (see text). Legend is the same as in Fig.~\ref{Fig:Histograms}.}
    \label{Fig:nH}%
\end{figure*}

According to the Unified Model \citep{Lawrence1991,Antonucci1993}, most of the obscuring material around AGN is concentrated close to the equatorial 
plane. This obscuring region presents very high column densities at edge-on views (n$_{\rm H} \gg$ 10$^{24}$~cm$^{-2}$, \citealt{Matt2004b}), and 
the amount of hydrogen does not deviate strongly from a Compton-thick state until the observer's line-of-sight starts to graze the circumnuclear 
dust horizon. The lower column density of obscuring material allows the partial transmission of type-1 characteristics such as broad optical lines 
(e.g., Fairall~51, \citealt{Smith2002} or 3C~68.1, \citealt{Brotherton1998}), indicating that the system is seen at an intermediate inclination. 
The resulting hydrogen column density is between 10$^{23}$ $\le$ n$_{\rm H} \le$ 10$^{24}$~cm$^{-2}$ \citep{Risaliti2005}, a range that 
corresponds to the Changing-look AGN class mentioned in Sect.~\ref{Comp:Inclination}. At inclinations closer to the pole are ionized outflows with 
low (n$_{\rm H} \le$ 10$^{23}$~cm$^{-2}$) hydrogen column densities \citep{Wilkes2013}. However, recent works suggest that n$_{\rm H}$ evolves rather
smoothly from the edge to the pole. High-resolution, hydrodynamic, numerical simulations by \citet{Wada2009} and \citet{Wada2012}, looking at the 
inner parsecs around SMBH, have recently found that the total gas and H2 column densities evolve smoothly from the Compton-thick equatorial structure 
to the Compton-thin pole, with the transition angle between the two regimes lying around 50$^\circ$. According to \citet{Wada2009} and \citet{Wada2012}, 
AGN are likely to be surrounded by a non-uniform shell of gas with inclination-dependent column densities (see their Fig.~4-a). This result is supported
by the recent exploration of the correlation between the optical classification of Seyfert galaxies and their observed X-ray absorption by \citet{Burtscher2015}.
Plotting the estimated X-ray absorbing columns of 25 local AGN against their Seyfert sub-classes (1, 1.5, 1.8, 1.9, 1i, 1h and 2, see \citealt{Burtscher2015} 
for additional information), they found a good agreement between optical and X-ray classification, indicating a correlation between $i$ and n$_{\rm H}$.

The collection of intrinsic n$_{\rm H}$ found in the literature are summarized in Tab~\ref{Table:nH}, and Fig~\ref{Fig:nH} presents the different 
results obtained for the full sample as well as for each orientation indicator. Note that the estimation of hydrogen column densities along the 
observerd line-of-sight is always model-dependent and potential deviations can be found between two authors. Yet, it clearly seems, as expected, 
that type-2 AGN have much higher n$_{\rm H}$ values, with a transition value between type-1s and type-2s being dependent on the method. There is 
a rather large data dispersion in hydrogen column density in all samples, which reflects the diversity of AGN even at a given inclination 
\citep{Wada2009,Wada2012}. Most of the type-2 n$_{\rm H}$ are lower limits, as the procedure for data fitting is often limited to values lower 
than 10$^{25}$~cm$^{-2}$ due to to small signal-to-noise ratios \citep[e.g.][]{Bianchi2005b} or to computing limitations 
\citep[e.g.][]{Balokovic2014}. Nevertheless, it is possible to look for correlations within the different samples using statistical rank 
correlation tests \citep{Spearman1904}, while accounting for upper and lower limits \citep{LaValley1992}. There are two efficient estimators 
used to measure the relationship between rankings of different ordinal variables: the Spearman's rank correlation coefficient $\rho$ and the 
Kendall non-parametric hypothesis test $\tau$ for statistical dependence. By normal standards for the sample sizes presented in this paper, 
a $\mid\rho\mid$ value between 0 and 0.29 represents an absence of association, $\mid\rho\mid$ between 0.30 and 0.49 a possible correlation 
and $\mid\rho\mid~>$~0.50 is a highly significant correlation. $\tau$ has usually lower values, and high statistical significance is reached
when $\mid\tau\mid~>$~0.40. Note that those thresholds depend on the field of study; the $\mid\rho\mid~>$~0.50 criterion to reach high 
statistical significance is the one commonly used in physical and social sciences \citep{Cohen1988,Haukoos2005,Curran2014}. The sign of the 
coefficient indicates whether it is a correlation (positive) or an anticorrelation (negative).

The $\rho$ and $\tau$ values for the four inclination indicators, along with their two-tailed p-values, are summarized in Tab~\ref{Table:FinalResults}. 
Both the IR and NLR methods have rank correlation coefficients greater that 0.5, which by normal standards indicates that the association 
between $i$ and n$_{\rm H}$ is highly statistically significant (at 95\% confidence level for rejecting null hypothesis). The X-ray and 
M-$\sigma$ methods present weak rank coefficients ($\rho$ = 0.36 and $\tau$ = 0.25 for the former, $\rho$ = 0.31 and $\tau$ = 0.26 for 
the later) suggesting a possible correlation. It is interesting to note that, despite being able to reproduce a correlation between inclination 
and X-ray absorption, the IR, NLR and (possibly) the X-ray methods show very different normalizations. The NLR method suggests that the 
hydrogen column density is still of the order of 10$^{22}$~cm$^{-2}$ at inclinations close to 10$^\circ$, while the IR and reflection 
spectroscopy methods suggest values about two orders of magnitude lower. The former result is high for the Unified Model standard, as for 
a Galactic reddening curve, a column density of 10$^{22}$~cm$^{-2}$ corresponds to an extinction in the V-band of A$_V$ = 5. This can 
only be explained by a hollow structure of the biconical NLR wind, where the inner funnel is relatively free of gas while the hot 
flow sustains a much higher column density mixed with dust. The later methods are more aligned with the predictions of the AGN scheme, 
where the hydrogen column density would drop to almost zero at perfect polar orientations.

\subsection{Balmer H$\beta$ line widths}
\label{Exploiting:Balmer}

The morphology of the region responsible for Doppler broadening of AGN emission lines, directly visible in pole-on quasars but only 
revealed by scattering-induced polarization in edge-on objects, is still debated. The discovery of double-peaked Balmer line profiles 
in a dozen of radio-loud AGN by \citet{Eracleous1994} favors a disk-like geometry dominated by rotational motions. Even more striking 
evidence comes from the investigation of \citet{Wills1986}, \citet{Brotherton1996}, or \citet{Jarvis2006}, who found a highly significant 
correlation between the ratio of the radio core flux density to the extended radio lobe flux density, $R$, and the FWHM of broad H$\beta$ 
lines. The Doppler width of Balmer lines was found to be unimportant at high $R$, i.e. when the system is seen close to being face-on. 
Since Doppler broadening is inclination-dependent, increasing at large viewing angles, the picture is consistent with a disk-like structure 
of the LIL BLR. However, there is no direct evidence yet that this picture also applies to radio-quiet objects, despite that a small fraction 
of radio-quiet AGN also shows double-peaked profiles \citep{Marziani1992,Shapovalova2004}. Several authors tried to reveal this 
inclination-dependent behavior using simulations \citep[e.g.][]{Zhang2002} even if the width of the point response function at half 
the maximum intensity remains difficult to estimate due to the observed line variability \citep{Asatrian2012}. As noted by \citet{Antonucci1989},
the absence of L$\alpha$ continuum absorption in any type-1 AGN requires that the clouds producing the broad emission lines are hidden 
from our line-of-sight by the circumnuclear region.

\begin{figure*}
    \begin{center}
      \begin{tabular}{cc}
	\includegraphics[trim = 0mm 0mm 0mm 0mm, clip, width=8cm]{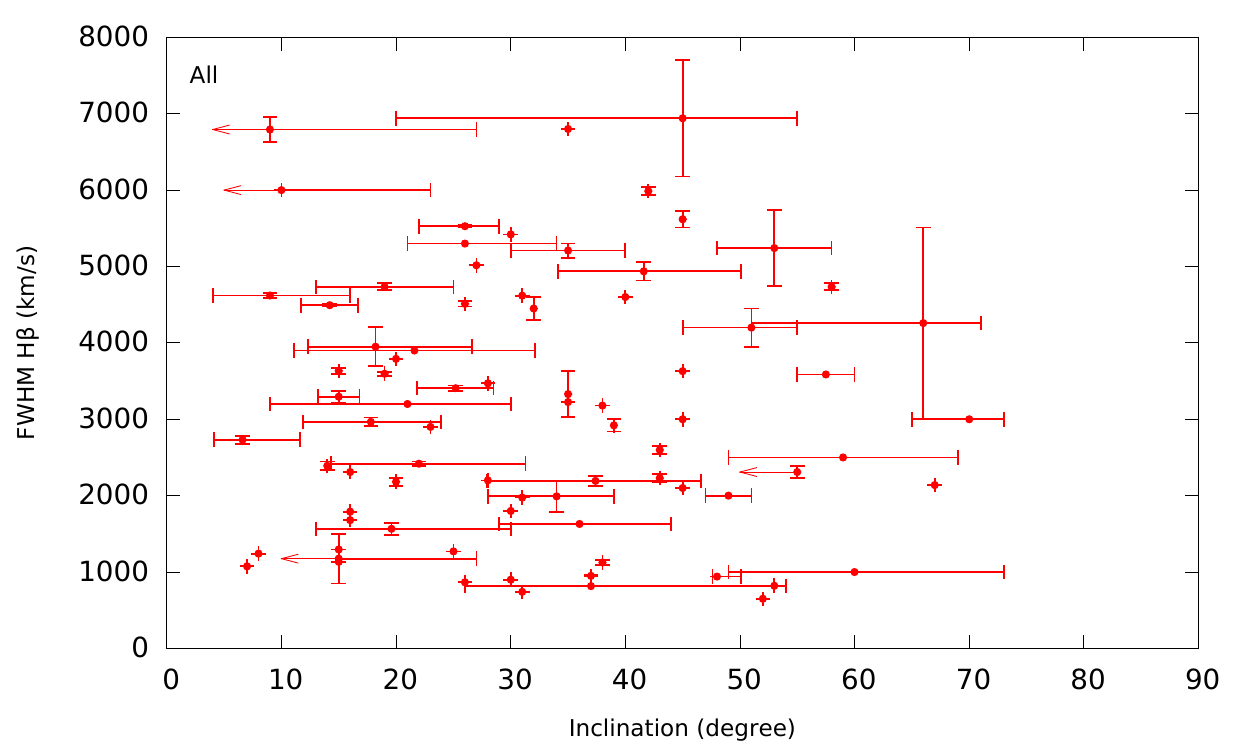} \\
      \end{tabular}
      \begin{tabular}{cc}
	\includegraphics[trim = 0mm 0mm 0mm 0mm, clip, width=8cm]{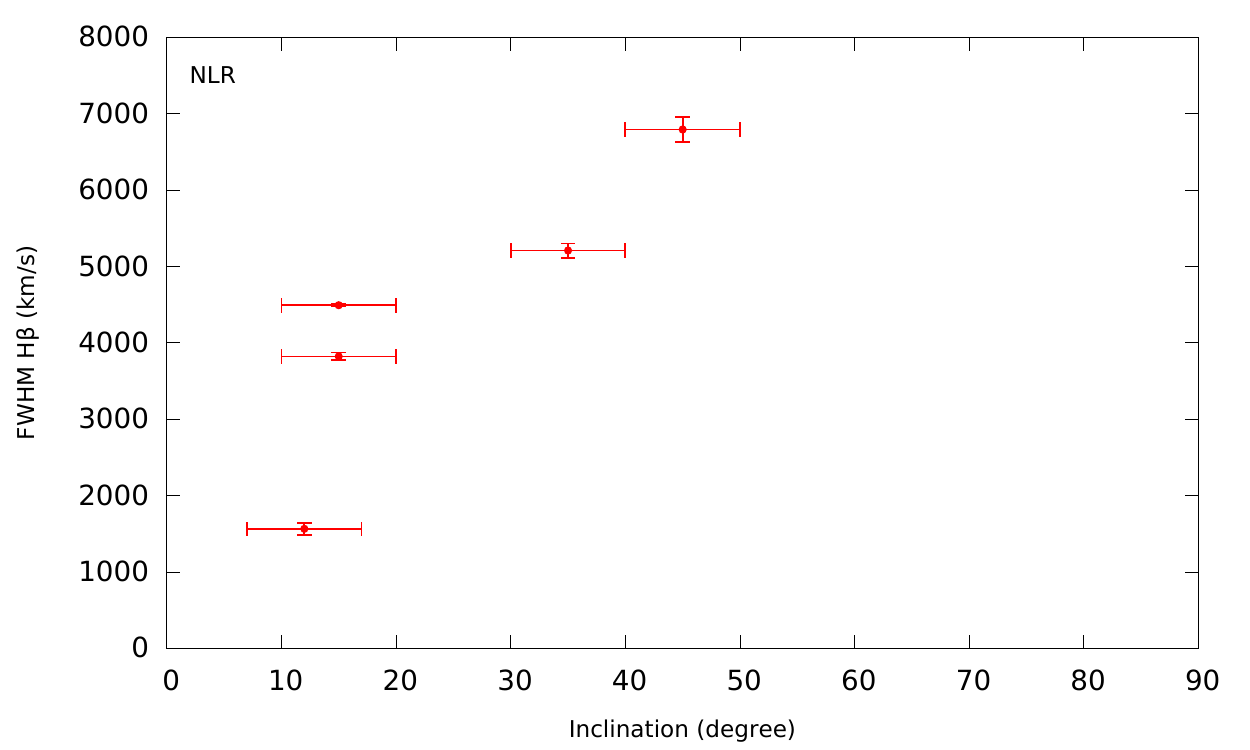} & \includegraphics[trim = 0mm 0mm 0mm 0mm, clip, width=8cm]{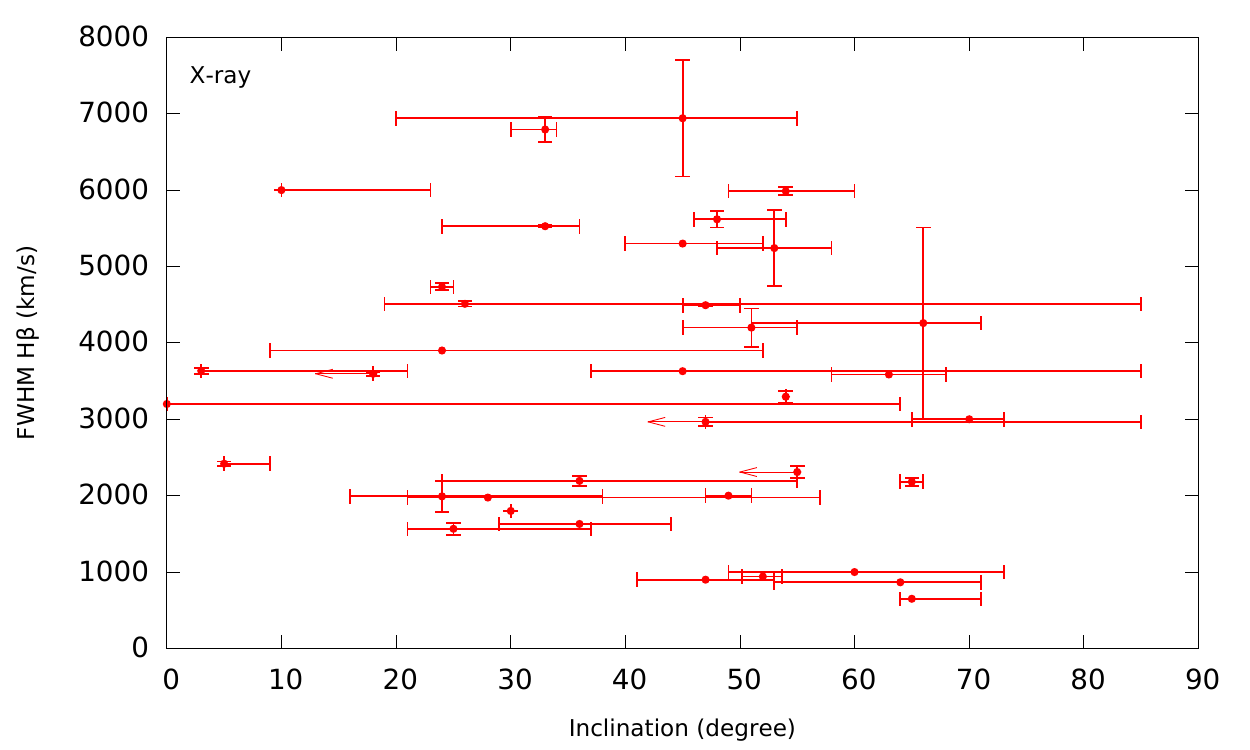} \\
      \end{tabular}
      \begin{tabular}{cc}
	\includegraphics[trim = 0mm 0mm 0mm 0mm, clip, width=8cm]{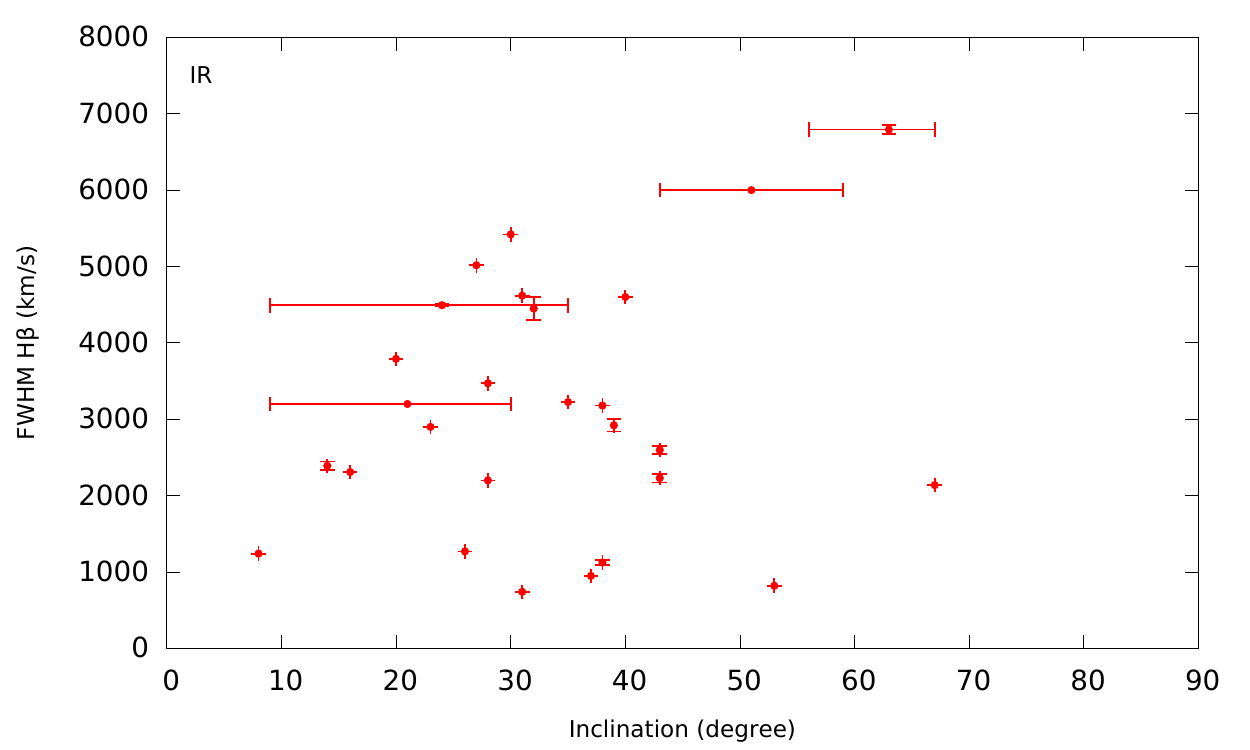} & \includegraphics[trim = 0mm 0mm 0mm 0mm, clip, width=8cm]{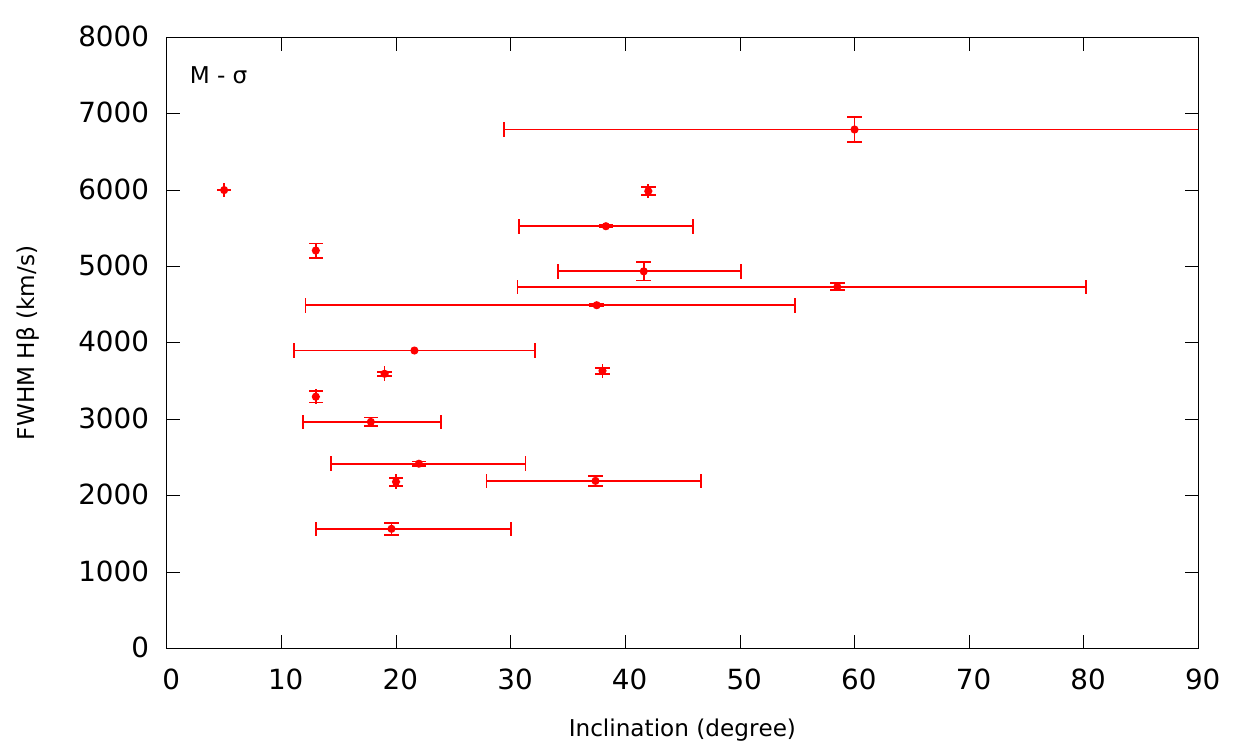} \\
      \end{tabular}
    \end{center}
    \caption{Broad H$\beta$ FWHM as a function of AGN inclination 
	     derived from the indicated method (see text). Legend 
	     is the same as in Fig.~\ref{Fig:Histograms}.}
    \label{Fig:Hbeta}%
\end{figure*}

Archival optical FWHM measurements of the H$\beta$~$\lambda$4861 line were retrieved from literature for 74 type-1 AGN. 
They are listed in Tab.~\ref{Table:Hbeta} and plotted against inclination in Fig~\ref{Fig:Hbeta}. The expected relation between
the velocity field of the disk-like LIL BLR and the inclination of the system observed by \citet{Wills1986} in the case of 
radio-loud AGN is proportional to $(v_r^2 + v_p^2 \sin^2 i)^\frac{1}{2}$, where $v_r$ is a random isotropic velocity and 
$v_p$ a Keplerian component only in the plane of the disk. According to \citet{McLure2001} and \citet{Gaskell2013}, $v_r$ is 
small in comparison with $v_p$, and $v_p$ is of the order of several thousands of kilometers per second. The expected increase 
of FWHM with inclination is visible in the plots of the NLR and M-$\sigma$ methods, as confirmed by the $\rho$ and $\tau$ rank 
correlation coefficients (see Tab~\ref{Table:FinalResults}), but the later method is intrinsically biased. Indeed, \citet{Wu2001}
use the H$\beta$ FWHM as a parameter in their equations to retrieve the inclination of their AGN sample (see their 
Eq.~2, 3 and 4), so it is logical that there is a good correlation between H$\beta$ FWHM and $i$. Disregarding the M-$\sigma$ method
from this analysis, only the NLR fitting method by \citet{Fischer2013} is able to retrieve the expected disk-like signature
of the LIL BLR region (as it was already shown in \citealt{Fischer2014}). However, due to the small number of type-1s matched 
with H$\beta$ FWHM for the NLR technique, additional data are needed to confirm the validity of this correlation. Finally, 
signs of an anticorrelation between H$\beta$ FWHM and $i$ appeared in the case of the X-ray indicator ($\rho$ = -0.15, 
$\tau$ = -0.11), a singular characteristic already mentioned by \citet{Nishiura1998}. Such anticorrelation, if real, would 
indicate that AGN with face-on accretion disks have larger BLR velocities. This consequence will be discussed in 
Sect.~\ref{Discussion:Coplanar}.

\subsection{Optical continuum polarization}
\label{Exploiting:Pol}

\begin{figure*}
    \begin{center}
      \begin{tabular}{cc}
	\includegraphics[trim = 0mm 0mm 0mm 0mm, clip, width=8cm]{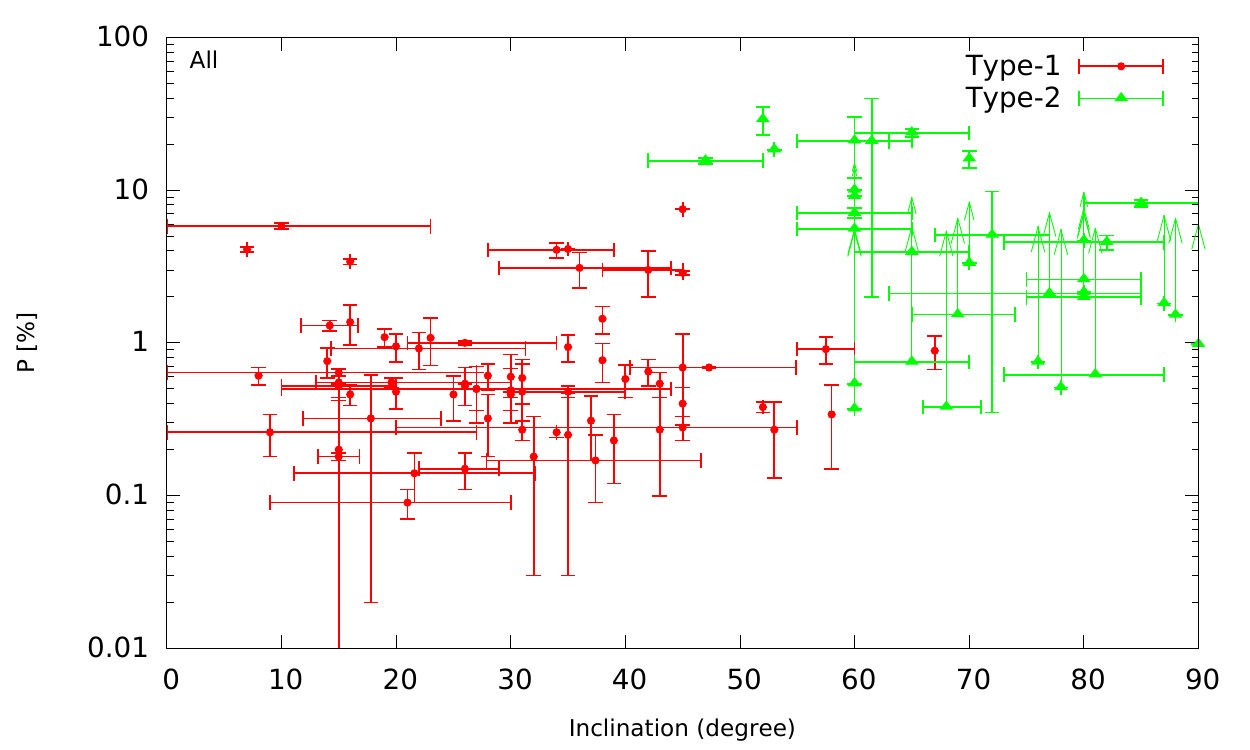} \\
      \end{tabular}
      \begin{tabular}{cc}
	\includegraphics[trim = 0mm 0mm 0mm 0mm, clip, width=8cm]{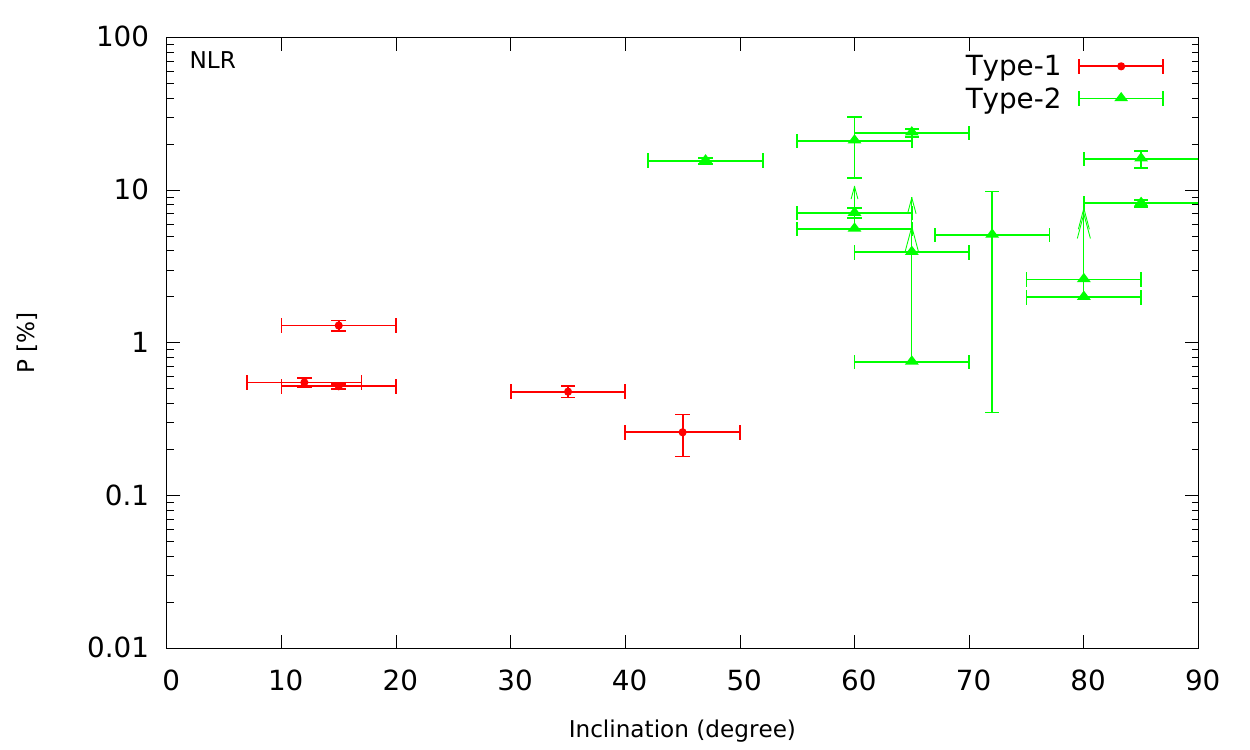} & \includegraphics[trim = 0mm 0mm 0mm 0mm, clip, width=8cm]{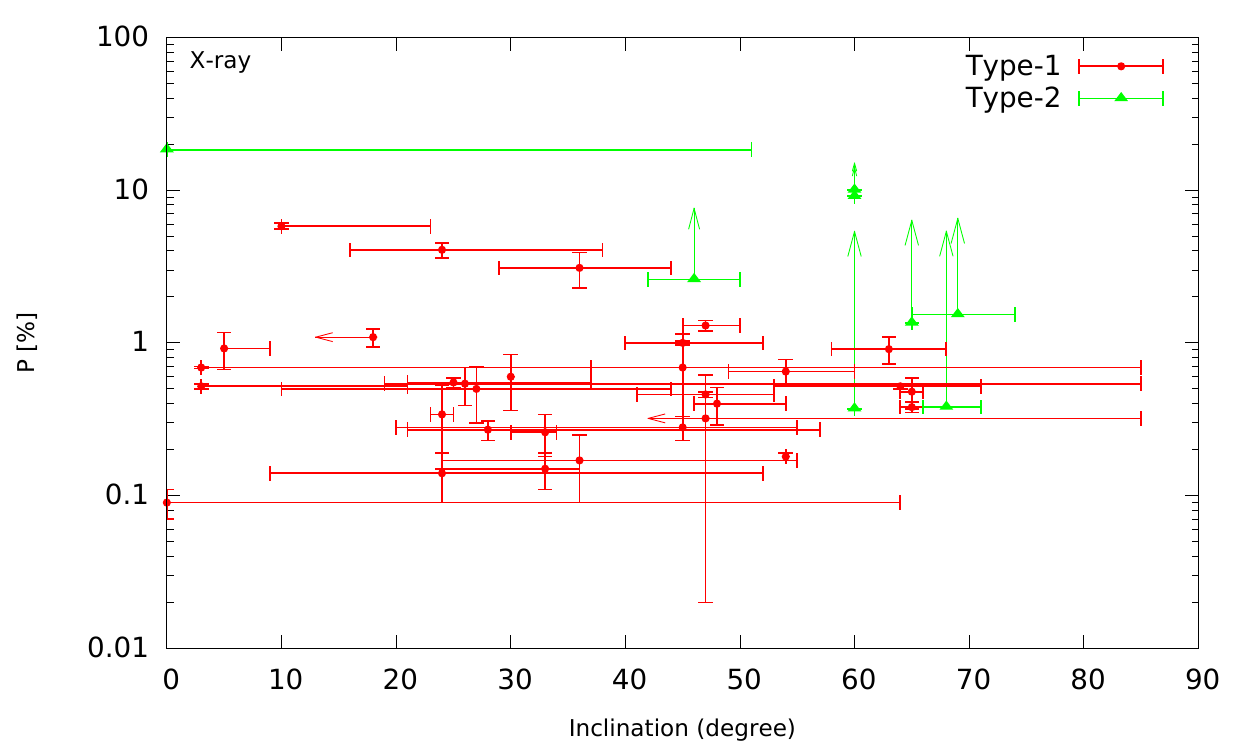} \\
      \end{tabular}
      \begin{tabular}{cc}
	\includegraphics[trim = 0mm 0mm 0mm 0mm, clip, width=8cm]{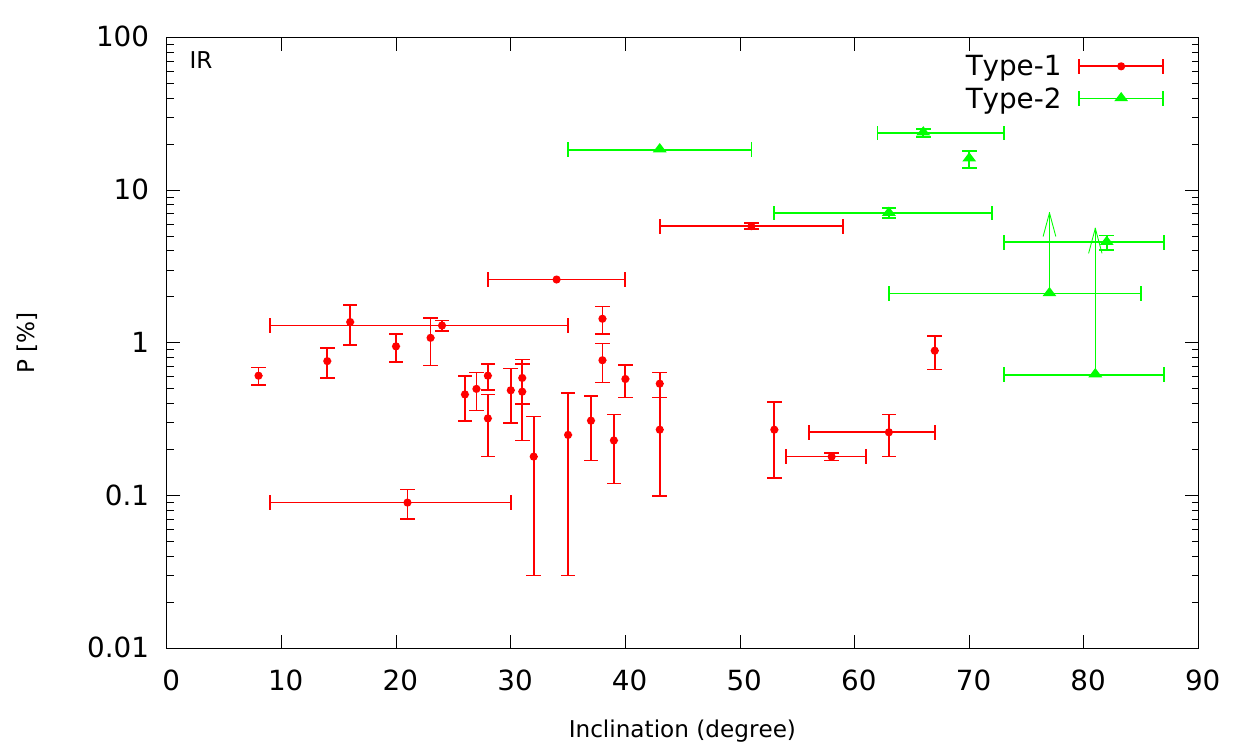} & \includegraphics[trim = 0mm 0mm 0mm 0mm, clip, width=8cm]{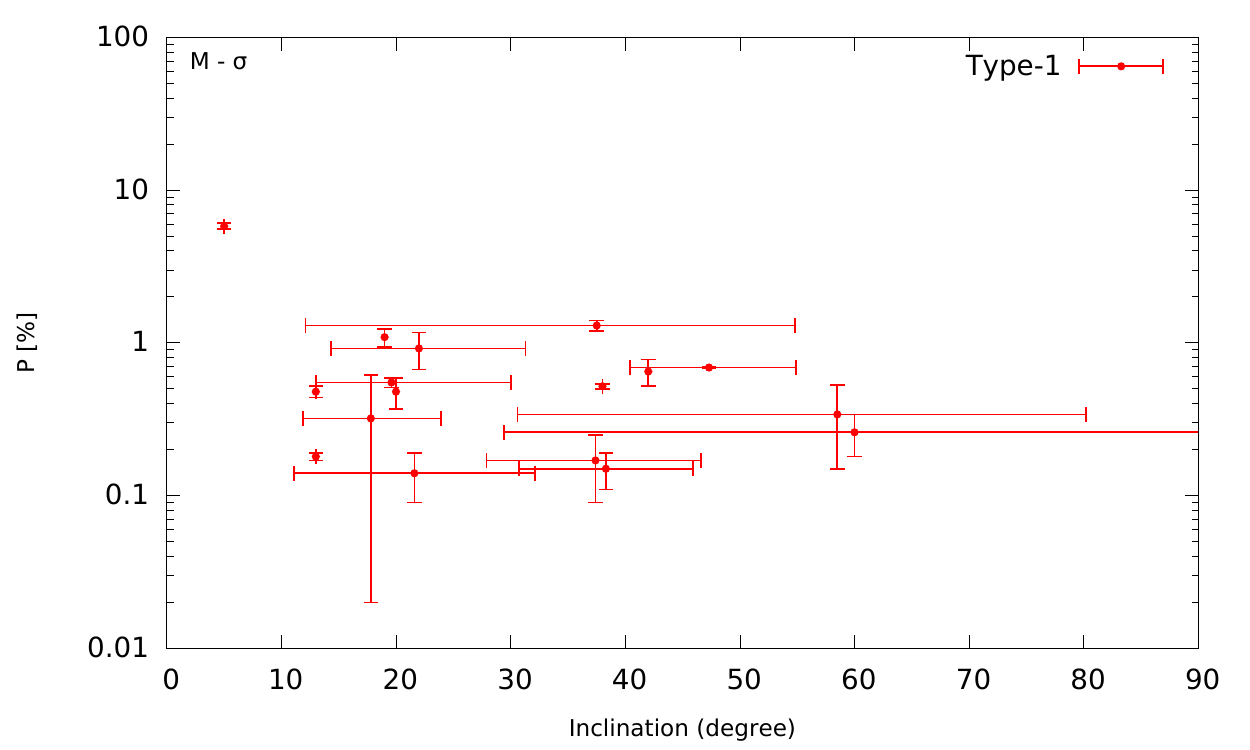} \\
      \end{tabular}
    \end{center}
    \caption{Optical continuum polarization degree $P$ as a function 
	     of AGN inclination derived from the indicated method 
	     (see text). Legend is the same as in Fig.~\ref{Fig:Histograms}.}
    \label{Fig:Polarization}%
\end{figure*}

The AGN structure can be probed with great precision by using the geometry-sensitive technique of polarimetry. Optical 
polarimetry laid the ground for the Unified Scheme, first by revealing the predominance of polarization position angles 
parallel to the projected radio axis of type-1s, while type-2s only show perpendicular polarization position angles 
\citep{Antonucci1984}, but also by uncovering broad Balmer lines in the polarized flux spectra of type-2 Seyferts 
\citep{Miller1983,Antonucci1985}. This was one of the strongest assertion in favor of equatorial obscuration, an argument 
that still holds firmly. By looking at the optical polarization of AGN, it is possible to estimate the composition, kinematics 
and geometry of the Seyfert constituents. This has been theoretically and numerically shown in a number of papers 
\citep[e.g.][]{Kartje1995,Young2000,Goosmann2007,Marin2012a,Marin2015}, where the linear continuum polarization was 
found to vary continuously with inclination. The pairing between observed optical polarization measurements and Seyfert types 
has been examined for a sample of 53 objects in \citet{Marin2014} and this paper will now extend this investigation to 
a larger number of AGN, including a diagnostic of four different methods used to retrieve inclination estimations.

The polarimetric data are listed in Tab~\ref{Table:Polarization}, Tab~\ref{Table:PolScattDomAGN} and Tab~\ref{Table:HighPolS1},
and the plots of optical, continuum, linear polarization $P$ versus inclination $i$ are shown in Fig.~\ref{Fig:Polarization}.
In the case of the full sample, there is a clear dichotomy between type-1s and type-2s in terms of polarization degrees, with 
Seyfert-2s showing much larger $P$. However, there appears to be no clear correlation within each individual group. This is 
particularly relevant for type-1 objects, showing a rather large $P$ dispersion for a given $i$. This could result from the 
competition of parallel (arising from from the accretion flow between the torus and the accretion disk) and perpendicular (from 
the torus funnel - depending on its half-opening angle - and the polar outflows) polarization components, as type-1s sometimes 
show perpendicular polarization position angles \citep{Smith2002}. These peculiar objects are called ``polar scattering dominated 
AGN'' and are identified and listed in Tab.~\ref{Table:PolScattDomAGN}. The small number of polar scattering dominated AGN in 
this sample is unlikely to be the explanation for the dispersion; some of their inclination angles are probably misestimated. This 
appears clearly when plotting of $P$ versus $i$ for the four methods: the M-$\sigma$ methods shows no correlation, and the X-ray 
and IR methods are only weakly correlated ($\rho$ = 0.28, $\tau$ = 0.16 -- 0.19). The only strong correlation arises from the NLR 
orientation indicator ($\tau$ = 0.59), where $P$ is almost zero at pole-on inclinations, then rises to about 1\% at 20$^\circ$  
before decreasing until $i$ reaches $\sim$~40$^\circ$, where the polarization starts to rise to tens of percents at type-2 inclinations. 
This behavior is in excellent agreement with the predictions arising from numerical modeling of the Unified Scheme. The 
polarization degree is expected to rise with increasing viewing angles, then decrease at intermediate orientations due to the 
competition between parallel and perpendicular polarizations, and finally rise again at type-2 angles due to perpendicular 
scattering in the polar outflows \citep{Marin2012a,Marin2015}.

\subsection{Flux ratios}
\label{Exploiting:Fluxes}

The anisotropic arrangement of obscuring matter around AGN, with most of the dust grains and gases located along the equatorial 
plane, can be used as a strong proxy to estimate whether the object is seen through the circumnuclear dust funnel (pole-on view), or 
if the radiation is severely obscured (edge-on view). This will result in different fluxes, the former being up to orders of 
magnitude higher (depending on the waveband considered). However, as stated in Sect.~\ref{Exploiting:nH}, the distribution of 
matter around AGN probably varies with inclination, rather than being a binary function, and thus should result in 
inclination-dependent fluxes. Hence, in the following subsections, two wavebands will be investigated to test this hypothesis: 
the 2-10~keV X-ray and the 6~$\mu$m IR fluxes. All fluxes are extracted from the NASA/IPAC Extragalactic Database (HEASARC) 
and corrected for redshift.

To normalize the X-ray and NIR fluxes, the $IRAS$ 25~$\mu$m fluxes were chosen. Based on tight MIR/X-ray correlation,
\citet{Gandhi2009} and \citet{Asmus2015} showed that the MIR radiation (at least at 25~$\mu$m, as the MIR definition also includes 
shorter bands) is emitted almost isotropically by dust-reemission (see also \citealt{Ichikawa2012} or \citealt{Honig2011} for high 
redshift radio-galaxies). The isotropy of MIR emission is supported by interferometric results, where dust re-emission is not only 
found to originate from the dusty circumnuclear region, but also (and probably predominantly) from the polar outflows 
\citep{Honig2012,Honig2013,Tristram2014}. In that case, the MIR anisotropy between face-on and edge-on systems is possibly much 
lower and thus $IRAS$ 25~$\mu$m fluxes can be chosen as a valid normalization parameter\footnote{\citet{Asmus2015} discuss this results 
in the context of the torus scenario and present a number of alternatives to explain the MIR emission isotropy.}.

\subsubsection{XMM-Newton 2-10~keV fluxes}
\label{Exploiting:Fluxes:SoftX}

\begin{figure*}
    \begin{center}
      \begin{tabular}{c}
	\includegraphics[trim = 0mm 0mm 0mm 0mm, clip, width=8cm]{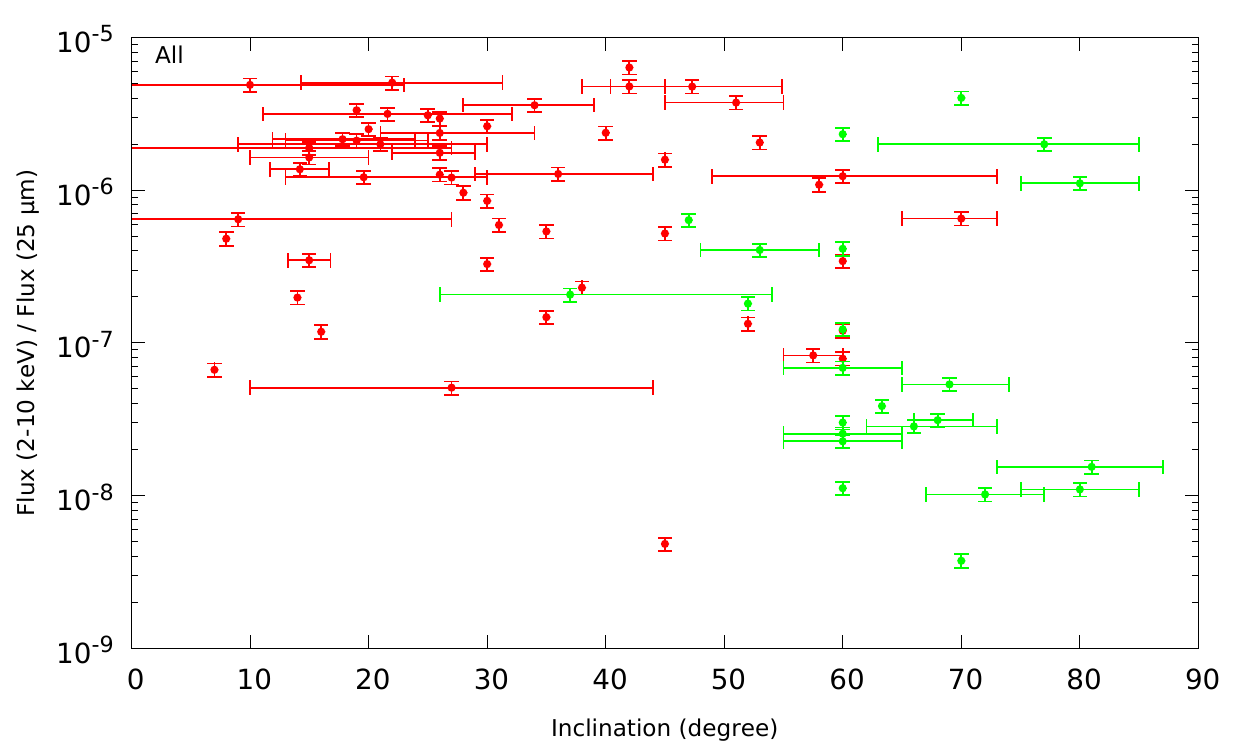} \\
      \end{tabular}
      \begin{tabular}{cc}
	\includegraphics[trim = 0mm 0mm 0mm 0mm, clip, width=8cm]{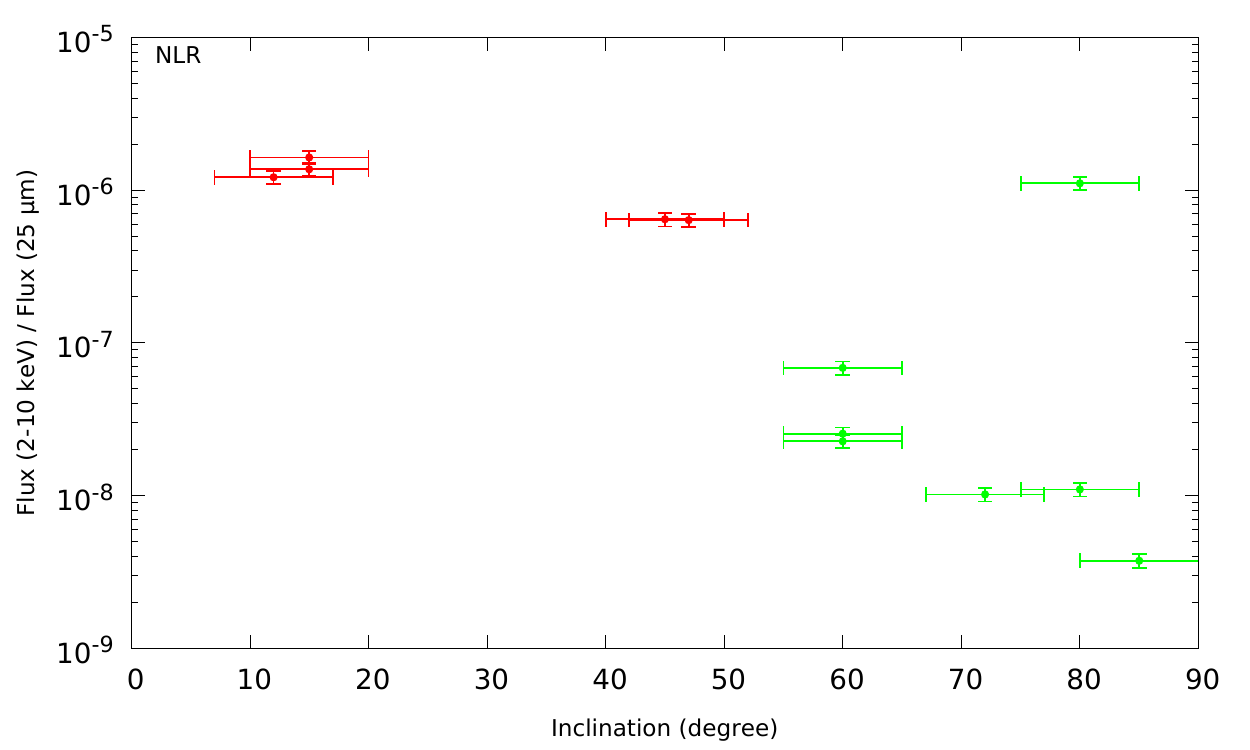} & \includegraphics[trim = 0mm 0mm 0mm 0mm, clip, width=8cm]{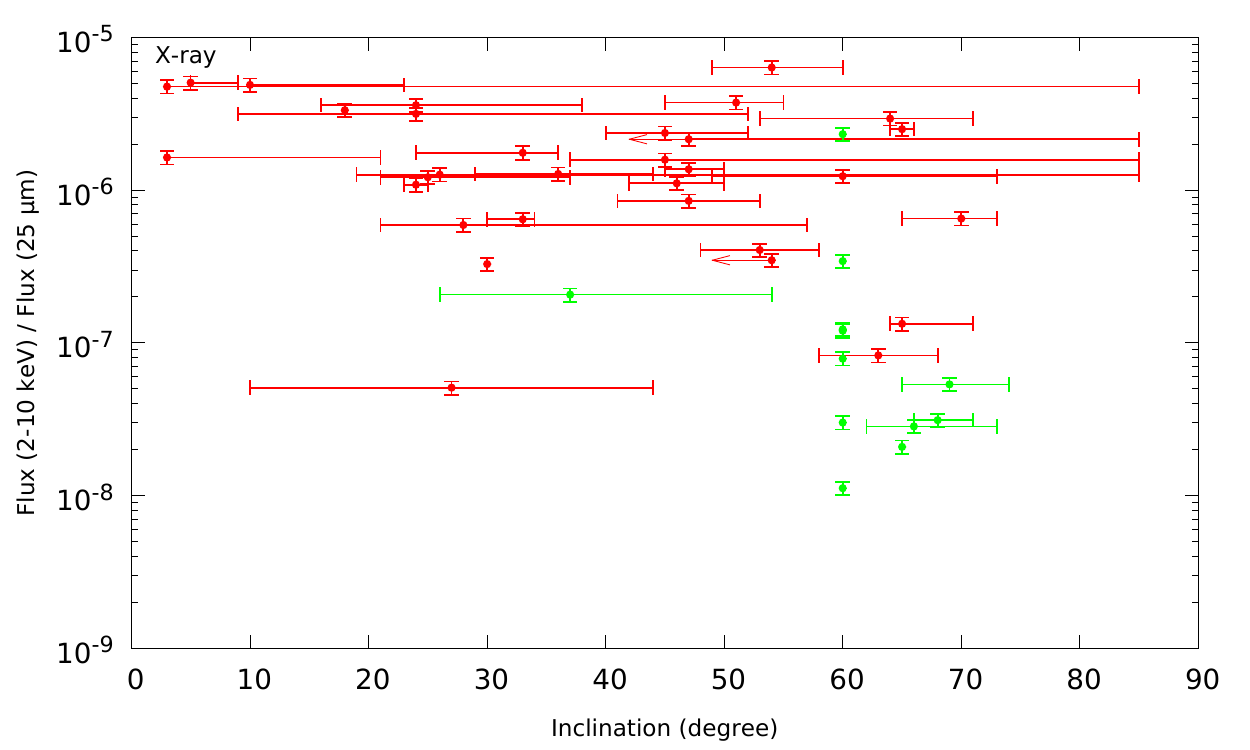} \\
      \end{tabular}
      \begin{tabular}{cc}
	\includegraphics[trim = 0mm 0mm 0mm 0mm, clip, width=8cm]{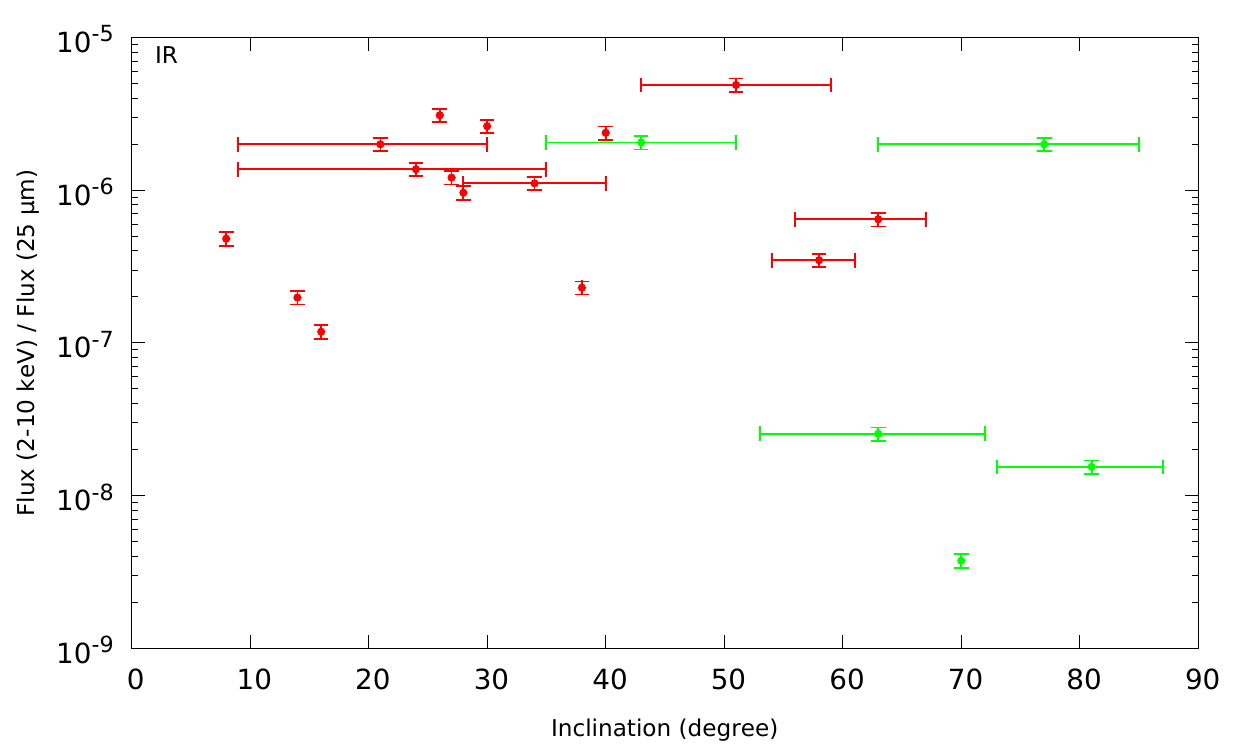} & \includegraphics[trim = 0mm 0mm 0mm 0mm, clip, width=8cm]{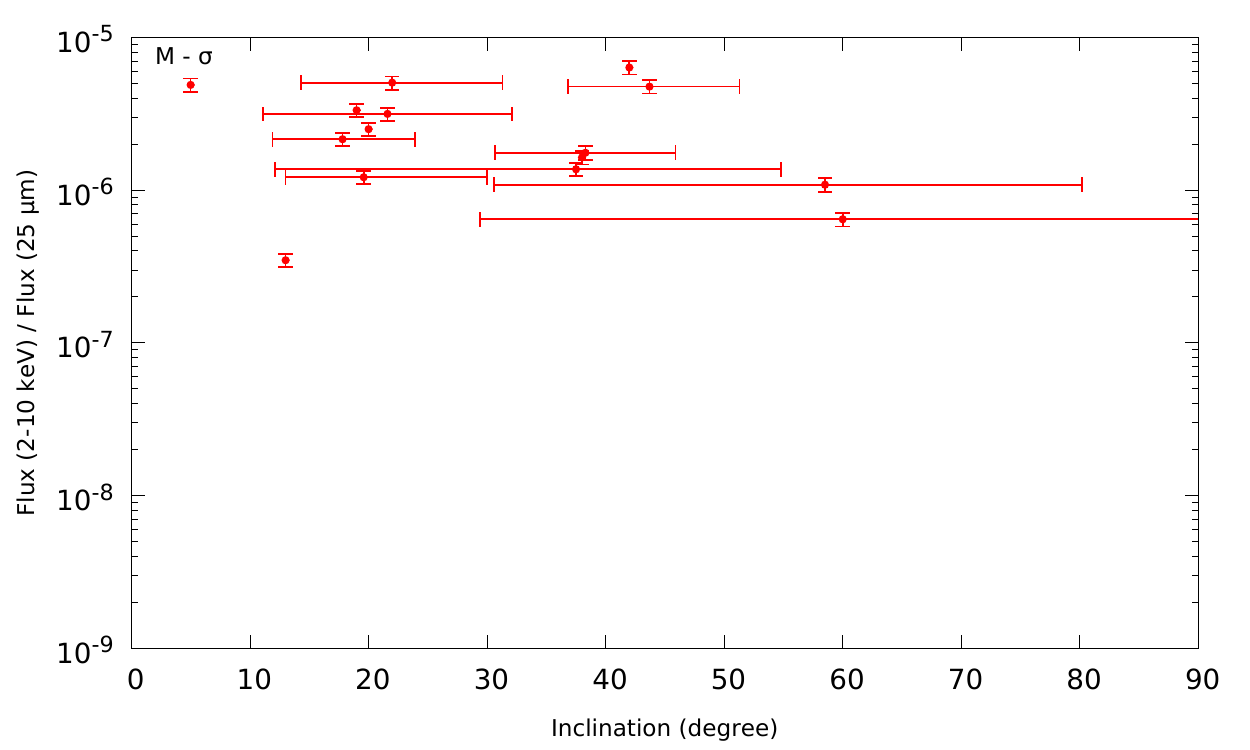} \\
      \end{tabular}
    \end{center}
    \caption{Ratio of the $XMM-Newton$ 2-10~keV and $IRAS$ 25~$\mu$m 
	     fluxes as a function of AGN inclination derived from 
	     the indicated method (see text). Legend is the same as 
	     in Fig.~\ref{Fig:Histograms}.}
    \label{Fig:IRSoftX}%
\end{figure*}

The X-ray radiation of AGN is generally thought to result from Compton up-scattering of thermal photons in a hot corona. 
This electron plasma is usually thought to be located in a compact region (few tens of gravitational radii) above the accretion 
disk \citep[e.g.][]{Bisnovatyi1976,Haardt1991,Haardt1993,Dovciak2004,Wilkins2012,Wilkins2014}, which means that the X-ray source 
is fairly close to the equatorial plane. If the Compton-thick matter that obscures the view of an observer along type-2 inclinations 
has a height larger than the disk-corona distance, a coplanar observer is thus not likely to see direct X-ray radiation from 
the corona. The observed X-ray radiation from AGN is therefore expected to be anisotropic. The 2-10~keV band was selected for the 
numerous XMM-Newton observations that were available in the literature. Harder photons have sufficient energies to pass through 
the equatorial dust and gas, and softer photons would be too much attenuated by photoelectric absorption by the interstellar 
and intergalactic media. 

As can be seen in Fig.~\ref{Fig:IRSoftX}, the ratio of X-ray-to-MIR fluxes versus inclination is complicated in the case of the 
complete sample. Only the most extreme type-2 objects show a net starvation of photons due to equatorial obscuration, 
and the difference between type-1 and type-2 AGN is not clear. The same conclusions apply to the IR torus-fitting and 
M-$\sigma$ methods, where no correlation is found. The X-ray reflection spectroscopy technique shows a weakly significant 
anticorrelation ($\rho$ = -0.44, $\tau$ = -0.32) while the [O~{\sc iii}]-mapping indicator clearly stand out: the AGN flux 
ratio shows a net weakening with increasing inclinations (Fig.~\ref{Fig:IRSoftX}, middle-left), an anticorrelation supported by 
large $\rho$ and $\tau$ values (-0.79 and -0.68, respectively). There is only one type-2 outsider in this method, NGC~5506, a 
peculiar case that will be discussed in Sect.~\ref{Discussion:best}. Between pole-on and edge-on views, the flux ratio differs 
by a factor 100, which is consistent with a circumnuclear material with a half-opening angle of 50$^\circ$ -- 60$^\circ$ with 
respect to the torus symmetry axis. This threshold value is in agreement with the torus half-opening angles that have been 
found by \citet{Shen2010}, \citet{Marin2014} and \citet{Sazonov2015}

\subsubsection{Spitzer 6~$\mu$m fluxes}
\label{Exploiting:Fluxes:6um}

\begin{figure*}
    \begin{center}
      \begin{tabular}{c}
	\includegraphics[trim = 0mm 0mm 0mm 0mm, clip, width=8cm]{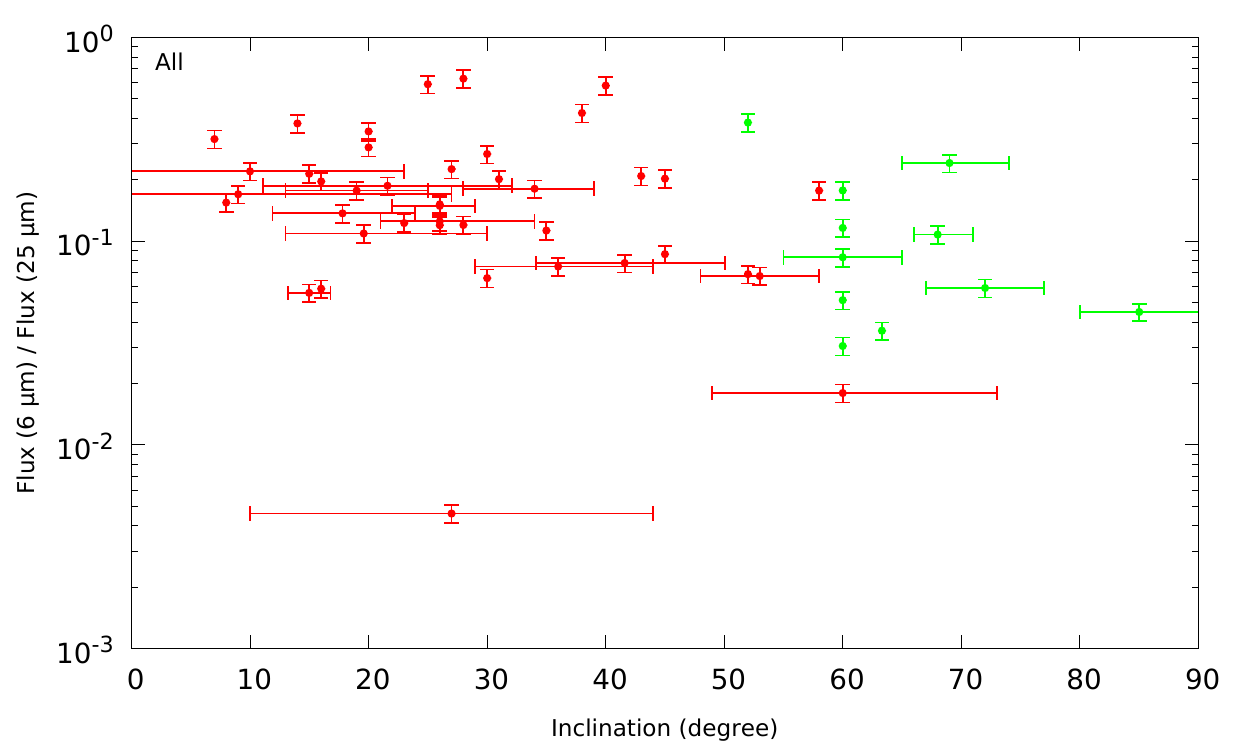} \\
      \end{tabular}
      \begin{tabular}{cc}
	\includegraphics[trim = 0mm 0mm 0mm 0mm, clip, width=8cm]{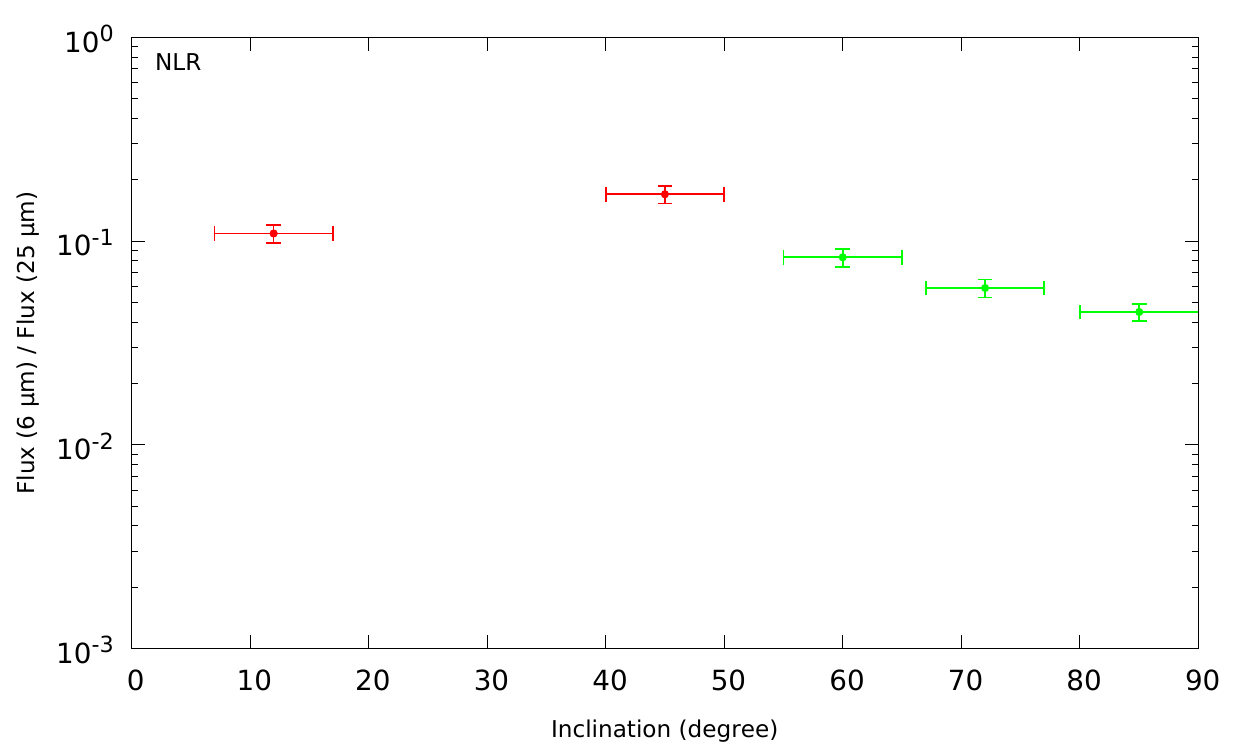} & \includegraphics[trim = 0mm 0mm 0mm 0mm, clip, width=8cm]{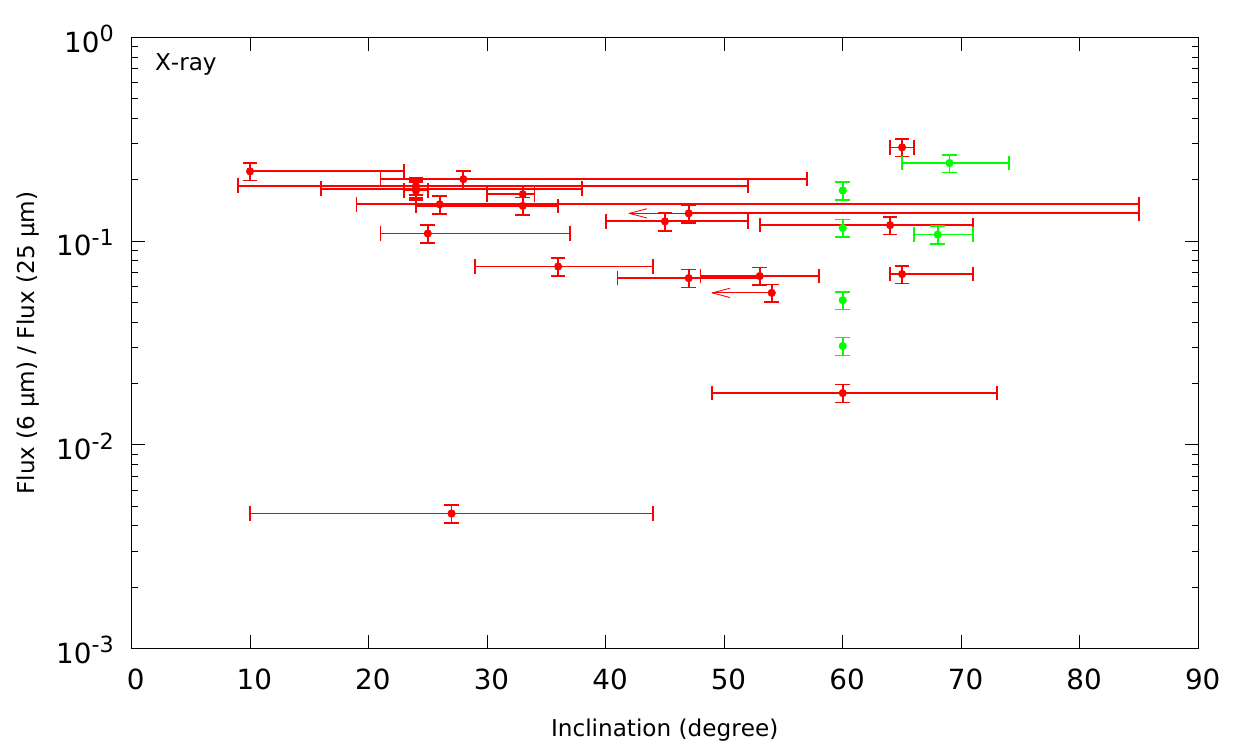} \\
      \end{tabular}
      \begin{tabular}{cc}
	\includegraphics[trim = 0mm 0mm 0mm 0mm, clip, width=8cm]{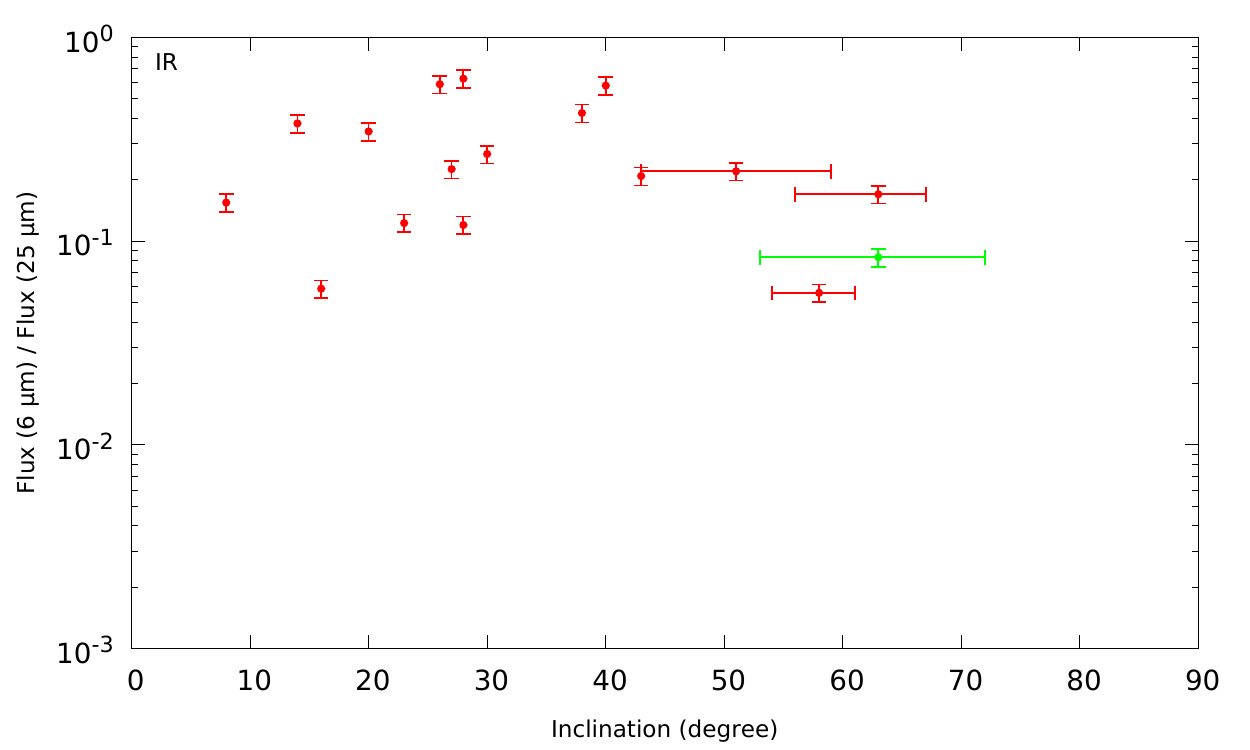} & \includegraphics[trim = 0mm 0mm 0mm 0mm, clip, width=8cm]{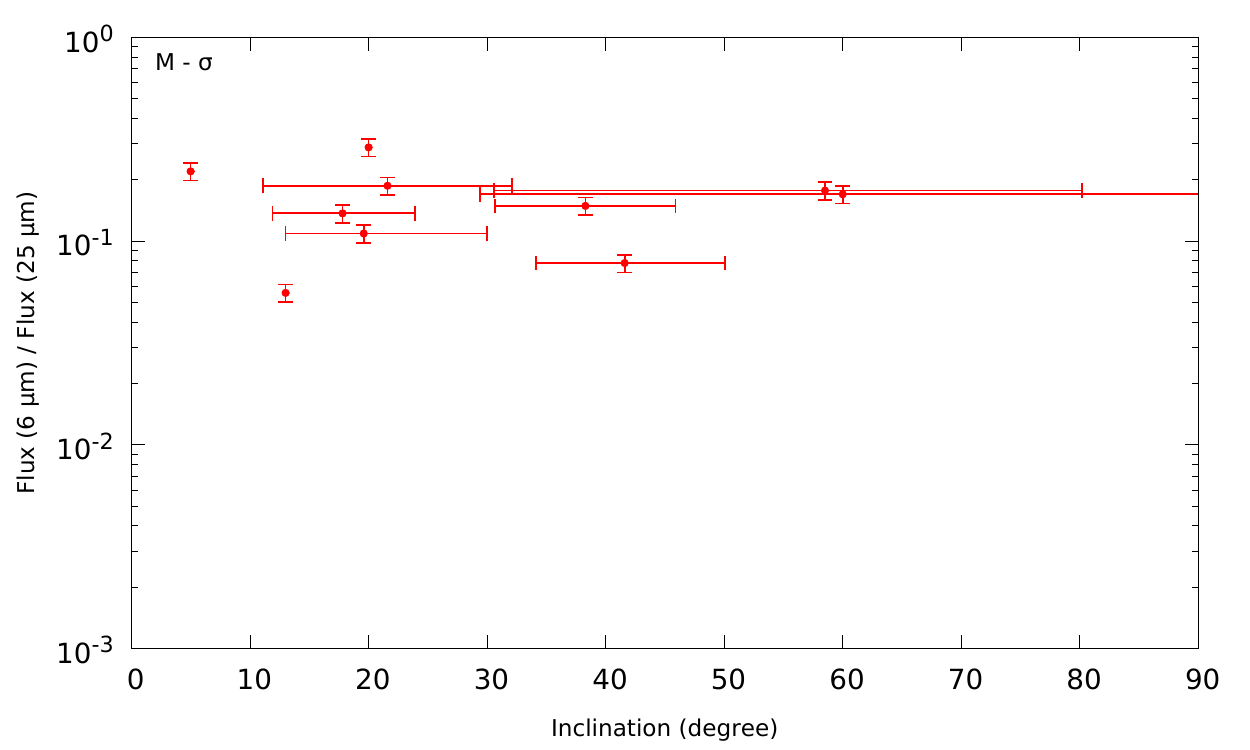} \\
      \end{tabular}
    \end{center}
    \caption{Ratio of the $Spitzer$ 6~$\mu$m and $IRAS$ 25~$\mu$m 
	     fluxes as a function of AGN inclination derived from 
	     the indicated method (see text). Legend is the same 
	     as in Fig.~\ref{Fig:Histograms}.}
    \label{Fig:Spitzer6}%
\end{figure*}

Dust clouds situated at the outer rim of the accretion disk are thought to re-radiate the disk emission in the IR band, from 0.5 to a 
couple of microns \citep{Phinney1989}. At longer wavelengths, isotropic MIR dust re-emission will dominate while the NIR disk signature ends. 
This effect has been observed by \citet{Deo2009}, who have found a deficit of 5.5~$\mu$m continuum flux density in Seyfert-2 AGN with 
respect to comparable Seyfert 1s. This confirms the hypothesis that the accretion disk is obscured at type-2 viewing angles and that 
near- and mid-IR flux ratio can be related to the system inclination. This method was used by \citet{Fischer2014} to test the robustness 
of their inclination predictions, using measurements from the $Spitzer$ satellite. In this section, the $Spitzer$ 6~$\mu$m fluxes from 
literature are normalized by the $IRAS$ 25~$\mu$m fluxes for consistency with the previous investigation.

Results are shown in Fig.~\ref{Fig:Spitzer6}. The flux ratio versus inclination is not conclusive in the case of the full AGN sample, as 
type-1 and type-2 objects are almost indistinguishable in terms of fluxes. Similarly to Sect.~\ref{Exploiting:Fluxes:SoftX}, the 
M-$\sigma$, X-ray and IR methods fail to show the expected correlation, but the NLR method by \citet{Fischer2013} remarkably 
stands out. The progressive diminution of flux with inclination is clearly visible and supported by the Spearman and Kendall rank 
correlation coefficients\footnote{Despite the limited number of points from the NLR method in the case of the 6~$\mu$m/25~$\mu$m fluxes 
ratio, the Kendall rank correlation and Spearman correlation remain reliable bivariate analyses as they are also adapted to small 
populations.} ($\rho$ = -0.90, $\tau$ = -0.80, see Tab.~\ref{Table:FinalResults}). The type-1/type-2 difference is more subtle than 
in the case of X-ray radiation, as the anisotropic contribution of the outer part of the disk is weak in comparison with the isotropic 
torus emission that also contributes to the total infrared flux. Nevertheless, the inclinations derived by the NLR method are precise 
enough to reveal the correlation between the IR flux ratio and inclinations.

\begin{table*}
    \caption{Spearman $\rho$ (top) and Kendall $\tau$ (bottom) rank 
	     correlation coefficients evaluated for the four methods 
	     tested in this paper. The values in parenthesis are the 
	     two-tailed p-values. Cells showing a blue color are highly 
	     statistically significant correlations (at $>$~95\% confidence
	     level for rejecting null hypothesis). The number of 
	     sources for each sample is indicated in Tab~\ref{Table:VEL}, 
	     \ref{Table:X}, \ref{Table:IR}, and \ref{Table:NLR}.}
     \begin{threeparttable}
     \centering
        \begin{tabular}{|l|c|c|c|c|}
            \hline \textbf{Spearman correlation $\rho$} 	& \textbf{NLR} 				& \textbf{X-ray} 	& \textbf{IR} 				& \textbf{M - $\sigma$}\\
            \hline n$_{\rm H}$ vs $i$ 				& \cellcolor{blue!25}0.60 (0.02)	& 0.36 (0.0096)		& \cellcolor{blue!25}0.60 (0.00041) 	& 0.31 (0.21)\\
            H$\beta$ FWHM vs $i$ 				& \cellcolor{blue!25}0.97 (0.0048)	& -0.15 (0.38) 		& 0.034 (0.87) 				& 0.44\tnote{a} (0.066)\\
            $P$ vs $i$ 						& \cellcolor{blue!25}0.59 (0.012)	& 0.28 (0.09) 		& 0.28 (0.092) 				& -0.14 (0.58)\\
            F$_{\rm 2-10~keV}$/F$_{\rm 25~{\mu}m}$ vs $i$ 	& \cellcolor{blue!25}-0.79 (0.0023)	& -0.44 (0.002) 	& -0.15 (0.51) 				& -0.11 (0.69)\\
            F$_{\rm 6~{\mu}m}$/F$_{\rm 25~{\mu}m}$ vs $i$ 	& \cellcolor{blue!25}-0.90 (0.037)	& -0.11 (0.60)	 	& -0.17 (0.51) 				& 0.055 (0.88)\\
	    \hline ~ & ~ & ~ & ~ & ~ \\
            \hline \textbf{Kendall correlation $\tau$} 		& \textbf{NLR}				& \textbf{X-ray} 	& \textbf{IR} 				& \textbf{M - $\sigma$}\\
            \hline n$_{\rm H}$ vs $i$ 				& \cellcolor{blue!25}0.45 (0.02)	& 0.25 (0.0095)		& \cellcolor{blue!25}0.41 (0.0015) 	& 0.26 (0.14)\\
            H$\beta$ FWHM vs $i$ 				& \cellcolor{blue!25}0.95 (0.043) 	& -0.11 (0.35) 		& 0.032 (0.83) 				& 0.34\tnote{a} (0.053)\\
            $P$ vs $i$ 						& 0.39 (0.038) 				& 0.19 (0.09) 		& 0.16 (0.17) 				& -0.10 (0.59)\\
            F$_{\rm 2-10~keV}$/F$_{\rm 25~{\mu}m}$ vs $i$ 	& \cellcolor{blue!25}-0.68 (0.0035)	& -0.32 (0.003) 	& -0.13 (0.46) 				& -0.067 (0.77)\\
            F$_{\rm 6~{\mu}m}$/F$_{\rm 25~{\mu}m}$ vs $i$ 	& \cellcolor{blue!25}-0.80 (0.086)	& -0.11 (0.47)		& -0.14 (0.46) 				& 0.022 (1.0)\\
            \hline
        \end{tabular}
    \begin{tablenotes}
            \item[a] Biased values. See text for details.
    \end{tablenotes}
    \label{Table:FinalResults} 
    \end{threeparttable}
\end{table*}

\section{Discussion}
\label{Discussion}

\subsection{Which one is the best inclination indicator?}
\label{Discussion:best}
The investigations presented in this paper focused on several inclination-dependent indicators, namely n$_{\rm H}$, H$\beta$ linewidth, 
optical polarization, F$_{\rm 2-10~keV}$/F$_{\rm 25~{\mu}m}$, and F$_{\rm 6~{\mu}m}$/F$_{\rm 25~{\mu}m}$, to test the reliability 
of four techniques (M-$\sigma$ relation, NLR modeling, X-ray fitting, and IR fitting) used to retrieve/estimate the nuclear 
orientation $i$ of AGN. Each method focuses on a specific AGN component: the accretion disk in the case of the X-ray method, 
the dusty torus in the IR fitting technique, the NLR in \citet{Fischer2013} and the internal regions of the host galaxy in 
the case of the M-$\sigma$ relation, and they are found to have different reliability. The Spearman and Kendall rank correlation 
coefficients, presented in Tab.~\ref{Table:FinalResults}, highlight the valid and invalid indicators. 

The method based on the empirical correlation found between M$_{\rm BH}$ and $\sigma$ by \citet{Gebhardt2000} and \citet{Ferrrarese2000}
proved to be ineffective to reproduce the expected correlations between $i$ and the observed properties. It means that the M-$\sigma$ 
relationship, valid for estimating the black hole mass in non-active galaxies, cannot be applied to AGN to infer the inclination. 
The fact that the derived inclinations agree with the mean angle obtained by fitting the iron K$\alpha$ lines of Seyfert~1 
galaxies observed with $ASCA$ \citep{Wu2001} is probably an occurrence based on chance. However, this conclusion does not affect 
the findings of \citet{Xiao2011} and \citet{Woo2015}, who explored the low-mass end of the M-$\sigma$ relation using narrow-line 
Seyfert 1 galaxies and found that the NLS1s do not significantly deviate from the expected black hole mass - stellar velocity 
dispersion trend, despite an observed offset with the host galaxy morphology.

The X-ray fitting method, taking into account a curved space-time, fails to reproduce the expected inclination-dependent trends 
at high statistical significance, though several weak (anti)correlations have been found when looking at dependencies between 
$i$ and n$_{\rm H}$ and F$_{\rm 2-10~keV}$/F$_{\rm 25~{\mu}m}$. A non-isotropically emitting X-ray corona could reinforce 
the weak correlation between $i$ and F$_{\rm 2-10~keV}$/F$_{\rm 25~{\mu}m}$ \citep{Yang2015}, yet none of the evaluated Kendall 
rank correlation coefficients exceed $\tau$ = 0.40, which means fitting the broad, asymmetric, iron K$\alpha$ line is not a 
clear indicator of the global AGN inclination. In particular, measuring the AGN inclination using X-ray spectroscopy is hampered 
by the fact that the method is biased towards low inclinations: the equivalent width and the reflection fraction decrease with the 
inclination angle, so highly inclined disks are more difficult to detect \citep{Fabian2000}. Moreover, the larger the inclination 
the broader the line, which also plays against detectability.

The third method, based on IR fitting and modeling of the dusty torus, only succeeds to confirm the expected correlation between 
n$_{\rm H}$ and $i$ at $>$~95\% confidence level for rejecting null hypothesis ($\rho$ = 0.60, $\tau$ = 0.41). This is not 
surprising as the number density of obscuring clouds is a key parameter in the IR models, and it is used to fine tune the final 
inclination of the AGN. However, as noted by \citet{Feltre2012}, the dust morphology, either smooth or clumpy, has little impact 
of the modeled SED in modern simulations. Degeneracies may then arise and this would explain why the IR torus modeling fails to reproduce 
the other inclination-dependent trends \citep{Honig2010}. The lack of statistical correlations can also be reinforced by the 
fact that optical and UV radiation are probably emitted anisotropically, with fewer photons transmitted in the direction closer 
to the equatorial plane \citep{Kawaguchi2010}, a feature that is not ubiquitously simulated in all IR models.

It is clear from Tab.~\ref{Table:FinalResults} that the method by \citet{Fischer2013}, based on kinematic models matching 
the radial velocities of the [O~{\sc iii}]-emitting NLR, is the best inclination indicator tested so far. It succeeded to 
reveal highly statistically significant correlations (at $>$~95\% confidence level for rejecting null hypothesis) between $i$ 
and n$_{\rm H}$, H$\beta$ FWHM, $P$, F$_{\rm 2-10~keV}$/F$_{\rm 25~{\mu}m}$ and F$_{\rm 6~{\mu}m}$/F$_{\rm 25~{\mu}m}$. 
\citet{Fischer2014} noticed that Mrk~279 and NGC~5506 were almost always outliers in the trends they investigated. The former
because column densities from several of its absorbers are yet to be determined, the latter because the modeled inclination angle 
is certainly degraded by a highly inclined host disk. The optical classification of NGC~5506 is somewhat debated as 
\citet{Goodrich1994} found that the Pa$\beta$ line profile is consistent with the type-2 category, while there is also evidence for 
permitted O~{\sc i} $~\lambda 1.1287~\mu$m line (with FWHM $<$ 2000 km/s) and several Fe~{\sc ii} lines in the 0.9-1.4 $\mu$m 
spectrum observed by \citet{Nagar2002}. \citet{Fischer2013} used the type-2 classification in their paper and several other
authors \citep[e.g.][]{Nikolajuk2009,Matt2015} followed the narrow-line Seyfert-1 from \citet{Nagar2002}; in our case removing 
NGC~5506 from the type-2 category and labeling it as a type-1 would increase the value of the rank correlation coefficients. 
Using a maximum-likelihood estimation to get the best values of $\rho$ and $\tau$ from the n$_{\rm H}$ and 
F$_{\rm 2-10~keV}$/F$_{\rm 25~{\mu}m}$ observables, it is possible to estimate the real orientation of NGC~5506: 40$^\circ$ $\pm$ 4$^\circ$. 
This value is in agreement with the corrected inclination derived by \citet{Fischer2014}: 40$^\circ$.

The best inclination indicator is thus provided by \citet{Fischer2013}, but the necessity to have a well resolved (distinct 
knots of emission visible over several arcseconds) NLR structure is a major limitation to the method. It constrains the 
analysis to nearby, bright Seyferts were long-slit spectroscopy can be applied, but this still represents hundreds of 
objects \citep{Bennert2006a,Bennert2006b,Crenshaw2000,Crenshaw2000b,Crenshaw2000c,Fischer2013}. However, it remains 
unclear how the kinematic model of the authors may account for the changes and misalignments in the polar outflows between 
the torus and the inner and outer NLR components that have been observed in a couple of AGN. In the case of NGC~1068,
\citet{Raban2009} have noted the misalignment of the HST-revealed ionization cone and NLR region with respect to the compact
and equatorial dust component. Despite lying $\sim$ 40 degrees further North from the torus symmetry axis, the observations 
are still consistent with the kinematic modeling of \citet{Das2006} if the ionized wind is partially obscured by large gas 
clouds, as suggested by \citet{Kishimoto1999}. The question remains opened for the treatment of the base of the polar wind,
where \citet{Muller2006} located the coronal line region (CLR). The CLR appears to be a medium where forbidden fine-structure 
transitions in the ground level of highly ionized atoms are responsible for the emission of highly ionized lines. \citet{Muller2011}
observed changes between the kinematics of the CLR and NLR for 6 radio-quiet AGN, among which the Circinus galaxy,
NGC~1068, and NGC~4151 have been analyzed by \citet{Fischer2013}. The statistically probable inclination derived for those 
three AGN indicates that the NLR orientation indicator remains valid at the scales of the CLR, but this needs to be confirmed 
for the remaining half of the sample of \citet{Muller2011}.

\subsection{Orientation duplicates and uncertainty}
\label{Discussion:Duplicates}

\begin{figure}
  \centering
    \includegraphics[trim = 0mm 0mm 0mm 0mm, clip, width=8.5cm]{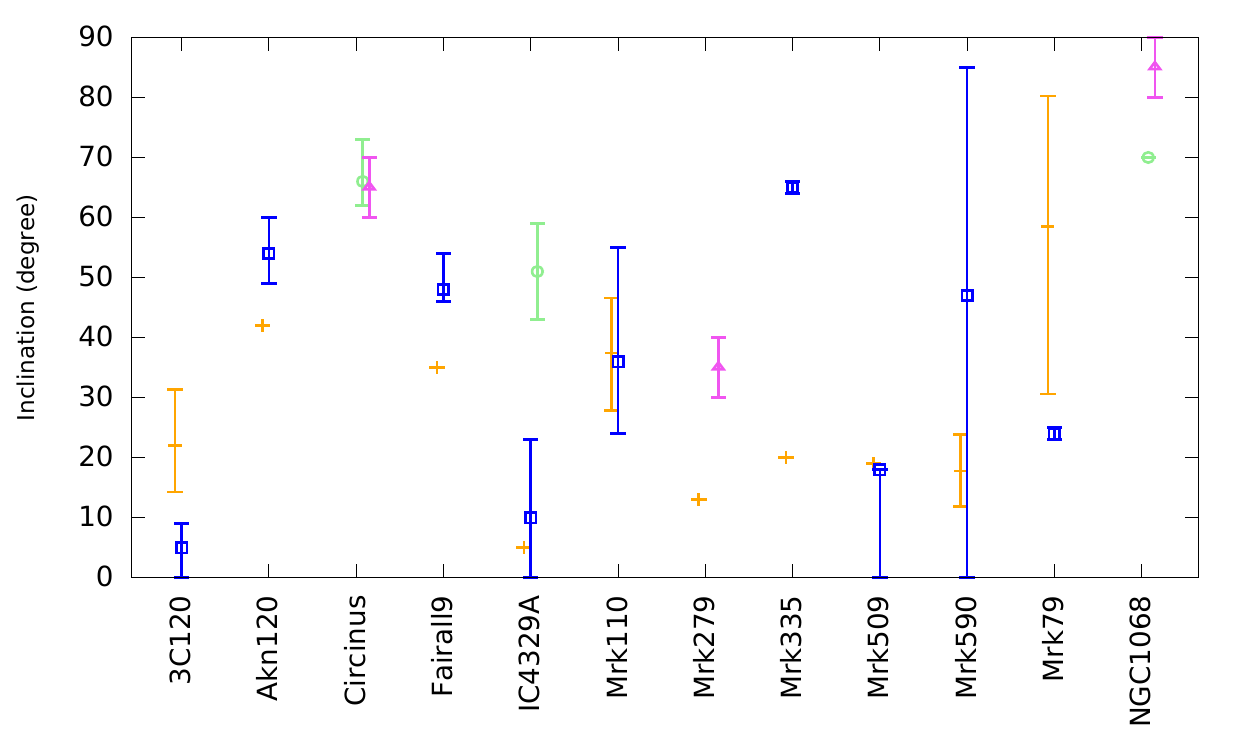}
    \includegraphics[trim = 0mm 0mm 0mm 0mm, clip, width=8.5cm]{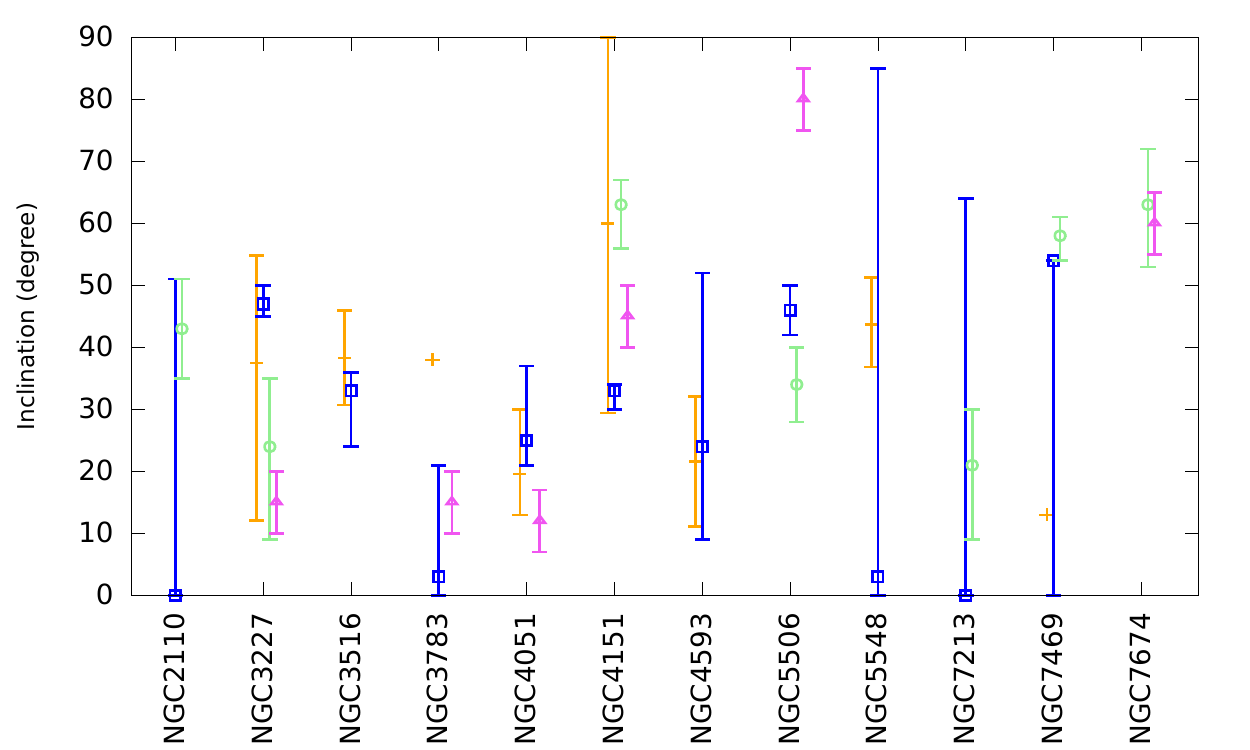}
    \caption{Comparison of the inclination duplicates. M-$\sigma$ method: 
	     orange crosses, X-ray: blue squares, IR: green circles, 
	     NLR: pink triangles.}
   \label{Fig:Duplicates}
\end{figure}

As stated in Sect.~\ref{Comp:Inclination}, our 124 Seyfert sample takes into account only one inclination per object despite the 
potential presence of duplicates. However but it has been shown through this paper that the global sample is not reliable. In this case, 
it is worth investigating if the sample could be fine tuned by selecting more reliable inclinations. Fig.~\ref{Fig:Duplicates} presents 
the orientation angles for 24 Seyferts that have duplicate estimations from at least two different methods. The M-$\sigma$ 
approach is marked with orange crosses, the X-ray fitting indicator with blue squares, the IR-torus fitting method using green 
circles, and the NLR technique with pink triangles. No dependency between the indicators is found: the angles estimated by 
\citet{Fischer2013} are not predominantly higher nor lower than the angles estimated from another method. In 66\% of the cases the 
NLR-derived angle is consistent within uncertainty with another technique, but this technique is not always the same. Overall, the 
data dispersion around the inclinations given by the NLR method shows no trend, and the same conclusion is valid for each of the three 
remaining methods.

It is interesting to note that the uncertainties retrieved by the 6.4~keV fitting technique are often either very large or unspecified. 
In this context, it is more difficult to estimate the reliability of the method. The lack of errors is due to the parametrization of 
the Kerr and Schwarzschild relativistic reflection models, which can be programmed with a fixed inclination. This results in an 
observer-biased choice that is often set to 30$^\circ$ (e.g., Mrk~509 \citep{Pounds2001}, Mrk~590 \citep{Longinotti2007}, 
NGC~5548 \citep{Brenneman2012}, NGC~7213 \citep{Ursini2015} or NGC~7469 \citep{Chiang2002}), or frozen at 45$^\circ$ (e.g., 
Mrk~509 \citep{Mehdipour2011} or NGC~7213 \citep{Emmanoulopoulos2013}) for type-1 objects. Type-2 inclinations are often set to 
60$^\circ$ \citep[e.g.][]{Noguchi2010}. However, if the inclination angle of the disk is left to vary, it will be retrieved by the 
model using reduced $\chi^2$ statistics to evaluate the goodness of fit. The inclinations found through this less-biased method 
are almost always dominated by large errors due to the considerable number of free parameters \citep[e.g.][]{Nandra1997,Nandra2007}. 

Improving numerical tools is necessary to narrow the uncertainties around the inclination angles derived from X-ray spectral 
fitting \citep[e.g.][]{Walton2013,Middleton2016} and test, once and for all, the reliability of the X-ray spectroscopic method.
In this regard, the help of X-ray polarimetry will be crucial. By fitting the measured spectroscopic and polarimetric information 
with a unique model (either using reflection spectroscopy or partial covering scenarios, see e.g. \citealt{Iso2016}), the constraints 
on the nuclear inclination of AGN will be much stronger as the number of free parameters will be drastically reduced. Information 
about the disk inclination, the black hole spin, the luminosity in the thermal flux, and the optical depth, electron temperature 
and scale height of the corona will all become available \citep{Dovciak2004,Schnittman2010,Dovciak2011,Marin2012b,Marin2013}.

\subsection{Coplanarity within the first parsec}
\label{Discussion:Coplanar}
AGN are usually depicted as axisymmetric structures, where the accretion disk, the BLR and the optically thick equatorial 
torus are coplanar, and the analyses of this paper have been carried out according to this hypothesis. This is the standard 
procedure applied by almost every author \citep[e.g.][]{Fine2008,Fine2010}, but alternative scenarios exist. \citet{Pringle1996} 
suggested that a powerful radiation source such as the center of AGN can modify the dynamics of the accretion disk and induce 
a warp that could explain the IR emission of Seyferts. If the equatorial rotating region is subject to a strong outflow, 
its surface might be even more unstable to warping and lead to detectable signatures in luminous accreting objects
that generate energetic winds \citep{Quillen2001}. The idea of twisted accretion disks is not new 
\citep{Petterson1977a,Petterson1977b,Petterson1978} but remains little studied despite the observational evidence
of warps in maser structures\footnote{Radio observations of NGC~4258 \citep{Miyoshi1995} and the Circinus galaxy 
\citep{Greenhill2000} have revealed that the 22~GHz (1.35~cm) water maser line emission arises from a twisted thin disk 
that could correspond to the outer rim of the AGN accretion structure.}, that are often related to the outer part of 
accretion disks \citep{Maloney1996}. If the dense, thin, rotating accretion disk is sufficiently twisted with 
respect to the BLR or torus regions, the inclination estimations produced by fitting the observed X-ray spectroscopic 
features would only apply to the innermost AGN regions as all the actual models use coplanar, geometrically flat disks. 
This would explain why the iron K$\alpha$ fitting method fails to produce reliable global inclinations.

This warping would naturally cause another effect, the non-alignment of the BLR-torus structures with the inner 
parts of the accretion disk. \citet{Bian2005} already mentioned that the lack of correlation found between the H$\beta$ 
linewidth measurements and the inclination of the accretion disk derived from X-ray spectral fitting (visible in 
Fig.~\ref{Fig:Hbeta}) seems to indicate that the BLR-torus structures are not coplanar with the accretion disk. A similar 
conclusion is shared by \citet{Nishiura1998}, who additionally found hints of a negative correlation between H$\beta$
and $i$, which is also what is found in this paper (but with weak rank correlation coefficients: $\rho$ = -0.15 
and $\tau$ = -0.11). They derived a radial distance of 0.01~pc from the central SMBH where the BLR and the outer parts 
of the disk should be still coplanar. It means that the inclinations derived by the NLR method are valid from kiloparsec 
scales to a fraction of a parsec, and when the dimensions to be probed meet the inner parts of the accretion disk, another 
method such as X-ray spectral fitting might become valid.

The situation seems to be the same in radio-loud quasars. \citet{Risaliti2011} looked at the [O~{\sc iii}] equivalent 
width of thousands of SDSS quasars and reached the same conclusion about the lack of coplanarity between the accretion 
disk and the circumnuclear absorber. This could be due to either a random alignment between the disk and the torus, 
or a very small torus covering factor in the case of quasars. If radio-quiet and radio-loud quasars are similar, with
the exception of the jet, the former hypothesis appears stronger. This is supported by the optical measurements of 
the polarization position angle of nearby AGN that almost always deviate from perfect parallel or perpendicular orientation
(see, e.g., Tab.~3 in \citealt{Antonucci1984}). This could be the result from non coplanar structures within the equatorial dust 
funnel, but resolution effects must be taken into account as different telescope apertures result in different measurements
of the polarization position angle or radio position angle (see, e.g., \citealt{Bailey1988} in the case of NGC~1068).

\section{Conclusions}
\label{Conclusions}

The conclusions of this paper can be summarized by the following points:

\begin{enumerate}
\item The M-$\sigma$ relationship, valid for non-active galaxies, cannot be applied to AGN 
to retrieve the nuclear inclination. 

\item Spectral fitting of the AGN SED in the IR band is not a good AGN orientation indicator, 
except when compared to estimated hydrogen column densities.

\item X-ray spectral fitting of the broad and asymmetric fluorescent iron line is too model-dependent 
and subject to degeneracies to be a valid method to determine the inclinations of AGN. However, it might 
work well if the disk is warped by radiation or outflows at small scales, leading to a non-coplanarity 
of the disk and the BLR-torus structures. 

\item The method developed by \citet{Fischer2013}, based on the original study of \citet{Crenshaw2000b}, 
has proven to be very effective in reproducing the expected inclination-dependent signatures of all the 
observables investigated in this paper. When targeting the extended polar winds, this technique 
shows that the derived inclination angles are valid at much smaller physical scales (down to a fraction of 
a parsec where the Balmer line signature originates). This would indicate that this orientation indicator might
work on multiple scales, from the extended NLR to the outer parts of the accretion disk. However, the 
misalignment observed for a few AGN between the torus and the inner and outer NLR parts might weaken 
this conclusion.

\item The expected hydrogen column densities at low inclinations are almost two orders of magnitude 
larger with the NLR technique in respect to the other orientation estimators. Further detailed observations
and modeling are needed to test this method.

\item The absence of correlation between the Balmer emission line FWHM and the X-ray-derived 
inclinations confirms that, if the inclinations are correct, the accretion disk is certainly 
not co-aligned with the BLR and torus regions.

\item The orientation of NGC~5506 has been evaluated using the inclination-dependent indicators 
and corresponds to 40$^\circ$ $\pm$ 4$^\circ$.
\end{enumerate}

Additional work is needed in the field of AGN to understand the structure and the three-dimensional arrangement 
of the innermost regions of quasars. We plan to try vetting these methods for radio-loud objects using the core 
dominance parameter, which at least for high redshift 3CRR objects separates the type-1s from the type-2s 
perfectly (Marin \& Antonucci, in prep). Optical polarimetry and [O~{\sc iii}] imaging are among the best tools 
to push forward these analyses, especially when coupled with numerical modeling. Testing the coplanarity (or 
the absence of coplanarity) between the equatorial structures is mandatory to validate or reject the X-ray spectral 
fitting method, leading to a potentially strong modification of the Unified Scheme that would need warped structures 
at its very center.

\section*{Acknowledgments}

I would like to acknowledge the anonymous referee for useful suggestions that helped to clarify 
this paper. I am also grateful to, in alphabetical order, Robert Antonucci, Michal Dov{\v c}iak, 
Ren\'e Goosmann, Vladimir Karas, Giorgio Matt, Delphine Porquet, Marvin Rose, Marko Stalevski, 
Francesco Tamborra, and Belinda Wilkes for their nice comments and helpful suggestions about 
the results of this work. The Torus2015 Workshop organized by Poshak Gandhi and Sebastian 
Hoenig was a great mine of information to achieve this work. Mari Kolehmainen was of a 
great help for English editing. Finally, I am grateful to Jules Garreau (\textcolor{cyan} 
{jul.garreau@wanadoo.fr}) for his artwork of the various components AGN.


\appendix
\setcounter{table}{0}
\renewcommand{\thetable}{A\arabic{table}}

\onecolumn

\begin{longtable}{|c|c|c|c|c|c|c|c|c|} 
  \hline {\bf Object}	& {\bf Type} & {\bf Redshift}	& {\bf BH mass (log)}	& {\bf L$_{\rm bol}$ (log erg/s)}	& {\bf Inclination ($^\circ$)}	& {\bf Ref.}	& {\bf Method}\\
      \hline 0019+0107	&BAL QSO	&2.123	&...	&...		&90.0	&Bor10	&	OTHER\\	
	  0145+0416	&BAL QSO	&2.03	&...	&...		&80.0	&Bor10	&	OTHER\\	
	  0226-1024	&BAL QSO	&2.256	&...	&...		&87.0	&Bor10	&	OTHER\\	
	  0842+3431	&BAL QSO	&2.13	&...	&...		&78.0	&Bor10	&	OTHER\\	
	  1235+1453	&BAL QSO	&2.686	&...	&...		&76.0	&Bor10	&	OTHER\\	
	  1333+2840	&BAL QSO	&1.91	&...	&...		&80.0	&Bor10	&	OTHER\\	
	  1413+1143	&BAL QSO	&2.560	&...	&...		&88.0	&Bor10	&	OTHER\\	
	  1H0419-577	&1.5	&0.104000	&8.30	&46.38	&51.0$^{+4}_{-6}$	&Wal13	&	X\\		 
	  1H0707-495	&NLS1	&0.040568	&6.85	&44.48	&48.8$^{+1.3}_{-1.2}$	&Dau12	&	X\\		
	  3C~120	&1.5	&0.033573	&7.74	&45.34	&22.0$^{+9.3}_{-7.7}$	&Wu01	&	BH-$\sigma$\\		
	  4C~13.41	&1.0	&0.24064	&...	&46.3	&35.0	&Mor09	&	IR\\		
	  Akn~120	&BLS1	&0.0323	&8.07	&44.91	&42.0	&Zha02	&	BH-$\sigma$\\		
	  Akn~564	&NLS1	&0.024917	&6.41	&44.77	&26.0	&Zha02	&	BH-$\sigma$\\		
	  Arp~151	&BLS1	&0.021091	&6.62	&43.7	&25.2$^{+3.3}_{-3.4}$	&Pan14	&	OTHER\\		
	  Circinus	&2.0	&0.001449	&6.42	&43.59	&65.0	&Fis13	&	NLR\\		
	  ESO~323-G077	&NLS1	&0.014904	&7.40	&43.9	&45.0	&Sch03	&	OTHER\\		
	  ESO~362-G18	&1.5	&0.012445	&7.65	&44.11	&53.0	$\pm$	5	&Agi14	&	OTHER\\
	  ESO~511-G30	&1.0	&0.022389	&8.40	&44.41	&59	$\pm$	10	&Lah14	&	OTHER\\
	  Fairall~51	&1.5	&0.014361	&8.00	&43.95	&45.0	&Sch01	&	OTHER\\			
	  Fairall~9	&BLS1	&0.048175	&8.20	&45.23	&35.0	&Zha02	&	BH-$\sigma$\\	
	  I~Zw~1	&NLS1	&0.060875	&7.24	&44.98	&8.0	&Mor09	&	IR\\		
	  IC~2560	&2.0	&0.009757	&6.48	&42.7	&66.0$^{+7}_{-4}$	&Bal14	&	X\\		
	  IC~4329A	&BLS1	&0.01613	&6.77	&44.78	&10.0$^{+13.0}_{-10.0}$	&Nan97	&	X\\		
	  IC~5063	&2.0	&0.011274	&7.74	&44.53	&82.0$^{+5}_{-9}$	&Alo11	&	IR\\		
	  IRAS~00521-7054	&2.0	&0.068900	&...	&49.43	&37$^{+4/+13}_{-4/-7}$	&Tan12	&	X\\		
	  IRAS~13224-3809	&1.0	&0.0658	&7.00	&44.95	&52.0	&Pon10	&	X\\		
	  IRAS~13349+2438	&2.0	&0.10853	&8.75	&46.3	&52.0	&Wil92	&	OTHER\\		
	  K~348-7	&1.0	&0.2341	&8.58	&46.16	&35.0	&Mor09	&	IR\\		
	  MCG-2-8-39	&2.0	&0.029894	&7.85	&42.57	&60.0	&Nog10	&	X\\		
	  MCG-3-34-64	&1.5	&0.017092	&7.69	&44.8	&27.0	$\pm$	17	&Min07	&	X\\
	  MCG-3-58-7	&2.0	&0.031462	&...	&44.7	&60.0	&Nog10	&	X\\		
	  MCG-6-30-15	&1.5	&0.00758	&6.46	&43.85	&34.0$^{+5.0}_{-6.0}$	&Nan97	&	X\\		
	  MCG+8-11-11	&1.5	&0.02004	&8.08	&44.43	&45.0	&Bha11	&	X\\		
	  Mrk~1014	&1.0	&0.16274	&8.03	&46.26	&16.0	&Mor09	&	IR\\		
	  Mrk~1018	&1.0	&0.042436	&8.6	&44.9	&45$^{+14}_{-10/-15}$	&Wal13	&	X\\		
	  Mrk~1066	&2.0	&0.012082	&7.01	&44.55	&80.0	&Fis13	&	NLR\\		
	  Mrk~110	&1.5	&0.03552	&7.29	&44.71	&37.4$^{+9.2}_{-9.5}$	&Wu01	&	BH-$\sigma$\\		
	  Mrk~1239	&NLS1	&0.0196	&6.38	&44.65	&7.0	&Zha02	&	BH-$\sigma$\\		
	  Mrk~1298	&BLS1	&0.06	&5.00	&45.54	&28.0	&Mor09	&	IR\\		
	  Mrk~1310	&1.0	&0.019560	&8.10	&43.5	&6.6$^{+5.0}_{-2.5}$	&Pan14	&	OTHER\\		
	  Mrk~1383	&BLS1	&0.087	&8.92	&45.78	&30.0	&Mor09	&	IR\\		
	  Mrk~176	&2.0	&0.02646	&8.00	&45.84	&60.0	&Nog10	&	X\\		
	  Mrk~231	&BLS1	&0.04147	&7.94	&46.18	&45.0	&Car98	&	OTHER\\		
	  Mrk~273	&2.0	&0.03734	&8.22	&47	&60.0	&Nog10	&	X\\		
	  Mrk~279	&BLS1	&0.030601	&7.43	&44.36	&35.0	&Fis13	&	NLR\\		
	  Mrk~3	&2.0	&0.013443	&8.26	&44.54	&85.0	&Fis13	&	NLR\\		
	  Mrk~304	&BLS1	&0.066293	&...	&44.56	&40.0	&Mor09	&	IR\\		
	  Mrk~335	&NLS1	&0.025418	&7.23	&44.69	&20.0	&Zha02	&	BH-$\sigma$\\		
	  Mrk~34	&2.0	&0.05095	&7.80	&44.78	&65.0	&Fis13	&	NLR\\		
	  Mrk~348	&2.0	&0.015034	&7.21	&44.27	&60.0	&Smi01	&	X\\		
	  Mrk~359	&NLS1	&0.01684	&6.23	&43.55	&30.0	&Obr01	&	X\\		
	  Mrk~463	&2.0	&0.050382	&7.88	&45.28	&60.0	&Nog10	&	X\\		
	  Mrk~478	&NLS1	&0.079	&7.33	&45.56	&25.0	&Zha02	&	BH-$\sigma$\\		
	  Mrk~486	&BLS1	&0.039	&7.03	&45.04	&16.0	&Zha02	&	BH-$\sigma$\\		
	  Mrk~50	&1.0	&0.023433	&7.57	&44.34	&9$^{+7}_{-5}$	&Pan14	&	OTHER\\		
	  Mrk~509	&BLS1	&0.03501	&8.05	&45.03	&19.0	&Zha02	&	BH-$\sigma$\\		
	  Mrk~573	&2.0	&0.017285	&7.58	&44.44	&60.0	&Fis13	&	NLR\\		
	  Mrk~590	&BLS1	&0.02609	&7.57	&44.63	&17.8$^{+6.1}_{-5.9}$	&Wu01	&	BH-$\sigma$\\		
	  Mrk~6	&1.5	&0.018676	&8.10	&44.56	&26.0	&Bha11	&	X\\		
	  Mrk~705	&NLS1	&0.0288	&6.92	&44.74	&16.0	&Zha02	&	BH-$\sigma$\\		
	  Mrk~707	&NLS1	&0.05026	&6.63	&44.79	&15.0	&Zha02	&	BH-$\sigma$\\		
	  Mrk~766	&NLS1	&0.01271	&6.63	&44.23	&36.0$^{+8.0}_{-7.0}$	&Nan97	&	X\\		
	  Mrk~78	&2.0	&0.03715	&8.14	&44.59	&60.0	&Fis13	&	NLR\\		
	  Mrk~79	&BLS1	&0.022185	&7.61	&44.57	&58.0	&Zha02	&	BH-$\sigma$\\		
	  Mrk~817	&1.5	&0.031455	&7.60	&44.99	&41.6$^{+8.5}_{-7.5}$	&Wu01	&	BH-$\sigma$\\		
	  Mrk~841	&1.5	&0.03642	&7.90	&45.84	&26.0$^{+8.0}_{-5.0}$	&Nan97	&	X\\		
	  Mrk~876	&BLS1	&0.138512	&8.95	&45.81	&27.0	&Mor09	&	IR\\		
	  Mrk~877	&BLS1	&0.112	&8.44	&45.33	&20.0	&Mor09	&	IR\\		
	  Mrk~896	&NLS1	&0.026784	&6.58	&43.89	&15.0	&Zha02	&	BH-$\sigma$\\		
	  NGC~1068	&2.0	&0.00381	&7.59	&44.3	&70.0	&Hon07	&	IR\\		
	  NGC~1097	&LINER	&0.004218	&8.15	&47.59	&34.0	&Sto97	&	OTHER\\		
	  NGC~1320	&2.0	&0.0092	&7.18	&43.86	&68.0$^{+3}_{-2}$	&Bal14	&	X\\		
	  NGC~1365	&1.8	&0.005476	&8.20	&43.1	&57.5	$\pm$	2.5	&Ris13	&	X\\
	  NGC~1386	&1.9/2.0	&0.002905	&7.42	&43.38	&81.0$^{+6}_{-8}$	&Rus14	&	IR\\		
	  NGC~1566	&1.5	&0.005036	&6.92	&44.45	&30.0	&Kaw13	&	X\\		
	  NGC~1667	&2.0	&0.015204	&7.62	&44.69	&72.0	&Fis13	&	NLR\\		
	  NGC~2110	&1.9/2.0	&0.007579	&8.30	&43.7	&53.0	&Sto99	&	OTHER\\		
	  NGC~2655	&LINER	&0.004670	&8.50	&42.08	&60.0	&Nog10	&	X\\		
	  NGC~2992	&2.0	&0.007296	&7.72	&43.92	&70.0	&Mar98	&	OTHER\\		
	  NGC~3227	&1.5	&0.00365	&6.77	&43.86	&14.2	$\pm$	2.5	&Hic08	&	NLR\\
	  NGC~3281	&2.0	&0.010674	&8.60	&43.8	&69$^{+11}_{-11}$	&Sal11	&	IR\\		
	  NGC~3516	&1.5	&0.008816	&7.39	&44.29	&26.0$^{+3}_{-4}$	&Nan97	&	X\\		
	  NGC~3783	&1.5	&0.009755	&7.37	&44.41	&15.0	&Fis13	&	NLR\\		
	  NGC~4051	&NLS1	&0.00216	&6.13	&43.56	&19.6$^{+10.4}_{-6.6}$	&Wu01	&	BH-$\sigma$\\		
	  NGC~4151	&1.5	&0.003262	&7.56	&43.73	&9.0$^{+18}_{-9}$	&Nan97	&	X\\		
	  NGC~424	&2.0	&0.01184	&7.78	&44.85	&69.0$^{+5}_{-4}$	&Bal14	&	X\\		
	  NGC~4388	&2.0	&0.00862	&7.23	&44.1	&60.0	--	63	&Bec04	&	OTHER\\
	  NGC~4395	&1.8	&0.00106	&5.45	&41.37	&15.0$^{+12}_{-15}$	&Nan07	&	X\\		
	  NGC~4507	&1.9/2.0	&0.011907	&7.65	&44.4	&47.0	&Fis13	&	NLR\\		
	  NGC~4593	&BLS1	&0.008344	&6.88	&44.09	&21.6	$\pm$	10.5	&Wu01	&	BH-$\sigma$\\
	  NGC~4941	&2.0	&0.00369	&6.90	&43.0	&70.0	&Kaw13	&	X\\		
	  NGC~4945	&2.0	&0.001878	&6.15	&43.4	&62.0	&Cho07	&	OTHER\\		
	  NGC~5506	&NLS1	&0.00589	&7.95	&44.3	&80.0	&Fis13	&	NLR\\		
	  NGC~5548	&1.5	&0.01627	&7.72	&44.83	&47.3$^{+7.6}_{-6.9}$	&Wu01	&	BH-$\sigma$\\		
	  NGC~5643	&2.0	&0.00399	&7.40	&42.3	&65.0	&Fis13	&	NLR\\		
	  NGC~6240	&2.0	&0.024480	&8.94	&44.3	&63.3	&Fan07	&	OTHER\\		
	  NGC~7172	&2.0	&0.008616	&7.67	&43.3	&77.0$^{+8}_{-14}$	&Alo11	&	IR\\		
	  NGC~7213	&LINER	&0.005869	&7.74	&44.3	&21.0$^{+9.0}_{-12.0}$	&Rus14	&	IR\\		
	  NGC~7314	&1.9	&0.004771	&6.70	&43.23	&42.0$^{+3}_{-4}$	&Nan07	&	X\\		
	  NGC~7469	&1.5	&0.01588	&6.96	&45.28	&15.0	$\pm$	1.8	&Hic08	&	NLR\\
	  NGC~7582	&2.0	&0.00525	&7.74	&43.3	&65.0	&Riv15	& X\\
	  NGC~7674	&2.0	&0.02998	&7.56	&45.47	&60.0	&Fis13	&	NLR\\		
	  PDS~456	&1.0	&0.184000	&9.00	&47.00	&70$^{+3}_{-5}$	&Wal13	&	X\\		
	  PG~0026+129	&NLS1	&0.142	&8.49	&45.39	&43.0	&Mor09	&	IR\\		
	  PG~1001+054	&NLS1	&0.16012	&7.65	&45.76	&38.0	&Mor09	&	IR\\		
	  PG~1211+143	&NLS1	&0.0809	&7.37	&45.81	&31.0	&Zha02	&	BH-$\sigma$\\		
	  PG~1244+026	&NLS1	&0.04813	&6.11	&44.13	&31.0	&Mor09	&	IR\\		
	  PG~1302-102	&1.0	&0.2784	&8.94	&46.33	&32.0	&Mor09	&	IR\\		
	  PG~1411+442	&BLS1	&0.09	&8.54	&45.58	&14.0	&Mor09	&	IR\\		
	  PG~1435-067	&BLS1	&0.126	&8.24	&45.5	&38.0	&Mor09	&	IR\\		
	  PG~1448+273	&NLS1	&0.06451	&6.92	&45.02	&53.0	&Mor09	&	IR\\		
	  PG~1626+554	&BLS1	&0.133	&8.37	&45.85	&31.0	&Mor09	&	IR\\		
	  PG~1700+518	&NLS1	&0.292	&8.79	&46.56	&43.0	&Mor09	&	IR\\		
	  PG~2251+113	&1.0	&0.325252	&8.96	&46.56	&67.0	&Mor09	&	IR\\		
	  RBS~1124	&BLS1	&0.208000	&8.26	&45.53 	&66$^{+5}_{-12}$	&Wal13	&	X\\		
	  SBS~1116+583A	&1.0	&0.027872	&6.99	&...	&18.2$^{+8.4}_{-5.9}$	&Pan14	&	OTHER\\		
	  Swift~J2127.4+5654	&NLS1	&0.014400	&7.18	&44.54	&49.0 $\pm$ 2.0	&Mar14	&	X\\	
	  TON~1388	&1.0	&0.1765	&8.50	&45.92	&39.0	&Mor09	&	IR\\		
	  TON~1542	&BLS1	&0.06355	&7.93	&45.27	&28.0	&Mor09	&	IR\\		
	  TON~1565	&1.0	&0.18291	&8.21	&45.89	&37.0	&Mor09	&	IR\\		
	  Ton~S180	&NLS1	&0.061980	&7.30	&45.70	&60$^{+3/+10}_{-1/-10}$	&Wal13	&	X\\		
	  UGC~3973	&1.5	&0.022189	&8.10	&44.31	&19.0	$\pm$	6	&Lah14	&	OTHER\\
	  UGC~6728	&1.0	&0.006518	&6.30	&43.0	&$<$55	&Wal13	&	X\\	  
	  VII~Zw~244	&BLS1	&0.131344	&...	&45.35	&23.0	&Mor09	&	IR\\		
	\hline
        \caption{Archival data listing the 124 Seyfert galaxies with identified nuclear inclination measurements. 
	    The redshift are obtained through {\sc simbad}. The central black hole masses are in logarithmic 
	    units and taken from the ``AGN black hole database'' \citep{Bentz2015}, \citet{Kaspi2000}, 
	    \citet{Vestergaard2002}, \citet{Vasudevan2007}, \citet{Esquej2014}, and \citet{Feng2014}. 
	    The AGN classification types are also indicated for completeness. The different main methods (labeled 
	    as IR, X, NLR and VEL) used to determine the inclination of the system are presented in the text.
	    Legend: Wil92 - \citet{Wills1992}; Nan97 - \citet{Nandra1997}; Sto97 - \citet{Storchi1997}; 
	    Car98 - \citet{Carilli1998}; Mar98 - \citet{Marquez1998}; Sto99 - \citet{Storchi1999};
	    Obr01 - \citet{OBrien2001}; Sch01 - \citet{Schmid2001}; Smi01 - \citet{Smith2001}; 
	    Wu01 - \citet{Wu2001}; Zha02 - \citet{Zhang2002}; Sch03 - \citet{Schmid2003}; Bec04 - \citet{Beckmann2004}; 
	    Cho07 - \citet{Chou2007}; Fan07 - \citet{Fan2007}; Hon07 - \citet{Honig2007}; Min07 - \citet{Miniutti2007}; 
	    Nan07 - \citet{Nandra2007}; Hic08 - \citet{Hicks2008}; Mor09 - \citet{Mor2009}; Bor10 - \citet{Borguet2010}; 
	    Nog10 - \citet{Noguchi2010}; Pon10 - \citet{Ponti2010}; Alo11 - \citet{Alonso2011}; 
	    Bha11 - \citet{Bhayani2011}; Sal11 - \citet{Sales2011}; Dau12 - \citet{Dauser2012};  Tan12 - \citet{Tan2012}; 
	    Fis13 - \citet{Fischer2013}; Kaw13 - \citet{Kawamuro2013}; Ris13 - \citet{Risaliti2013}; Wal13 - \citet{Walton2013};  
	    Agi14 - \citet{Agis2014}; Bal14 - \citet{Balokovic2014}; Lah14 - \citet{Laha2014}; Mar14 - \citet{Marinucci2014}; 
	    Pan14 - \citet{Pancoast2014}; Rus14 - \citet{Ruschel2014} and Riv15 - \citet{Rivers2015}.}
    \label{Table:Data}
\end{longtable}

\begin{longtable}{|c|c|c|c|c|c|}
      \hline {\bf Object}	& {\bf Inclination ($^\circ$)} 	& {\bf Ref.}	& {\bf Object}	& {\bf Inclination ($^\circ$)} 	& {\bf Ref.}	\\
      \hline 3C~120		& 22.0$^{+9.3}_{-7.7}$ 		&  Wu01 	& Mrk~817	& 41.6$^{+8.5}_{-7.5}$ 		&  Wu01 \\
	    Akn~120		& 42.0				&  Zha02 	& NGC~3227	& 37.5$^{+17.3}_{-25.4}$ 	&  Wu01 	\\
	    Fairall~9		& 35.0				&  Zha02 	& NGC~3516	& 38.3 $\pm$ 7.6 		&  Wu01 	\\
	    IC~4329A		& 5.0				&  Zha02 	& NGC~3783	& 38.0				&  Zha02 \\
	    Mrk~110		& 37.4$^{+9.2}_{-9.5}$ 		&  Wu01 	& NGC~4051	& 19.6$^{+10.4}_{-6.6}$		&  Wu01 \\
	    Mrk~279		& 13.0				&  Zha02 	& NGC~4151	& 60.0$^{+30.0}_{-30.6}$ 	&  Wu01 \\
	    Mrk~335		& 20.0				&  Zha02 	& NGC~4593	& 21.6 $\pm$ 10.5 		&  Wu01 	\\
	    Mrk~509		& 19.0				&  Zha02 	& NGC~5548	& 43.7$^{+7.6}_{-6.9}$		&  Wu01 \\
	    Mrk~590		& 17.8$^{+6.1}_{-5.9}$ 		&  Wu01 	& NGC~7469	& 13.0				&  Zha02 \\
	    Mrk~79		& 58.5$^{+21.7}_{-27.9}$ 	&  Wu01 	& ~		& ~				&  ~ \\
      \hline
    \caption{Inclinations of 19 Seyfert nuclei derived from the BH mass - bulge 
	      velocity dispersion relation. Note that the data taken from 
	      \citet{Zhang2002} are restricted to AGN with measured BLR size 
	      and FWHM of H$\beta$ emission line. All AGN are type-1s.
	      Legend: Wu01 - \citet{Wu2001} and Zha02 - \citet{Zhang2002}.}
    \label{Table:VEL}
\end{longtable}

\begin{longtable}{|c|c|c|c|c|c|}
      \hline {\bf Object}	& {\bf Inclination ($^\circ$)} 	& {\bf Ref.}		& {\bf Object}	& {\bf Inclination ($^\circ$)} 	& {\bf Ref.}	\\ 
   \hline 1H~0419-577		& 51.0$^{+4}_{-6}$ 		& Wal13 		& Mrk~766	& 36.0$^{+8}_{-7}$		& Nan97 \\
	  1H~0707-495	& 52.0$^{+1.7}_{-1.8}$ 	& Dau12	& Mrk~79	& 24.0 $\pm$ 1 	& Gal11 \\
	  3C~120	& 5$^{+4}_{-5}$ 	& Loh13 	& Mrk~841	& 45.0$^{+7}_{-5}$	& Wal13 \\
	  Akn~120	& 54.0$^{+6}_{-5}$ 	& Wal13 	& NGC~1320	& 68.0$^{+3}_{-2}$	& Bal14 \\
	  Akn~564	& 64.0$^{+1/+6}_{-11}$ 	& Wal13 	& NGC~1365	& 63 $\pm$ 4 	& Wal14 \\
	  ESO~362-G18	& 53.0 $\pm$ 5 	& Agi14 	& NGC~1566	& 30.0	& Kaw13 \\
	  Fairall~9	& 48.0$^{+6}_{-2}$ 	& Loh12 	& NGC~2110	& 0.0$^{+51}_{-0}$	& Bha11 \\
	  IC~2560	& 66.0$^{+7}_{-4}$ 	& Bal14	& NGC~2655	& 60.0	& Nog10 \\
	  IC~4329A	& 10.0$^{+13}_{-10}$ 	& Nan97	& NGC~3227	& 47.0$^{+3}_{-2}$ 	& Pat12 \\
	  IRAS~00521-7054	& 37.0$^{+4/+13}_{-4/-7}$ 	& Tan12 	& NGC~3516	& 33.0$^{+3}_{-9}$ 	& Nan07 \\
	  IRAS~13224-3809	& 65.0$^{+1/+5}_{-1}$ 	& Chi15	& NGC~3783	& 3.0$^{+18}_{-3}$ 	& Nan07 \\
	  MCG-2-8-39	& 60.0	& Nog10	& NGC~4051	& 25.0$^{+12}_{-4}$ 	& Nan97 \\
	  MCG-3-34-64	& 27.0 $\pm$ 17 	& Min07 	& NGC~4151	& 33.0$^{+1}_{-3}$ 	& Nan07 \\
	  MCG-3-58-7	& 60.0 	& Nog10 	& NGC~424	& 69.0$^{+5}_{-4}$ 	& Bal14 \\
	  MCG-6-30-15	& 34.0$^{+5.0}_{-6.0}$ 	& Nan97	& NGC~4593	& 24.0$^{+28}_{-15}$ 	& Nan07 \\
	  MCG+8-11-11	& 45.0$^{+40}_{-8}$ 	& Bha11	& NGC~4941	& 70.0 	& Kaw13 \\
	  Mrk~1018	& 45.0$^{+14}_{-10/-15}$ 	& Wal13 	& NGC~5506	& 46.0$^{+4}_{-4}$ 	& Bha11 \\
	  Mrk~110	& 36.0$^{+19}_{-12}$ 	& Bha11	& NGC~5548	& 3.0$^{+82}_{-3}$ 	& Bha11 \\
	  Mrk~176	& 60.0 	& Nog10	& NGC~7213	& 0.0$^{+64}_{-0}$ 	& Bha11 \\
	  Mrk~273	& 60.0 	& Nog10 	& NGC~7469	& $<$ 54 	& Wal13 \\
	  Mrk~335	& 65.0 $\pm$ 1 	& Par14	& NGC~7582	& 65.0	& Riv15 \\
	  Mrk~348	& 60.0 	& Smi01	& PDS~456	& 70$^{+3}_{-5}$	& Wal13 \\
	  Mrk~359	& 47.0 $\pm$ 6	  	& Wal13 	& PG~1211+143	& 28.0$^{+7/+22}_{-7}$ 	& Zog15 \\
	  Mrk~463	& 60.0 	& Nog10	& RBS~1124	& 66$^{+5}_{-12}$	  	& Wal13 \\
	  Mrk~509	& $<$ 18	  	& Wal13 	& Swift~J2127.4+5654	& 49.0 $\pm$ 2.0	  	& Mar14 \\
	  Mrk~590	& 47.0$^{+38}_{-47}$ 	& Bha11	& Ton~S180	& 60$^{+3/+10}_{-1/-10}$	  	& Wal13 \\
	  Mrk~6	& 26.0$^{+59}_{-7}$ 	& Bha11	& UGC~6728	& $<$ 55  	& Wal13 \\
      \hline
    \caption{Inclinations of 54 Seyfert nuclei, as determined by fitting
	      their X-ray spectra with disk reflection models in
	      special and general relativistic environments.
 	      See also \citet{Middleton2016} and reference therein.	      
 	      Legend: Nan97 - \citet{Nandra1997}; Smi01 - \citet{Smith2001}; 
 	      Min07 - \citet{Miniutti2007}; Nan07 - \citet{Nandra2007}; 
 	      Nog10 - \citet{Noguchi2010}; Bha11 - \citet{Bhayani2011}; 
 	      Gal11 - \citet{Gallo2011}; Dau12 - \citet{Dauser2012}; 
 	      Loh12 - \citet{Lohfink2012}; Pat12 - \citet{Patrick2012}; 
 	      Tan12 - \citet{Tan2012}; Kaw13 - \citet{Kawamuro2013}; 
 	      Loh13 - \citet{Lohfink2013}; Wal13 - \citet{Walton2013}; 
 	      Agi14 - \citet{Agis2014}; Bal14 - \citet{Balokovic2014}; 
 	      Mar14 - \citet{Marinucci2014}; Par14 - \citet{Parker2014}; 
 	      Wal14 - \citet{Walton2014}; Chi15 - \citet{Chiang2015};
 	      Riv15 - \citet{Rivers2015} and Zog15 - \citet{Zoghbi2015}.}
    \label{Table:X}
\end{longtable}

\begin{longtable}{|c|c|c|c|c|c|}
      \hline {\bf Object}	& {\bf Inclination ($^\circ$)} 	& {\bf Ref.}	& {\bf Object}	& {\bf Inclination ($^\circ$)} 	& {\bf Ref.}	\\ 
	\hline Circinus	&	66.0$^{+7}_{-4}$		& Alo11 	& NGC~7172	& 77.0$^{+8}_{-14}$		& Alo11 \\
	  I~Zw~1	&	8				& Mor09 	& NGC~7213	& 21.0$^{+9}_{-12}$		& Rus14 \\
	  IC~4329A	&	51.0$^{+8}_{-8}$		& Alo11 	& NGC~7469	& 58.0$^{+3}_{-4}$		& Alo11 \\
	  IC~5063	&	82.0$^{+5}_{-9}$		& Alo11 	& NGC~7674	& 63.0$^{+9}_{-10}$		& Alo11 \\
	  K~348-7	&	35				& Mor09 	& PG~0026+129	& 43				& Mor09 \\
	  Mrk~304	&	40				& Mor09		& PG~1001+054	& 38				& Mor09 \\
	  Mrk~478	&	26				& Mor09 	& PG~1244+026	& 31				& Mor09 \\
	  Mrk~876	&	27				& Mor09 	& PG~1302-102	& 32				& Mor09 \\
	  Mrk~877	&	20				& Mor09 	& PG~1411+442	& 14				& Mor09 \\
	  Mrk~1014	&	16				& Mor09 	& PG~1435-067	& 38				& Mor09 \\
	  Mrk~1298	&	28				& Mor09 	& PG~1448+273	& 53				& Mor09 \\
	  Mrk~1383	&	30				& Mor09 	& PG~1626+554	& 31				& Mor09 \\
	  NGC~1068	&	70				& Hon07		& PG~1700+518	& 43				& Mor09 \\
	  NGC~1386	&	81.0$^{+6}_{-8}$		& Rus14 	& PG~2251+113	& 67				& Mor09 \\
	  NGC~2110	&	43.0$^{+8}_{-8}$		& Alo11 	& TON~1388	& 39				& Mor09 \\
	  NGC~3227	&	24.0$^{+11}_{-15}$		& Alo11 	& TON~1542	& 28				& Mor09 \\
	  NGC~3281	&	69.0$^{+11}_{-11}$		& Sal11 	& TON~1565	& 37				& Mor09 \\
	  NGC~4151	&	63.0$^{+4}_{-7}$		& Alo11 	& VII~Zw~244	& 23				& Mor09 \\
	  NGC~5506	&	34.0$^{+6}_{-6}$		& Alo11 	& ~		& ~				& ~ \\
      \hline
    \caption{Inclinations of 37 Seyfert nuclei, as determined by fitting
	      their IR spectra with radiative transfer in clumpy
	      environments; see Sect.~\ref{Comp}.
	      Legend: Hon07 - \citet{Honig2007}; Mor09 - \citet{Mor2009}; 
	      Alo11 - \citet{Alonso2011}; Sal11 - \citet{Sales2011} and 
	      Rus14 - \citet{Ruschel2014}.}
    \label{Table:IR}
\end{longtable}

\begin{longtable}{|c|c|c|c|c|c|}
      \hline {\bf Object}	& {\bf Inclination ($^\circ$)} 	& {\bf Ref.}	& {\bf Object}	& {\bf Inclination ($^\circ$)} 	& {\bf Ref.}	\\  
        \hline Circinus		& 	65			&  Fis13 	& NGC~3227	&	15			&  Fis13 \\
	Mrk~3			&	85			&  Fis13 	& NGC~3783	&	15			&  Fis13 \\
	Mrk~34			&	65			&  Fis13 	& NGC~4051	&	12			&  Fis13 \\
	Mrk~78			&	60			&  Fis13 	& NGC~4151	&	45			&  Fis13 \\
	Mrk~279			&	35			&  Fis13 	& NGC~4507	&	47			&  Fis13 \\
	Mrk~573			&	60			&  Fis13 	& NGC~5506	&	80			&  Fis13 \\
	Mrk~1066		&	80			&  Fis13 	& NGC~5643	&	65			&  Fis13 \\
	NGC~1068		&	85			&  Fis13 	& NGC~7674	&	60			&  Fis13 \\
	NGC~1667		&	72			&  Fis13 	& ~		&	~			&  ~ \\
	\hline
    \caption{Inclinations of 17 Seyfert nuclei, as determined by their NLR 
	      kinematics. A standard $\pm$~5$^\circ$ uncertainty is 
	      added to the inclinations by the authors (see, 
	      e.g., \citealt{Rose2015}). Legend: 
	      Fis13 - \citet{Fischer2013}.}
    \label{Table:NLR}
\end{longtable}

\begin{longtable}{|c|c|c|c|c|c|}
      \hline {\bf Object}	& {\bf n$_{\rm H}$ ($\times$ 10$^{22~}$cm$^{-2})$}	& {\bf Ref.} &  {\bf Object}	& {\bf n$_{\rm H}$ ($\times$ 10$^{22~}$cm$^{-2})$}	& {\bf Ref.}\\
      \hline 
         \hline 1H0419-577	& 4.3 $\pm$ 0.4	& Pou04 &	Mrk~877	& 0.008$^{+0.033}_{-0.008}$	& Sch96 \\
3C~120	& 0.16	& Win09 &	Mrk~896	& 0.034 $\pm$ 0.004	& Bol96 \\
Akn~120	& 0.02	& Win09 &	NGC~1068	& 15 -- 1000	& Bau14 \\
Akn~564	& 0.073 $\pm$ 0.004	& Bol96 &	NGC~1097	& 0.023	& Era10 \\
Arp~151	& 0.2173$^{+0.0059}_{-0.0199}$	& Win09 &	NGC~1320	& 400$^{+20}_{-10}$	& Bal14 \\
Circinus	& 600 -- 1000	& Are14 &	NGC~1365	& 450	& Win09 \\
ESO~323-G077	& 5.85$^{+0.12}_{-0.11}$	& Jim08a &	NGC~1386	& 140$^{+10}_{-20}$	& Rus14 \\
ESO~362-G18	& 26.6	& Win09 &	NGC~1566	& 0.007 $\pm$ 0.011	& Wal93 \\
ESO~511-G30	& 0.098	& Win09 &	NGC~1667	& $>$ 100	& Bia05b \\
Fairall~51	& 1.6 $\pm$ 0.2	& Jim08b &	NGC~2110	& 2.84	& Win09 \\
Fairall~9	& 0.023	& Win09 &	NGC~2655	& 30.2$^{+39.47}_{-24.21}$	& Gon08 \\
I~Zw~1	& 0.065 $\pm$ 0.007	& Bol96 &	NGC~2992	& 1.19	& Win09 \\
IC~2560	& $>$ 1000	& Bal14 &	NGC~3227	& 0.35 $\pm$ 0.18	& Mar09 \\
IC~4329A	& 0.61	& Win09 &	NGC~3281	& 86.3	& Win09 \\
IC~5063	& 25	& Win09 &	NGC~3516	& 0.353	& Win09 \\
IRAS~00521-7054	& 7.0 $\pm$ 0.8	& Tan12 &	NGC~3783	& 3.6 $\pm$ 0.5	& Cre12 \\
IRAS~13224-3809	& 0.0534	& Kam15 &	NGC~4051	& 2.1 $\pm$ 1.1	& Cre12 \\
IRAS~13349+2438	& 2.5 $\pm$ 1.5	& Sak01 &	NGC~4151	& 9.4 $\pm$ 2.8	& Cre12 \\
MCG-2-8-39	& 31.6	& Lir13 &	NGC~424	& 300 $\pm$ 10	& Bal14 \\
MCG-3-34-64	& 40.7	& Win09 &	NGC~4388	& 36.2	& Win09 \\
MCG-3-58-7	& 25.1	& Lir13 &	NGC~4395	& 3.3	& Win09 \\
MCG-6-30-15	& 0.19	& Win09 &	NGC~4507	& 43.9 $\pm$ 5.7	& Mat04a \\
MCG+8-11-11	& 0.25	& Win09 &	NGC~4593	& 0.031	& Win09 \\
Mrk~1018	& 0.01 $\pm$ 0.016	& Wal93 &	NGC~4941	& 0.2412$^{+0.0012}_{-0.0017}$	& Vas12 \\
Mrk~1066	& $>$ 100	& Ris99 &	NGC~4945	& 355	& Puc14 \\
Mrk~110	& 1.78	& Win09 &	NGC~5506	& 3.7 $\pm$ 0.8	& Ris02 \\
Mrk~1239	& 0.083 $\pm$ 0.016	& Bol96 &	NGC~5548	& 1.2 -- 9.6	& Meh15 \\
Mrk~1310	& 0.242$^{+0.0024}_{-0.0018}$	& Win09 &	NGC~5643	& 70.7$^{+30}_{-10}$	& Gua04 \\
Mrk~1383	& 0.021 $\pm$ 0.011	& Wal93 &	NGC~6240	& 150	& Puc15 \\
Mrk~231	& 12 $\pm$ 1	& Ten14 &	NGC~7172	& 8.19	& Win09 \\
Mrk~273	& 43.8$^{+9.5}_{-5.7}$	& Ten15 &	NGC~7213	& 50$^{+20}_{-16}$	& Urs15 \\
Mrk~279	& 0.013	& Win09 &	NGC~7314	& 1.16	& Win09 \\
Mrk~3	& 136$^{+3}_{-4}$	& Bia05a &	NGC~7469	& 0.041	& Win09 \\
Mrk~304	& 0.0145	& Kar96 &	NGC~7582	& 300	& Riv15 \\
Mrk~335	& 0.03$^{+0.05}_{-0.03}$	& Wal92 &	NGC~7674	& $>$ 1000	& Bia05b \\
Mrk~34	& 250 -- 1000	& Gan14 &	PDS~456	& 12.1 $\pm$ 1	&  Nar15 \\
Mrk~348	& 16	& Win09 &	PG~0026+129	& 0.0522 $\pm$ 0.0105	& Rac96 \\
Mrk~359	& 0.05 $\pm$ 0.007	& Bol96 &	PG~1001+054	& 0.0233	& Wan96 \\
Mrk~463	& 0.2382 $\pm$ 0.0003	& Vas12 &	PG~1211+143	& 0.03 $\pm$ 0.01	& Wal93 \\
Mrk~478	& 0.02 $\pm$ 0.003	& Bol96 &	PG~1244+026	& 0.0311 $\pm$ 0.0049	& Wan96 \\
Mrk~50	& 0.006	& Win09 &	PG~1302-102	& 0.027 $\pm$ 0.0076	& Rac96 \\
Mrk~509	& 0.015	& Win09 &	PG~1411+442	& 0.0118 $\pm$ 0.0094	& Rac96 \\
Mrk~573	& $>$ 100	& Shu07 &	PG~1448+273	& 0.044 $\pm$ 0.016	& Wal93 \\
Mrk~590	& 0.027	& Win09 &	PG~1626+554	& 0.03$^{+0.019}_{-0.013}$	& Sch96 \\
Mrk~6	& 3.26	& Win09 &	PG~1700+518	& $<$ 0.12	& Sae12 \\
Mrk~705	& 0.039 $\pm$ 0.013	& Wal93 &	RBS~1124	& 6.0$^{+3}_{-2}$	& Min10 \\
Mrk~766	& 0.525	& Win09 &	Swift~J2127.4+5654	& 0.213 $\pm$ 0.005	& Mar14 \\
Mrk~78	& 57.5 $\pm$ 5.8	& Gil10 &	TON~1542	& 0.037 $\pm$ 0.019	& Wal93 \\
Mrk~79	& $<$ 0.0063	& Ich12 &	Ton~S180	& 0.037$^{+0.023}_{-0.022}$	& Roz04 \\
Mrk~817	& 0.1285 $\pm$ 0.0008	& Win10 &	UGC~3973	& 0.734 $\pm$ 0.019	& Wal93 \\
Mrk~841	& 0.219	& Win09 &	UGC~6728	& 0.01$^{+0.02}_{-0.01}$	& Win08 \\
Mrk~876	& 0.043$^{+0.009}_{-0.008}$	& Sch96 &	VII~Zw~244	& 0.061$^{+0.13}_{-0.044}$	& Sch96 \\
      \hline
    \caption{Archival X-ray column density for 104/124 Seyfert galaxies. 
	    The hydrogen column density is ionized for type-1s and 
	    cold for type-2s. Legend: Wal92 - \citet{Walter1992}; 
	    Wal93 - \citet{Walter1993}; Bol96 - \citet{Boller1996}; Kar96 - \citet{Kartje1996}; 
	    Rac96 - \citet{Rachen1996}; Sch96 - \citet{Schartel1996}; Wan96 - \citet{Wang1996}; 
	    Ris99 - \citet{Risaliti1999}; 
	    Sak01 - \citet{Sako2001}; Ris02 - \citet{Risaliti2002}; 
	    Gua04 - \citet{Guainazzi2004}; Mat04a - \citet{Matt2004a}; 
	    Pou04 - \citet{Pounds2004}; Roz04 - \citet{Rozenska2004}; Bia05a - \citet{Bianchi2005a}; 
	    Bia05b - \citet{Bianchi2005b}; Shu07 - \citet{Shu2007}; Jim08a - \citet{Jimenez2008a}; 
	    Jim08b - \citet{Jimenez2008b}; Mar08 - \citet{Martin2008}; Win08 - \citet{Winter2008}; 
	    Mar09 - \citet{Markowitz2009}; Win09 - \citet{Winter2009}; Era10 - \citet{Eracleous2010}; 
	    Gal10 - \citet{Gallo2010}; Gil10 - \citet{Gilli2010}; Min10 - \citet{Miniutti2010}; 
	    Win10 - \citet{Winter2010}; Cre12 - \citet{Crenshaw2012}; Ich12 - \citet{Ichikawa2012}; 
	    Sae12 - \citet{Saez2012}; Tan12 - \citet{Tan2012}; Vas12 - \citet{Vasudevan2012}; 
	    Lir13 - \citet{Lira2013}; Are14 - \citet{Arevalo2014}; Bal14 - \citet{Balokovic2014}; Bau14 - \citet{Bauer2014}; Gan14 - \citet{Gandhi2014}
	    Mar14 - \citet{Marinucci2014}; Puc14 - \citet{Puccetti2014}; Rus14 - \citet{Ruschel2014}; Ten14 - \citet{Teng2014}; Kam15 - \citet{Kammoun2015}; 
	    Meh15 - \citet{Mehdipour2015}; Nar15 - \citet{Nardini2015}; Puc15 - \citet{Puccetti2015}; Ten15 - \citet{Teng2015} and Urs15 - \citet{Ursini2015}.}	    	    
    \label{Table:nH}
\end{longtable}

\begin{longtable}{|c|c|c|c|c|c|}
      \hline {\bf Object}	& {\bf FWHM H$\beta$ (km.s$^{-1}$)}	& {\bf Ref.} & {\bf Object}	& {\bf FWHM H$\beta$ (km.s$^{-1}$)}	& {\bf Ref.}\\
      \hline 
         \hline 1H0419-577	& 4200 $\pm$ 250 	& Tur99	&	Mrk~1014	& 2308	& Mac84	\\
1H0707-495	& 940	& Don15	&	Mrk~1018	& 6940 $\pm$ 760	& Mur14	\\
3C~120 	& 2419 $\pm$ 29 	& Fen14 	&	Mrk~1239	& 1075	& Ver01	\\
4C~13.41	& 6800	& Bro87	&	Mrk~1298	& 2200	& Goo89	\\
Akn~120$^a$	& 5987 $\pm$ 54	& Fen14	&	Mrk~1310	& 2731 $\pm$ 51 	& Fen14	\\
Akn~564	& 865	& Ver01	&	Mrk~1383	& 5420	& Sul89	\\
Arp~151	& 3407 $\pm$ 35 	& Fen14	&	NGC~1097$^a$	& ...	& ...	\\
ESO~323-G077	& 2100	& Sch03	&	NGC~1365	& 3586	& Sch99	\\
ESO~362-G18	& 5240 $\pm$ 500	& Agi14	&	NGC~1566	& 1800	& Win92	\\
ESO~511-G30	& 2500	& Win92	&	NGC~3227	& 4494 $\pm$ 19	& Fen14	\\
Fairall~51	& 3330 $\pm$ 300	& Ben06	&	NGC~3516	& 5527 $\pm$ 17	& Fen14	\\
Fairall~9	& 5618 $\pm$ 107 	& Fen14	&	NGC~3783	& 3634 $\pm$ 41 	& Fen14	\\
I~Zw~1	& 1240	& Bol96	&	NGC~4051 	& 1565 $\pm$ 80 	& Fen14	\\
IC~4329A$^a$	& 6000	& Mar92	&	NGC~4151	& 6794 $\pm$ 161	& Fen14	\\
IRAS~00521+7054	& 817	& You96	&	NGC~4395	& 1175 $\pm$ 325	& Edr12	\\
IRAS~13224-3809	& 650	& Bra97	&	NGC~4593	& 3900	& Kol97	\\
K~348-7	& 3225	& Til13	&	NGC~5548$^a$	& 12404 $\pm$ 20	& Fen14	\\
MCG-3-34-64	& ...	& ...	&	NGC~7213	& 3200 	& Win92 	\\
MCG-6-30-15	& 1990 $\pm$ 200	& Ben06	&	NGC~7314	& ...	& ...	\\
MCG+8-11-11	& 3630	& Ost82	&	NGC~7469	& 3296 $\pm$ 75	& Fen14	\\
Mrk~6	& 4512 $\pm$ 38 	& Fen14	&	PDS~456	& 3000	& Ree03	\\
Mrk~6	& 4512 $\pm$ 38 	& Fen14	&	PG~0026+129	& 2598 $\pm$ 57 	& Fen14	\\
Mrk~50	& 4621 $\pm$ 30 	& Fen14 	&	PG~1001+054	& 1125 $\pm$ 30	& Sha07	\\
Mrk~79	& 4735 $\pm$ 44 	& Kol06	&	PG~1211+143	& 1975	& Ver01	\\
Mrk~110	& 2194 $\pm$ 64 	& Fen14	&	PG~1244+026	& 740	& Ver01	\\
Mrk~231	& 3000	& Ver01	&	PG~1302-102	& 4450 $\pm$ 150	& Gra15	\\
Mrk~279	& 5208 $\pm$ 95 	& Fen14	&	PG~1411+442	& 2392 $\pm$ 56 	& Fen14	\\
Mrk~304	& 4600	& Sul89	&	PG~1435-067	& 3180	& San10	\\
Mrk~335	& 2182 $\pm$ 53 	& Fen14 	&	PG~1448+273	& 820	& Bol96	\\
Mrk~359	& 900	& Ver01	&	PG~1626+554	& 4618	& Til13	\\
Mrk~478	& 1270	& Ver01	&	PG~1700+518	& 2230 $\pm$ 57 	& Fen14	\\
Mrk~486	& 1680	& Ver01	&	PG~2251+113	& 2139	& Esp94	\\
Mrk~509	& 3595 $\pm$ 24 	& Fen14	&	RBS~1121	& 4260 $\pm$ 1250	& Min10	\\
Mrk~590	& 2966 $\pm$ 56	& Fen14	&	SBS~1116+583A	& 3950 $\pm$ 255 	& Fen14	\\
Mrk~705	& 1790	& Ver01	&	Swift~J2127.4+5654	& 2000	& Mal08	\\
Mrk~707	& 1295	& Ver01	&	TON~1388	& 2920 $\pm$ 80	& Gru99	\\
Mrk~766 	& 1630	& Ver01	&	TON~1542	& 3470	& San10	\\
Mrk~817	& 4937 $\pm$ 120 	& Fen14 	&	TON~1565	& 950$^{+10}_{-0}$	& Sha07	\\
Mrk~841	& 5300	& Bra97	&	Ton~S180	& 1000	& Ver01	\\
Mrk~876	& 5017	& Sul89	&	UGC~3973	& 4735 $\pm$ 44	& Fen14	\\
Mrk~877	& 3790	& Sul89	&	UGC~6728	& 2308.3 $\pm$ 79.6	& Win09	\\
Mrk~896	& 1135	& Ver01	&	VII~Zw~244	& 2899	& Til13	\\
      \hline
    \caption{Archival optical FWHM measurements of the H$\beta$~$\lambda$4861 line 
	    for 74/90 type-1 AGN. Some objects show large, double-peaked, Balmer 
	    line profiles and are identified with the superscript $^a$. A larger fraction 
	    of AGN, that need to be identified, might show similar characteristics.
	    Legend: Ost82 - \citet{Osterbrock1982}; Mac84 - \citet{MacKenty1984}; 
	    Bro87 - \citet{Browne1987}; Goo89 - \citet{Goodrich1989}; 
	    Sul89 - \citet{Sulentic1989}; Mar92 - \citet{Marziani1992}; Win92 - \citet{Winkler1992}; 
	    Esp94 - \citet{Espey1994}; Bol96 - \citet{Boller1996}; You96 - \citet{Young1996}; 
	    Bra97 - \citet{Brandt1997}; Kol97 - \citet{Kollatschny1997}; Gru99 - \citet{Grupe1999}; 
	    Sch99 - \citet{Schulz1999}; Tur98 - \citet{Turner1999}; Ver01 - \citet{Veron2001}; 
	    Ree03 - \citet{Reeves2003}; Sch03 - \citet{Schmid2003}; Ben06 - \citet{Bennert2006a}; 
	    Kol06 - \citet{Kollatschny2006}; Sha07 - \citet{Shang2007}; Mal08 - \citet{Malizia2008}; 
	    Win09 - \citet{Winter2009}; Min10 - \citet{Miniutti2010}; San10 - \citet{Sani2010}; 
	    Edr12 - \citet{Edri2012}; Til13 - \citet{Tilton2013}; Agi14 - \citet{Agis2014}; 
	    Fen14 - \citet{Feng2014}; Mur14 - \citet{LaMura2014}; Don15 - \citet{Done2015} 
	    and Gra15  - \citet{Graham2015}.}	    	    
    \label{Table:Hbeta}
\end{longtable}

\begin{longtable}{|c|c|c|c|c|}
      \hline {\bf Object}	& {\bf Waveband (\AA)}	& {\bf Pol. degree (\%)}	& {\bf Pol. angle ($^\circ$)}	& {\bf Ref.}\\
      \hline 
         \hline 3C~120	& 3800 -- 5600 	& 0.92 $\pm$ 0.25 	& 103.5 $\pm$ 7.9	& Mar83	\\
0019+0107	& 4000 -- 8600 	& $>$ 0.98 	& 35.0 $\pm$ 0.5	& Ogl99	\\
0145+0416	& 1960 -- 2260 	& $>$ 2.14 	& 126.0 $\pm$ 1.0	& Ogl99	\\
0226-1024	& 4000 -- 8600 	& $>$ 1.81 	& 167.1 $\pm$ 0.2	& Ogl99	\\
0842+3431	& 4000 -- 8600 	& $>$ 0.51 	& 27.1 $\pm$ 0.6	& Ogl99	\\
1235+1453	& 1600 -- 1840 	& $>$ 0.75 	& 175.0 $\pm$ 12.0	& Ogl99	\\
1333+2840	& 4000 -- 8600 	& $>$ 4.67 	& 161.5 $\pm$ 0.1	& Ogl99	\\
1413+1143	& 4000 -- 8600 	& $>$ 1.52 	& 55.7 $\pm$ 0.9	& Ogl99	\\
4C~13.41	& 3200 -- 8600	& 0.94 $\pm$ 0.19	& 87.0 $\pm$ 6.0	& Ber90	\\
Akn~120	& 3800 -- 5600 	& 0.65 $\pm$ 0.13	& 78.6 $\pm$ 5.7	& Mar83	\\
Akn~564	& 6000 -- 7500 	& 0.52 $\pm$ 0.02	& 87.0 $\pm$ 1.3	& Smi02	\\
Circinus	& 5650 -- 6800 	& 22.4 -- 25.0	& 45.0	& Ale00	\\
ESO~323-G077	& 3600 	& 7.5 	& 84	& Sch03	\\
Fairall~51	& 4700 -- 7200 	& 4.12 $\pm$ 0.03 	& 141.2 $\pm$ 0.2	& Smi02	\\
Fairall~9	& 3800 -- 5600 	& 0.4 $\pm$ 0.11 	& 2.4 $\pm$ 7.6	& Mar83	\\
I~Zw1	& 3200 -- 8600	& 0.61 $\pm$ 0.08	& 8.0 $\pm$ 3.0	& Ber90	\\ 
IC~4329A	& 5000 -- 5800 	& 5.80 $\pm$ 0.26	& 42.0 $\pm$ 1.0	& Bri90	\\
IC~5063	& 3800 -- 5600	& 4.05 -- 5.05	& 10.1 $\pm$ 3.2	& Mar83	\\
IRAS~13224-3809 	& 4445 -- 8150	& 0.38 $\pm$ 0.03	& 84.0 $\pm$ 2.0	& Kay99	\\
IRAS~13349+2438	& 3200 -- 8320 	& 23 -- 35	& 124.0 $\pm$ 5.0	& Wil92	\\
K~348-7	& 3200 -- 8600	& 0.25 $\pm$ 0.22	& 42.0 $\pm$ 25.0	& Ber90	\\
MCG-3-34-64	& 6015 -- 7270	& 0.50 $\pm$ 0.20	& 75.0 $\pm$ 25.0	& You96	\\
MCG-6-30-15	& 5000 -- 5800 	& 4.06 $\pm$ 0.45	& 120.0 $\pm$ 3.0	& Bri90	\\
MCG+8-11-11	& 3800 -- 5600	& 0.69 $\pm$ 0.46	& 166.4 $\pm$ 19.0	& Mar83	\\
Mrk~1014	& 3200 -- 8600	& 1.37 $\pm$ 0.40	& 21.0 $\pm$ 8.0	& Ber90	\\
Mrk~1018	& 4180 – 6903	& 0.28 $\pm$ 0.05	& 165.1 $\pm$ 5.2	& Goo89	\\
Mrk~1066	& 3200 -- 6200 	& $>$ 1.99	& 135.1 $\pm$ 2.6	& Kay94	\\
Mrk~110	& 3200 -- 8600 	& 0.17 $\pm$ 0.08	& 18.0 $\pm$ 15.0	& Ber90	\\
Mrk~1239	& 3800 -- 5600 	& 4.09 $\pm$ 0.14	& 136.0 $\pm$ 1.0	& Mar83	\\
Mrk~1298	& 3200 -- 8600	& 0.32 $\pm$ 0.14	& 76.0 $\pm$ 12.0	& Ber90	\\
Mrk~1383	& 3200 -- 8600	& 0.49 $\pm$ 0.19	& 58.0 $\pm$ 11.0	& Ber90	\\
Mrk~176	& 3800 -- 5600 	& $>$ 0.54	& 146.3 $\pm$ 8.2	& Mar83	\\
Mrk~231	& 3800 -- 5600	& 2.87 $\pm$ 0.08	& 95.1 $\pm$ 0.8	& Mar83	\\
Mrk~273	& 3800 -- 5600	& $>$ 0.37	& 66.7 $\pm$ 52.0	& Mar83	\\
Mrk~279	& 6000 -- 7500 	& 0.48 $\pm$ 0.04	& 58.9 $\pm$ 2.4	& Smi02	\\
Mrk~3	& 5000 	& 7.77 -- 8.61	& 167.0	& Tra95	\\
Mrk~304	& 3200 -- 8600 	& 0.58 $\pm$ 0.14	& 107.0 $\pm$ 7.0	& Ber90	\\
Mrk~335	& 3800 -- 5600 	& 0.48 $\pm$ 0.11	& 107.6 $\pm$ 6.9	& Mar83	\\
Mrk~34	& 3200 -- 6200 	& $>$ 3.92	& 53.0 $\pm$ 4.5	& Kay94	\\
Mrk~348	& 3200 -- 6200	& $>$ 9.09	& 78.9 $\pm$ 1.3	& Kay94	\\
Mrk~359	& 4214 -- 6937	& 0.46 $\pm$ 0.02	& 112.0 $\pm$ 1.2	& Goo89	\\
Mrk~463	& 3200 -- 6200	& $>$ 10	& 84.0	& Tra95	\\
Mrk~478	& 3800 -- 5600 	& 0.46 $\pm$ 0.15	& 44.9 $\pm$ 9.5	& Mar83	\\
Mrk~486	& 3800 -- 5600 	& 3.40 $\pm$ 0.14	& 136.8 $\pm$ 1.2	& Mar83	\\
Mrk~509	& 3800 -- 5600 	& 1.09 $\pm$ 0.15	& 146.5 $\pm$ 4.0	& Mar83	\\
Mrk~573	& 3200 -- 6200 	& $>$ 5.56	& 48.0 $\pm$ 2.0	& Kay94	\\
Mrk~590	& 3800 -- 5600	& 0.32 $\pm$ 0.30	& 105.9 $\pm$ 26.6	& Mar83	\\
Mrk~6	& 3800 -- 5600	& 0.54 $\pm$ 0.15	& 141.2 $\pm$ 8.0	& Mar83	\\
Mrk~705	& 4700 -- 7200 	& 0.46 $\pm$ 0.07	& 49.3 $\pm$ 6.5	& Smi02	\\
Mrk~707	& 3800 -- 5600 	& 0.20 $\pm$ 0.24	& 140.9 $\pm$ 52.0	& Mar83	\\
Mrk~766	& 4500 -- 7100 	& 3.10 $\pm$ 0.80	& 90.0	& Bat11 	\\
Mrk~78	& 3200 -- 6200 	& 21.0 $\pm$ 9.0	& 75.3 $\pm$ 11.2	& Kay94	\\
Mrk~79	& 3800 -- 5600 	& 0.34 $\pm$ 0.19	& 0.4 $\pm$ 16.2	& Mar83	\\
Mrk~841	& 4500 -- 7500	& 1.00 $\pm$ 0.03	& 103.4 $\pm$ 1.0	& Smi02	\\
Mrk~876	& 3200 -- 8600	& 0.50 $\pm$ 0.14	& 86.0 $\pm$ 8.0	& Ber90	\\
Mrk~877	& 3200 -- 8600	& 0.95 $\pm$ 0.20	& 69.0 $\pm$ 6.0	& Ber90	\\
Mrk~896	& 3800 -- 5600 	& 0.55 $\pm$ 0.13	& 1.9 $\pm$ 7.1	& Mar83	\\
NGC~1068	& 3500 -- 5200 	& 16.0 $\pm$ 2.0	& 95.0	& Mil83	\\
NGC~1097	& 5100 -- 6100 	& 0.26 $\pm$ 0.02	& 178 $\pm$ 2.0	& Bar99	\\
NGC~1320	& 3200 -- 6300	& $>$ 0.38	& 91.3 $\pm$ 3.0	& Kay94	\\
NGC~1365	& 5000 -- 5900 	& 0.91 $\pm$ 0.18	& 157 $\pm$ 6.0	& Bri90	\\
NGC~1386	& 3800 -- 5600	& $>$ 0.62	& 34.3 $\pm$ 7.1	& Mar83	\\
NGC~1566	& 3800 -- 5600	& 0.60 $\pm$ 0.24	& 52.6 $\pm$ 11.6	& Mar83	\\
NGC~1667	& 5100 -- 6100 	& 0.35 -- 9.8	& 94.0 $\pm$ 1.0	& Bar99	\\
NGC~2110	& 5200 -- 6200	& 18.4	& 70.0	& Mor07	\\
NGC~2992	& 3800 -- 5600	& $>$ 3.32	& 33.3 $\pm$ 1.6	& Mar83	\\
NGC~3227	& 5000 	& 1.3 $\pm$ 0.1	& 133 $\pm$ 3.0	& Sch85	\\
NGC~3516	& 4500 -- 7500	& 0.15 $\pm$ 0.04	& 30.1 $\pm$ 8.0	& Smi02	\\
NGC~3783	& 4500 -- 7500 	& 0.52 $\pm$ 0.02	& 135.5 $\pm$ 1.0	& Smi02	\\
NGC~4051	& 4500 -- 7500 	& 0.55 $\pm$ 0.04	& 82.8 $\pm$ 1.8	& Smi02	\\
NGC~4151	& 4600 -- 7400	& 0.26 $\pm$ 0.08 	& 62.8 $\pm$ 8.4	& Mar83	\\
NGC~424	& 5000 -- 5900	& $>$ 1.53	& 59.0 $\pm$ 5.0	& Kay94	\\
NGC~4388	& 3800 -- 5600	& 2.0 -- 39.7	& 93.0 $\pm$ 29.0	& Kay94	\\
NGC~4395	& 5100 -- 6100	& 0.64 $\pm$ 0.03	& 30.0 $\pm$ 2.0	& Bar99	\\
NGC~4507	& 5400 -- 5600 	& 14.8 -- 16.3	& 37.0 $\pm$ 2.0	& Mor00	\\
NGC~4593	& 6000 -- 7600 	& 0.14 $\pm$ 0.05	& 109.5 $\pm$ 10.8	& Smi02	\\
NGC~5506	& 3200 -- 6200 	& $>$ 2.6	& 72.8 $\pm$ 4.5	& Kay94	\\
NGC~5548	& 6000 -- 7500 	& 0.69 $\pm$ 0.01	& 33.2 $\pm$ 0.5	& Smi02	\\
NGC~5643	& 5000 -- 5900 	& $>$ 0.75	& 57.0 $\pm$ 9.0	& Bri90	\\
NGC~7172	& 5000 -- 5900	& $>$ 2.10	& 96.0 $\pm$ 3.0	& Bri90	\\ 
NGC~7213	& 6000 -- 7500	& 0.09 $\pm$ 0.02	& 146.0 $\pm$ 7.6	& Smi02	\\
NGC~7314	& 4500 -- 7500	& 3.00 $\pm$ 1.00	& 35.0	& Lum04	\\
NGC~7469	& 6000 -- 7500 	& 0.18 $\pm$ 0.01	& 76.8 $\pm$ 1.7	& Smi02	\\
NGC~7582	& 3800 -- 4900 	& $>$ 1.35		& 157 $\pm$ 5.0		& Bri90	\\
NGC~7674	& 3200 -- 6200 	& 6.54 -- 7.6	& 31.0	& Tra95	\\
PG~0026+129	& 3200 -- 8600	& 0.27 $\pm$ 0.17	& 83.0 $\pm$ 17.0	& Ber90	\\
PG~1001+054	& 3200 -- 8600	& 0.77 $\pm$ 0.22	& 74.0 $\pm$ 8.0	& Ber90	\\
PG~1211+143	& 4700 -- 7200 	& 0.27 $\pm$ 0.04	& 137.7 $\pm$ 4.5	& Smi02	\\
PG~1244+026	& 3200 -- 8600	& 0.48 $\pm$ 0.25	& 108.0 $\pm$ 15.0	& Ber90	\\
PG~1302-102	& 3200 -- 8600	& 0.18 $\pm$ 0.15	& 26.0 $\pm$ 24.0	& Ber90	\\
PG~1411+442	& 3200 -- 8600	& 0.76 $\pm$ 0.17	& 61.0 $\pm$ 6.0	& Ber90	\\
PG~1435-067	& 3200 -- 8600	& 1.44 $\pm$ 0.29	& 27.0 $\pm$ 6.0	& Ber90	\\
PG~1448+273 	& 3200 -- 8600	& 0.27 $\pm$ 0.14	& 67.0 $\pm$ 15.0	& Ber90	\\
PG~1626+554 	& 3200 -- 8600	& 0.59 $\pm$ 0.19	& 10.0 $\pm$ 9.0	& Ber90	\\
PG~1700+518	& 3200 -- 8600	& 0.54 $\pm$ 0.10	& 56.0 $\pm$ 5.0	& Ber90	\\
PG~2251+113 	& 3200 -- 8600	& 0.89 $\pm$ 0.22	& 50.0 $\pm$ 7.0	& Ber90	\\
TON~1388	& 3200 -- 8600	& 0.23 $\pm$ 0.11	& 142.0 $\pm$ 13.0	& Ber90	\\
TON~1542	& 3200 -- 8600	& 0.61 $\pm$ 0.12	& 118.0 $\pm$ 6.0	& Ber90	\\
TON~1565	& 3200 -- 8600	& 0.31 $\pm$ 0.14	& 42.0 $\pm$ 13.0	& Ber90	\\
VII~Zw244	& 3200 -- 8600	& 1.08 $\pm$ 0.37	& 144.0 $\pm$ 10.0	& Ber90	\\
      \hline
    \caption{Recorded average white light continuum polarization of 100/124 AGN. 
	    Legend: Mar83 - \citet{Martin1983}; Mil83 - \citet{Miller1983}; Sch85 - \citet{Schmidt1985}; 
	    Goo89 - \citet{Goodrich1989}; Ber90 - \citet{Berriman1990}; Bri90 - \citet{Brindle1990}; 
	    Wil92 - \citet{Wills1992}; Kay94 - \citet{Kay1994}; Tra95 - \citet{Tran1995}; 
	    You96 - \citet{Young1996}; Bar99 - \citet{Barth1999}; Kay99 - \citet{Kay1999}; 
	    Ogl99 - \citet{Ogle1999}; Ale00 - \citet{Alexander2000}; Mor00 - \citet{Moran2000};
	    Smi02 - \citet{Smith2002}; Sch03 - \citet{Schmid2003}; Lum04 - \citet{Lumsden2004}; 
	    Mor07 - \citet{Moran2007}; and Bat11 - \citet{Batcheldor2011}.}	    	    
    \label{Table:Polarization}
\end{longtable}

\begin{longtable}{|c|c|}
      \hline {\bf Object}	& {\bf Pol. degree (\%)}\\
   \hline ESO~323-G077		& 7.5\\
	  Fairall~51		& 4.12 $\pm$ 0.03\\
	  Mrk~231		& 2.87 $\pm$ 0.08\\
	  Mrk~486		& 3.40 $\pm$ 0.14\\
	  Mrk~766		& 3.10 $\pm$ 0.80\\
	  Mrk~1239		& 4.09 $\pm$ 0.14\\
	  NGC~3227		& 1.3 $\pm$ 0.1\\
	  NGC~4593		& 0.14 $\pm$ 0.05\\
	  NGC~5548		& 0.69 $\pm$ 0.01\\
      \hline
    \caption{Sub-list from Tab~\ref{Table:Polarization} of type-1 Seyfert galaxies 
	    exhibiting optical polarization spectra similar to those of type 
	    2 objects (polar scattering dominated AGN).}
    \label{Table:PolScattDomAGN}
\end{longtable}

\begin{longtable}{|c|c|}
      \hline {\bf Object}	& {\bf Pol. degree (\%)}\\
      \hline ESO~323-G077	& 7.5\\
	  Fairall~51		& 4.12 $\pm$ 0.03\\
	  IC~4329A		& 5.80 $\pm$ 0.26\\
	  MCG-6-30-15		& 4.06 $\pm$ 0.45\\
	  Mrk~231		& 2.87 $\pm$ 0.08\\
	  Mrk~486		& 3.40 $\pm$ 0.14\\
	  Mrk~766		& 3.10 $\pm$ 0.80\\
	  Mrk~1239		& 4.09 $\pm$ 0.14\\
	  NGC~7314		& 3.00 $\pm$ 1.00\\
      \hline
    \caption{Sub-list from Tab~\ref{Table:Polarization} of type-1 Seyfert galaxies 
	    exhibiting high ($>$ 2~\%) optical polarization.}
    \label{Table:HighPolS1}
\end{longtable}

\bsp

\label{lastpage}

\end{document}